\documentclass[twocolumn]{aastex62}
\usepackage{amsmath}
\usepackage{longtable}
\usepackage[figuresright]{rotating} 
\usepackage{hyperref}
\usepackage{pifont}% http://ctan.org/pkg/pifont
\usepackage{enumerate}% http://ctan.org/pkg/enumerate
\usepackage{mathrsfs}

\begin{document}

\title{{\bf Stellar obliquities in exoplanetary systems}}

\correspondingauthor{Simon H.\ Albrecht}
\email{albrecht@phys.au.dk}

\author[0000-0003-1762-8235]{Simon H.\ Albrecht}
\affil{Stellar Astrophysics Centre, Department of Physics and Astronomy, Aarhus University, Ny Munkegade 120, 8000 Aarhus C, Denmark}

\author[0000-0001-9677-1296]{Rebekah I.\ Dawson}
\affiliation{Department of Astronomy \& Astrophysics, Center for Exoplanets and Habitable
Worlds, The Pennsylvania State University, University Park, PA 16802, USA}

\author[0000-0002-4265-047X]{Joshua N.\ Winn}
\affiliation{Department of Astrophysical Sciences, Princeton University, Princeton, NJ 08544, USA}

\begin{abstract}

The rotation of a star and the revolutions of its planets
are not necessarily aligned.
This article reviews the measurement techniques, key findings, and theoretical
interpretations related to the obliquities (spin-orbit angles)
of planet-hosting stars.
The best measurements are for stars with
short-period giant planets,
which have been found on prograde, polar, and retrograde orbits.
It seems likely that dynamical processes such as planet-planet scattering
and secular perturbations are responsible for tilting the orbits of close-in giant planets,
just as those processes are implicated in exciting orbital eccentricities.
The observed dependence of the obliquity
on orbital separation, planet mass, and stellar structure
suggests that in some cases, tidal dissipation damps a star's obliquity within its
main-sequence lifetime.
The situation is not as clear for stars with smaller or wider-orbiting
planets.  Although the earliest measurements of such systems tended to find
low obliquities, some glaring exceptions are now known in which the star's
rotation is misaligned with respect to the coplanar orbits of multiple planets.
In addition, statistical analyses based on projected rotation velocities and photometric
variability have found a broad range of obliquities for
F-type stars hosting compact multiple-planet systems.
The results suggest it is unsafe to assume that stars and their
protoplanetary disks are aligned. Primordial misalignments
might be produced by neighboring stars or more
complex events that occur during the epoch of planet formation.
\end{abstract}


\section{Introduction}
\label{sec:intro}

Soon after the earliest observations of sunspots by Galileo, Scheiner, Harriot, and Fabricius,
it became clear that the Sun's equator is nearly aligned with the ecliptic
\citep{Casanovas1997}. A modern value for the Sun's obliquity, based on
helioseismology, is $7.155\pm 0.002^\circ$ \citep{BeckGiles2005}. This relatively low
solar obliquity was part of the body of evidence that led Laplace to
his ``nebular theory'' for the formation of the Solar System, which was
incorrect but is remembered for the theoretical debut of the 
protoplanetary disk. Another fact that has inspired theorists
is that the Sun's obliquity seems significantly
higher than the root-mean-squared mutual inclination of 1.9$^\circ$ between
the orbits of the Sun's eight planets. Among the
proffered explanations are a close encounter with another star \citep{Heller1993},
a torque resulting from the motion of the protoplanetary disk through
the interstellar medium \citep{Wijnen+2017},
spin-axis precession driven by an undiscovered outer planet
\citep{BaileyBatyginBrown2016,Lai2016,GomezDeiennoMorbidelli2017},
an asymmetry of the solar wind \citep{Spalding2019},
and the imprint of a nearby supernova \citep{Zwart+2018}.

Exoplanetary systems show a wider range of orbital characteristics
than had been expected based on observations and interpretations of the properties of the Solar
System \citep[see, e.g.,][for reviews]{WinnFabrycky2015,ZhuDong2021}.
One of the goals of exoplanetary science is to understand
the physical processes responsible for this architectural
diversity.
Some examples of surprising systems are those with close-orbiting giant planets \citep{MayorQueloz1995},
planets on highly eccentric orbits \citep{Latham+1989,MarcyButler1996},
miniature systems of multiple planets on tightly packed orbits
\citep{Lissauer+2011,Fabrycky+2014} and, the subject of this review,
stars with large obliquities \citep{Hebrard+2008,Winn+2009_X03}.

\begin{figure}
  \begin{center}
    \includegraphics[width=5.5cm]{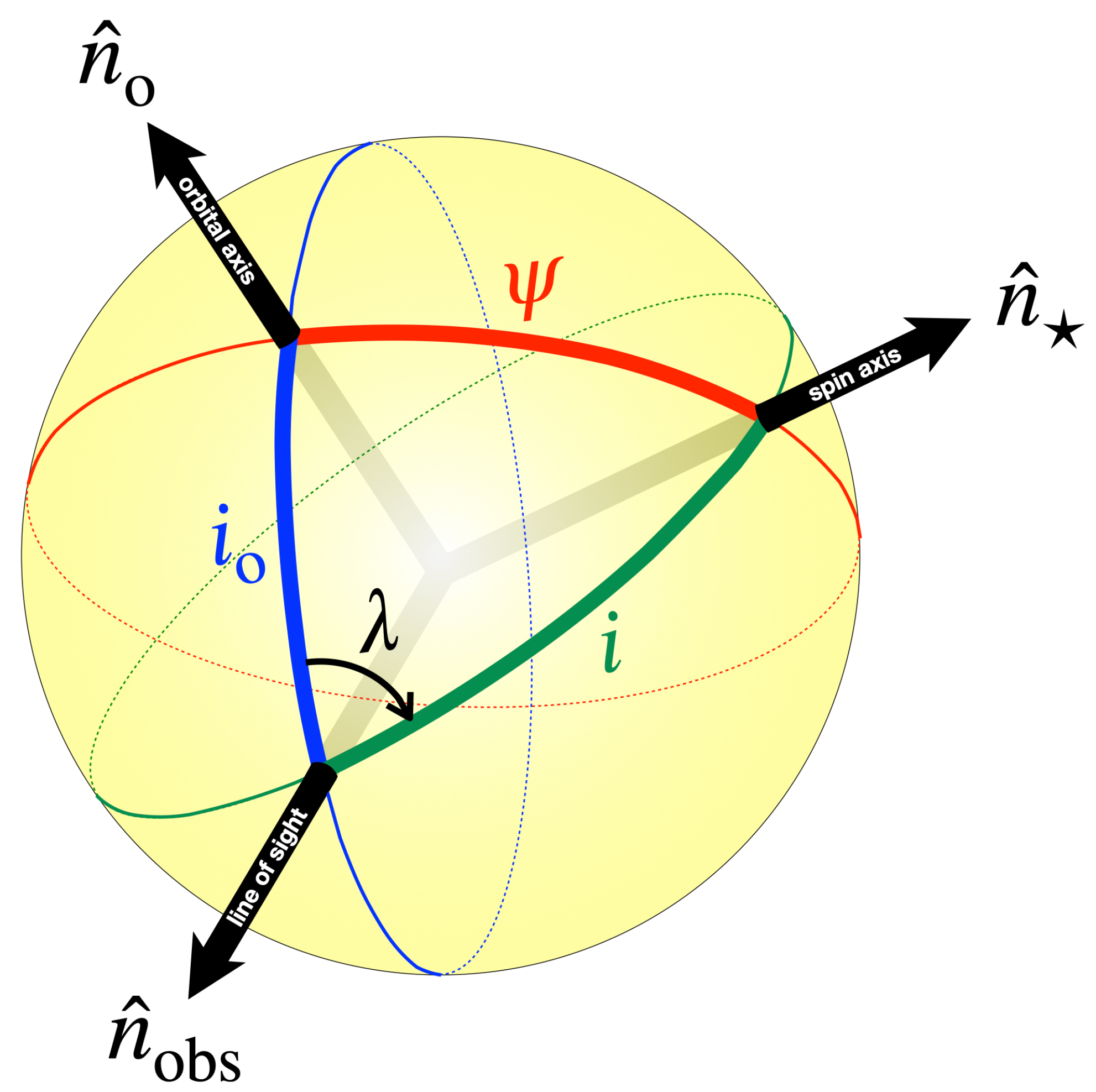}
    \caption  {\label{fig:geometry} Angles that specify
    the orientation of the spin and orbital angular momentum
    vectors.
    The obliquity is $\psi$,
    the orbital inclination is $i_{\rm o}$, and the inclination
    of the stellar rotation axis is $i$. Many authors define
    a Cartesian coordinate system with $\hat{z} = \hat{n}_{\rm obs}$
    and $\hat{y}$ aligned with the sky projection of $\hat{n}_{\rm o}$,
    although it is sometimes more
    convenient to align $\hat{y}$ with the sky projection of $\hat{n}_\star$.}
    \end{center}
\end{figure}

Measuring a star's obliquity is challenging
because ordinary observations lack the angular resolution to discern any details
on the spatial scale of the stellar surface. Nevertheless, using
an array of techniques, obliquity measurements are available for $\sim$10$^2$
stars, and statistical inferences about obliquity distributions
have been drawn from samples of $\sim$10$^3$ stars.
Prograde, polar, and retrograde orbits
have been found, and a few patterns have emerged relating obliquities
to stellar mass, planetary mass,
orbital distance, and transit multiplicity. 
There is unlikely to be a simple explanation for all the results.
Misalignments might occur before, during, or after the epoch
of planet formation. They might be linked to specific dynamical
events in a planet's history, such as planet-planet scattering
or high-eccentricity migration, or they might be
the outcome of general processes affecting stars and protoplanetary
disks irrespective of the planets that eventually form.

This article reviews the current status of the
observations and theories regarding the obliquities of stars with planets.
Section~\ref{sec:geometry} introduces the relevant geometry and terminology.
Section~\ref{sec:methods results} describes the measurement techniques
and key findings.
Section~\ref{sec:theory} discusses
the proposed physical mechanisms that can excite and damp obliquities,
and their success or failure in matching the observations.
Section~\ref{sec:summary} summarizes the main observational findings and their relationships to theories and gives some recommendations
for future work in this area.

%%%%%%%%%%%%%%%%%%%%%%%%%%%%%%%%%%%%%%%%%%%%%%%%%%%%%%%%%%%%%%%%%%%%%%%%%%

\tableofcontents

\begin{figure*}
  \begin{center}
    \includegraphics[width=0.95\textwidth]{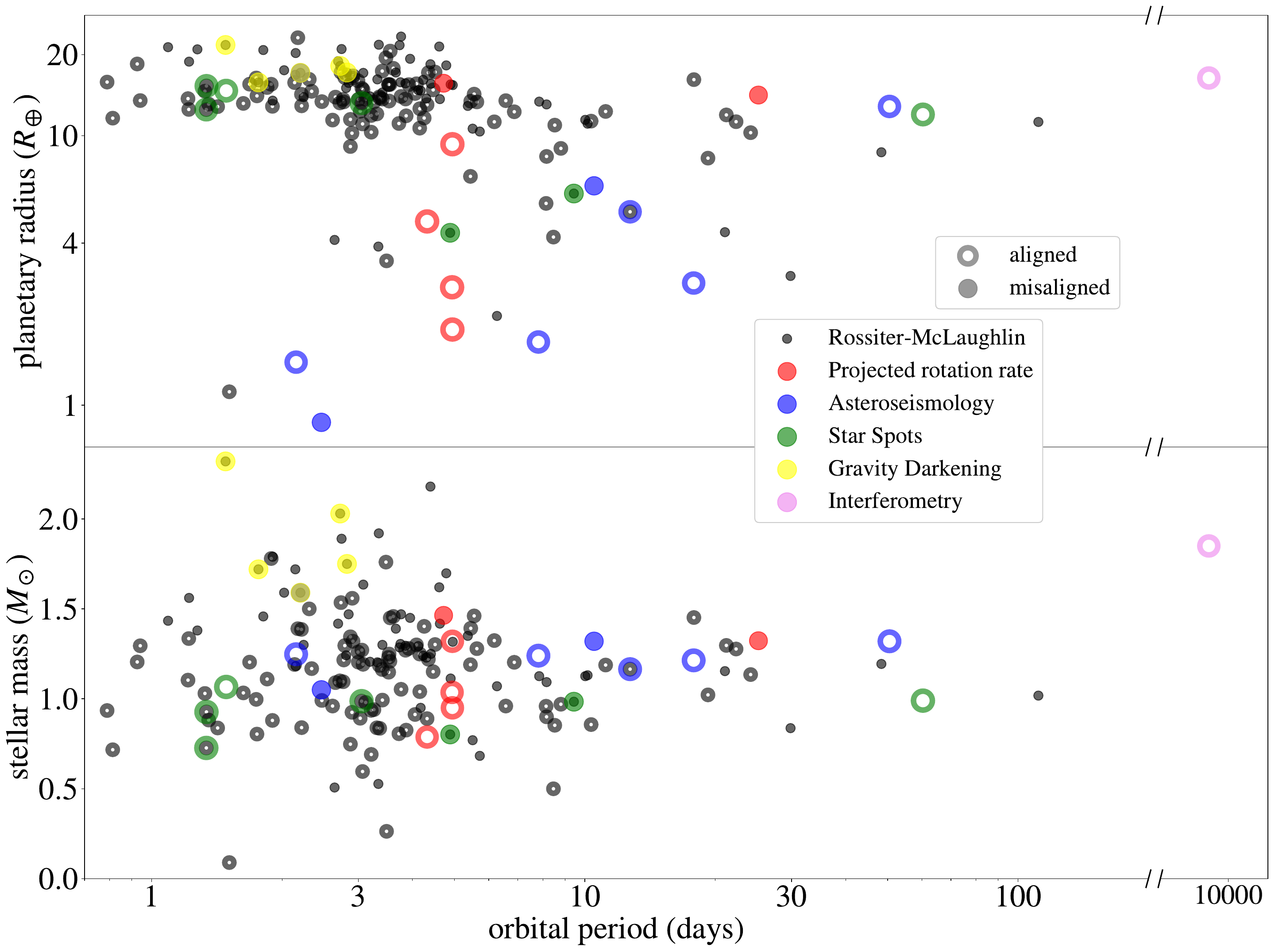}
     \caption {\label{fig:obli_techniques}
     {\bf Parameter space of obliquity measurement methods.} Each point
     represents an obliquity measurement,
     with a location that specifies the orbital period and
     either the planet's radius (top panel) or the star's mass (bottom panel).
     The points are color-coded by method.
     Solid symbols are for misaligned stars (by more than $10^\circ$ with 
     a $>$3-$\sigma$ departure from zero);
     open symbols are for well-aligned stars or ambiguous cases.
     The RM, starspot, and gravity-darkening methods require observations during transits, making them less applicable to systems with smaller planets
     or longer periods. The gravity-darkening method requires fast rotators, i.e., high-mass stars, while the starspot method is most applicable to low-mass stars with large and long-lived starspots.
     The asteroseismic and projected rotation velocity methods require
     a transiting planet but do not require intensive observations
     conducted during transits, making them applicable to planets of all
     types. The asteroseismic method requires moderately rapid rotation
     and long-lived pulsation
     modes, which generally occur for stars somewhat more massive than the Sun.
     Similarly, the projected rotation velocity method requires moderately rapid
     rotation, which is associated with more massive stars.
     The interferometric method requires very bright and rapidly
     rotating stars, as well as a constraint on the planetary orbital
     inclination. Also important, though not conveyed in this diagram,
     is that the methods differ in the achievable precision and
     the severity of parameter degeneracies.}
  \end{center}
\end{figure*}

\section{Geometry}
\label{sec:geometry}

Figure~\ref{fig:geometry} illustrates the unit vectors $\hat{n}_\star$, $\hat{n}_{\rm o}$,
and $\hat{n}_{\rm obs}$ that specify
the directions of the stellar angular momentum, the orbital angular momentum,
and the line of sight to the observer, respectively.
The obliquity $\psi$ is the angle between $\hat{n}_\star$ and $\hat{n}_{\rm o}$.
The angles $i$ and $i_{\rm o}$ are the line-of-sight inclinations of the stellar and
orbital angular momentum vectors, and $\lambda$ is the position angle between the sky projections
of those two vectors, measured clockwise from $\hat{n}_{\rm o}$ to $\hat{n}_\star$.
With these definitions,\footnote{Some authors use the opposite sign convention, measuring the position angle {\it counterclockwise}
from $\hat{n}_{\rm o}$ to $\hat{n}_\star$ and denoting the angle $\beta$ instead
of $\lambda$.}
\begin{align}
\hat{n}_\star \cdot \hat{n}_{\rm o} = \cos \psi &=&
  \cos i \cos i_{\rm o} + \sin i \sin i_{\rm o} \cos\lambda, \\
\left( \hat{n}_\star \times \hat{n}_{\rm o} \right) \cdot \hat{n}_{\rm obs} &=&
\sin i \sin i_{\rm o} \sin \lambda.
\end{align}

\begin{figure*}
  \begin{center}
    \includegraphics[width=1.0\textwidth]{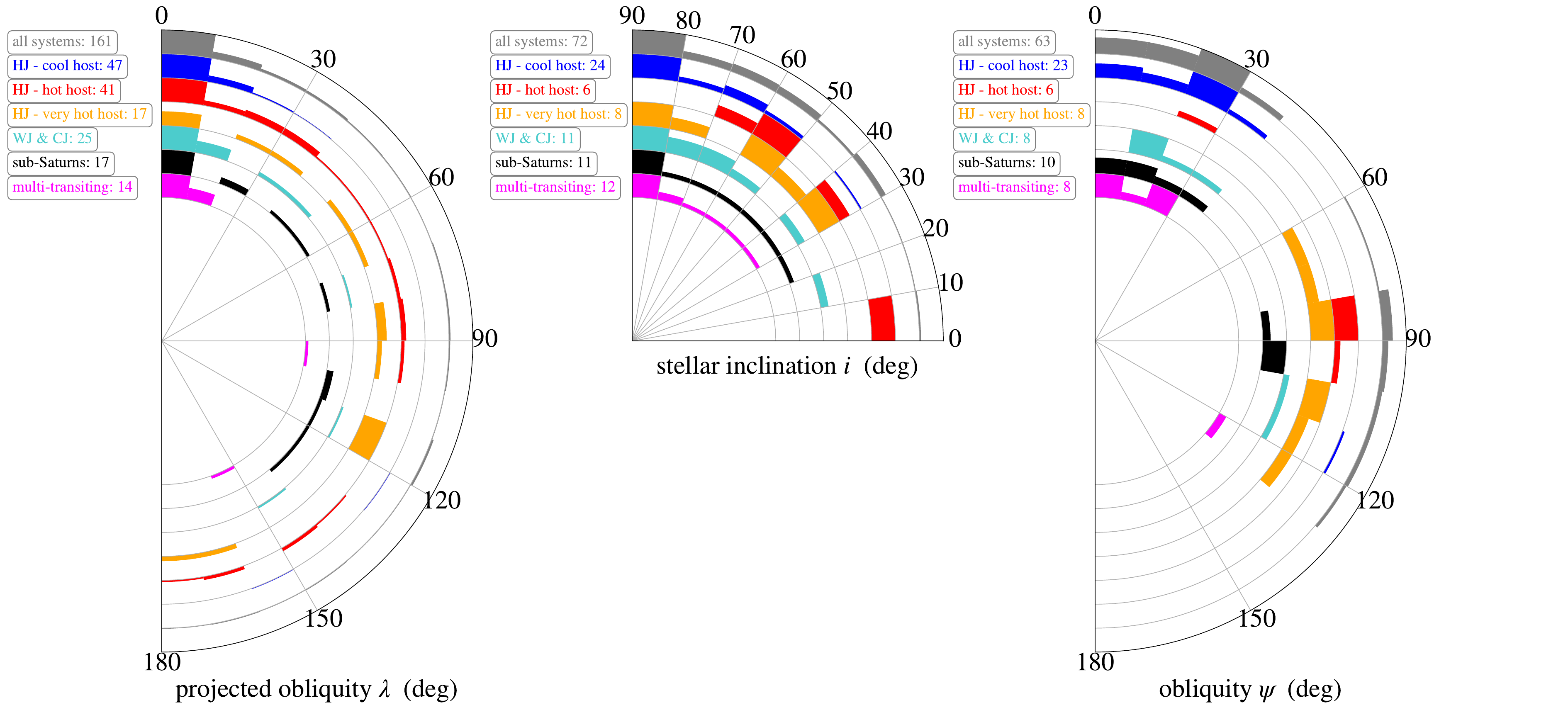}
    \caption {\label{fig:polar_hist}
    {\bf Histograms of host star obliquities.} {\it Left}:~Projected obliquities ($\lambda$), folded onto the range $[0,180^\circ]$. 
    {\it Middle}:~Stellar inclination measurements ($i$), folded onto the range $[0,90^\circ]$.  
    {\it Right:}~Three-dimensional obliquities ($\psi$), for the cases in which
    $\lambda$ and $i$ have both been measured. 
    The histograms are color-coded according to the system's characteristics.
    Stars are designated as cool, hot, or very hot, using effective temperature
    boundaries of 6250\,K and 7000\,K.
    Planets with masses exceeding 0.3\,M$_{\rm Jup}$ are designated hot Jupiters (HJ) if $a/R<10$, and warm/cold Jupiters (WJ/CJ) if $a/R>10$.
    Planets with masses $\lesssim$\,$0.3~M_{\rm Jup}$ are designated sub-Saturns.
    Readers preferring more traditional histograms are directed to Appendix~\ref{app:histrograms}.
    }
  \end{center}
\end{figure*}

As explained in Section~\ref{sec:methods results},
some techniques are capable of measuring
$\sin i$ and $\sin i_{\rm o}$ but provide
no information about $\lambda$.
Some other techniques are mainly sensitive to $\lambda$. 
Thus, to determine $\psi$ for an individual system,
it is usually necessary to combine the results
from more than one measurement technique.
On the other hand, to study the obliquity distribution of a
population of stars, it suffices to obtain information about either $\lambda$ or
$i$ for each star.
Current statistical studies rely on stars with transiting planets,
for which $i_{\rm o}\approx 90^\circ$ is guaranteed.
If a population of transit-hosting
stars were randomly oriented, then $\lambda$ and $\cos i$ would be
uniformly distributed. If, instead, the stars had low obliquities, then
only low values of $\lambda$ and $\cos i$ would be observed.
In between these two extremes, the measured width of the distribution of either
$\lambda$ or $\cos i$ can be used to determine
the width of the obliquity distribution.

For statistical analyses,
two useful references are \cite{FabryckyWinn2009} 
and \cite{MunozPerets2018}.
The former authors provided analytic
formulas for the
conditional probability densities $p(\psi|\lambda)$
and $p(\lambda|\psi)$ under the assumption of 
random orientations. They also showed how to use measurements
of $\lambda$ to model the obliquity distribution of a population
of stars as a von Mises-Fisher (vMF)
distribution,
\begin{eqnarray}
    p(\hat{n}_\star) &\propto& \exp(\kappa\,\hat{n}_\star\cdot \hat{n}_{\rm o}) \\
    \frac{dp}{d\psi} &=& \frac{\kappa}{2\sinh\kappa} \exp(\kappa\cos\psi) \sin\psi.
\end{eqnarray}
This is a widely used model in directional statistics, which resembles
a two-dimensional Gaussian distribution wrapped around a 
sphere. For small values of the concentration parameter $\kappa$,
the distribution approaches an isotropic distribution.
For large values of $\kappa$, the distribution of $\psi$
approaches a Rayleigh function with a width of
$\sigma = \kappa^{-1/2}$. 
\cite{MunozPerets2018} extended this framework to include
information about $i$ in addition to $\lambda$.

\section{Methods and Key Findings}
\label{sec:methods results}

\begin{table*}
\caption{{\bf Key results and systems of special interest.}
The first column describes an observational finding, and the second column indicates the main measurement technique. The third column specifies
the section of this article in which the trend is discussed, and the fourth column
provides the key citations to the literature.}
\label{tab:obs_trends}
    \begin{tabular}{lcll} 
    \hline 
  Result & Method & Section & Ref.   \\ 
    \hline
    \hline
Hot stars with hot Jupiters have high obliquities. & RM, VSI, QPV & \ref{sec:teff} & 1,2  \\ 
The highest-mass hot Jupiters have lower obliquities.  & RM & \ref{sec:oblique_mass} & 3 \\ 
Cool stars with Neptunes or warm Jupiters have high obliquities.  & RM & \ref{sec:oblique_ar} & 4     \\
Cool stars with hot Jupiters are sometimes aligned within $1^\circ$. & RM &  \ref{sec:well_aligned} &  \\
Tidal effects appear to damp obliquities. & RM & \ref{sec:tides} & 1,4 \\ 
Stars younger than 100\,Myr tend to be well-aligned?  & RM/VSI/INT &  \ref{sec:newborn} &  \\
Cool stars with compact multi-planet systems have $\psi \lesssim$\,$30^\circ$  & RM/SC/AS/QPV/VSI & \ref{sec:multitransits} & 5--11  \\
Hot stars with compact multi-planet systems have high obliquities & QPV/VSI &  \ref{sec:vsini} & 12 \\
A preponderance of perpendicular planets? & RM/VSI/QPV/GD/AS  &  \ref{sec:ppp} & 13 \\
HD\,80606: A high obliquity from Kozai-Lidov cycles and tidal friction? & RM & \ref{sec:orbit_plane_change}  &  14,15 \\
Kepler-56: A high obliquity from precession induced by an outer planet?  &  AS & \ref{sec:seismic}  & 16  \\
K2-290: A high obliquity from primordial disk misalignment?  &  RM/VSI & \ref{sec:primordial}  & 17  \\
    \hline 
    \end{tabular}
    \tablecomments{{\bf Techniques}:
    RM, Rossiter-McLaughlin effect.
    VSI, $v\sin i$ (projected rotation velocity).
    QPV, quasiperiodic photometric variability.
    AS, asteroseismology.
    SC, spot crossings.
    GD, gravity darkening.
    INT, interferometry.}
    \tablerefs{
    1.\ \cite{Winn+2010},
    2.\ \cite{Schlaufman2010},
    3.\ \cite{Hebrard+2011b},
    4.\ \cite{AlbrechtWinnJohnson+2012},
    5.\ \cite{AlbrechtWinnMarcy+2013},
    6.\ \cite{MortonWinn2014},
    7.\ \cite{Campante+2016},
    8.\ \cite{Mazeh+2015KOIs},
    9.\ \cite{LiWinn2016},
    10.\ \cite{Winn+2017},
    11.\ \cite{MunozPerets2018},
    12.\ \cite{Louden+2021},
    13.\ \cite{Albrecht+2021},
    14.\ \cite{WuMurray2003},
    15.\ \cite{Hebrard+2010},
    16.\ \cite{Huber+2013},
    17.\ \cite{Hjorth+2021}
 }
 \end{table*}

The main challenge in measuring any of the angles in Figure~\ref{fig:geometry}
is that stars are almost always spatially unresolved.  We can only
observe a star's flux and spectrum integrated over its entire visible hemisphere.
Fortunately, some characteristics of the disk-integrated flux and spectrum depend
on the star's orientation in space.
One such characteristic
is the {\bf rotational Doppler broadening} of its spectral absorption lines,
which is quantified by $v\sin i$, the projected equatorial rotation velocity  (\S\,\ref{sec:vsini}).
Another observable that is related to a star's orientation
is the {\bf amplitude of photometric variability}
due to rotating starspots, which is expected to vary roughly in proportion
to $\sin i$ (\S\,\ref{sec:QPV}).
A third type of data that bears information about a star's orientation
is the fine structure in its {\bf asteroseismic oscillation spectrum}.
The inclination $i$ affects the
relative amplitudes of the modes within each rotationally split multiplet
(\S\,\ref{sec:seismic}).
When we also have knowledge of $i_{\rm o}$
(based on transit observations, direct images, or astrometric
measurements), then the constraints on $i$ from
any of the three techniques described above
allow us to place constraints on the stellar obliquity.

These {\bf inclination-based methods} for obliquity determination
have important limitations.
Because of the north/south symmetry of the star,
they cannot distinguish between $i$ and $180^\circ - i$,
leading to a twofold degeneracy in the star's orientation.
In particular, we cannot tell whether a star has prograde or
retrograde rotation with respect
to the line of sight or with respect to the planetary orbit.\footnote{For transiting planets,
a similar degeneracy afflicts measurements of $i_{\rm o}$, although
this is usually a minor concern because
the geometrical requirement for transits implies that $i_{\rm o}$ is never
far from $90^\circ$.}
Another limitation is that the inclination-based methods
tend to give weak constraints at high
inclinations,
because of flattening of the sine function as $i\rightarrow 90^\circ$.
Even if $\sin i$ is constrained to be in the narrow range from 0.9 to 1,
the inclination can be any value between 64 and 116$^\circ$.
This problem arises often, because high inclinations are common.
In a sample of randomly oriented stars, we expect 44\% of the stars to have $\sin i>0.9$.

The other main class of methods for measuring the obliquity
relies on a transiting planet to provide
spatially resolved information, as its shadow scans across the stellar disk.
The intensity and emergent spectrum vary across
the star's photosphere in a manner that depends on the star's orientation.
For example, stellar rotation causes the radial velocity of its photosphere
to exhibit a gradient from the approaching side to the receding side.
When a transiting planet hides a portion of the stellar disk,
the corresponding radial-velocity component is diminished 
in the disk-integrated stellar spectrum, leading
to line-profile distortions known as 
the {\bf Rossiter-McLaughlin effect} (\S~\ref{sec:rm}).
Another technique is based on detecting the glitches in the
light curve whenever a transiting planet occults a starspot (or any kind of
inhomogeneity) on the stellar disk. Observations of these {\bf starspot anomalies}
can sometimes
be used to constrain the stellar obliquity (\S~\ref{sec:starspot-tracking}). 
A third technique is based on {\bf gravity darkening}.
The equatorial zone of a rapidly rotating
star is centrifugally lifted to higher elevation,
lowering its temperature and intensity
relative to the polar regions.  This effect breaks the usual
circular symmetry of the intensity profile across the
stellar disk, which in turn causes a distortion
of the transit light curve (\S~\ref{sec:gravity}).
The circular symmetry is also broken by relativistic effects known as
{\bf rotational Doppler boosting} (\S~\ref{sec:Rotational Doppler boosting}).

These {\bf transit-based methods} are usually more sensitive
to $\lambda$ than they are to $i$.\footnote{To be more precise, the transit-based techniques
are sensitive to $|\lambda|$, rather than $\lambda$.
This is because with transit data alone, we cannot distinguish a system
with $\lambda=a$ and $i_{\rm o}=b$ from an otherwise identical
system with $\lambda=-a$ and $i_{\rm o} = 180^\circ- b$.
Most authors arbitrarily assume $i_{\rm o}<90^\circ$ and report $\lambda$ in the
range from $-180$ to $+180^\circ$, or from $0$ to $360^\circ$.
We also note that when $i_{\rm o}\approx 90^\circ$, as is the
case for transiting planets, then 
$\psi>|\lambda|$ when $|\lambda|<90^\circ$,
and $\psi<|\lambda|$ when
$|\lambda|>90^\circ$.}
Indeed, in the best cases, $\lambda$ can be measured
with a precision on the order of $1^\circ$.
The disadvantages of these methods are that they require
time-critical observations of transits, and the signals are
generally proportional to the area of the planet's silhouette divided
by the area of the stellar disk. In practice, it has proven to be very challenging
to deploy these methods on planets smaller than Neptune
around Sun-like stars.

Finally, there is a technique that is mainly sensitive to $\lambda$ and
does {\it not} require a transiting planet: {\bf spectro-interferometry}.
For nearby bright stars,
optical interferometric observations with high spatial and spectral resolution
can partially resolve the stellar disk and reveal
the displacement on the sky between
the redshifted and blueshifted sides of the rotating star (\S~\ref{sec:inter_and_spec}).
This is still a highly specialized technique, though, and must be combined
with other data that specify the orientation of the planet's orbit.

Each technique works best in certain circumstances.
Figure~\ref{fig:obli_techniques} illustrates the
applicability of these different techniques to
systems with different planet sizes,
stellar masses, and orbital periods.
Below, we describe these techniques in more detail,
although not in the geometry-based order
described here. Instead, we devote the most attention to the techniques that
have delivered the most information.

\subsection{The Rossiter-McLaughlin effect}
\label{sec:rm}

\begin{figure*}
  \begin{center}
   \includegraphics[width=18.cm]{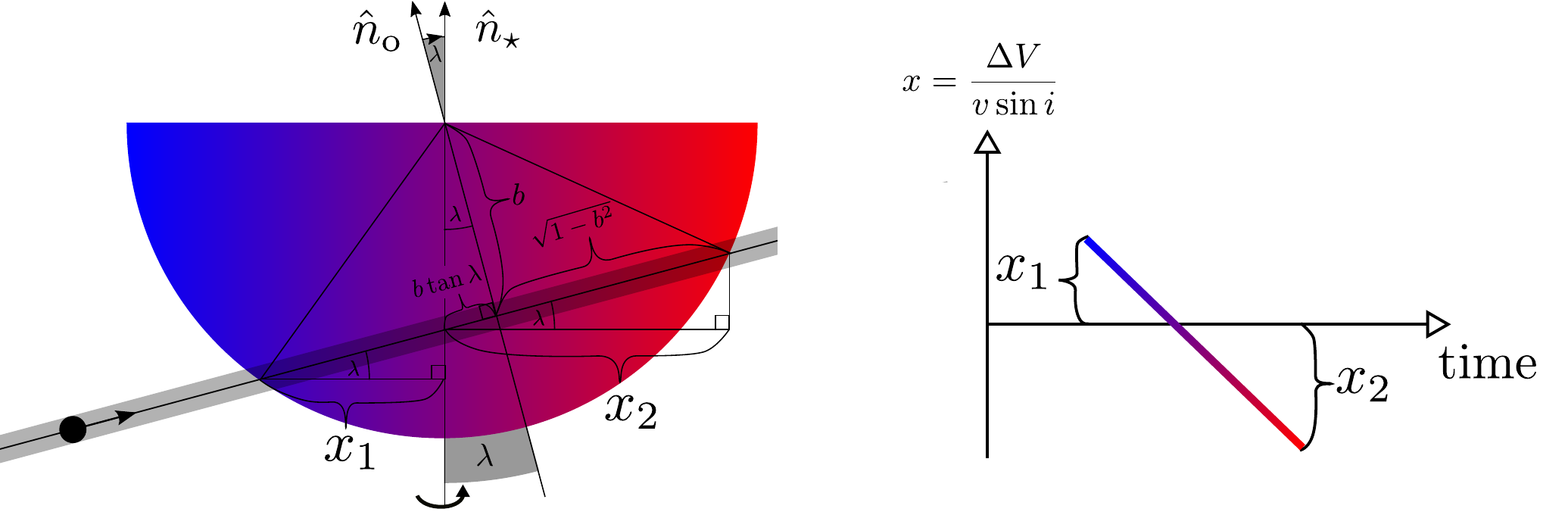}
   \caption {\label{fig:rm} {\bf Geometry of the Rossiter-McLaughlin
       effect.} The left panel illustrates a transit with $i_{\rm o}<90^\circ$.
       The planet crosses from left to right.
     Due to stellar rotation, the left
     side of the star is approaching the observer, and the right
     side is receding. 
     Expressing distance in units of the stellar radius
     with the $x$-axis parallel to the stellar equator,
     and neglecting differential rotation, the sub-planet 
     radial velocity is $(v\sin i)\,x$ and the extrema of the
     RM signal occur at ingress ($x_1$) and egress ($x_2$). The
     relations between $x_1$, $x_2$, $\lambda$ and the impact
     parameter $b$ are indicated.
     The right panel shows
     the corresponding velocity of the planet's ``Doppler shadow.''
     Adapted from \cite{AlbrechtWinnJohnson+2011}. }
  \end{center}
\end{figure*}

In a letter to the editor of the {\it Sidereal Messenger}, \cite{Holt1893}
pointed out that a star's rotation rate could be measured
by observing the time-variable distortions of its absorption spectrum during
an eclipse. We have not been able to learn anything more about this
insightful correspondent, nor have we found any earlier reference to what is now
called the Rossiter-McLaughlin (RM) effect.  The name honors the work of 
\cite{Rossiter1924} and \cite{McLaughlin1924}, who observed the effect
in the $\beta$~Lyrae and Algol systems,
respectively.\footnote{An earlier and less convincing
detection was reported by \cite{Schlesinger1910} for the $\delta$~Librae system.}

Due to rotation, light from the approaching half of the stellar disk
is blueshifted, light from the receding
half is redshifted, and the disk-integrated spectrum
shows a spread in Doppler shifts.
During an eclipse or transit,
a portion of the stellar disk is hidden from view,
weakening the corresponding radial-velocity components in the disk-integrated
spectral absorption lines.
The character and time-evolution of the spectral distortions
depend on $v\sin i$ and $\lambda$, in addition to the usual
eclipse parameters.

Observers have detected and modeled the RM effect in several
ways.  When the spectral lines are not well resolved,
the line-profile distortions are manifested as
shifts in the apparent central wavelength of the line.  When the blueshifted
half of the star is eclipsed, the lines exhibit an anomalous redshift, and vice versa.
This is the manner in which \cite{Rossiter1924} and \cite{McLaughlin1924}
displayed their data, as well as \cite{Queloz+2000}, who first
observed the RM effect for an exoplanet-hosting star.
Parametric models for the ``anomalous radial velocity''
and its relation to the positions and attributes of the two bodies have
been developed by many authors \cite[e.g.][]{Hosokawa1953, kopal1959,Sato1974,Ohta+2005,gimenez2006,Hirano+2011,ShporerBrown2011}.

Alternatively, the line-profile distortions can be detected and modeled directly
without the intermediate step of computing an anomalous radial velocity.
Models for the line-profile distortions 
have been extensively developed,
starting with a beautiful exposition by \cite{struve1931} for the Algol system and continuing to the present \citep{Albrecht+2007,CollierCameron+2010,AlbrechtWinnMarcy+2013,Johnson+2014,Cegla+2016,Zhou+2016,Johnson+2017}. This method is sometimes referred
to as Doppler tomography, although we prefer
the terms Doppler transit or Doppler shadow.\footnote{Tomography is
the reconstruction of a 3-d structure based on 2-d observations obtained over a wide range of viewing angles. In astrophysics, the term ``Doppler tomography''
was introduced in the 1980s to describe
the reconstruction of
a star's surface features or a binary's accretion geometry
based on spatially unresolved observations spanning an entire
rotational or orbital cycle.
In the case of a planetary transit, though, the range of viewing
angles is so narrow that there is no tomographic
quality to the analysis.}

The RM effect has been the basis of most
of the obliquity measurements of individual planet-hosting stars,
as reviewed by \cite{Triaud2018}.
Below, we describe the geometry of the RM effect (\S~\ref{subsec:RM}) and the results from
the two main methods for analyzing the RM effect: as a line-profile distortion (\S\,\ref{subsec:line_profile}) 
and as an anomalous radial velocity (\S\,\ref{subsec:RV_rm}).
Then, we review the key findings that have emerged
from RM observations (\S\,\ref{sec:teff}--\ref{sec:multitransits}).
Table~\ref{tab:obs_trends} gives an overview of these trends and highlights
some systems of particular interest. Appendix~\ref{app:data} describes the compilation
of data that was used to make the charts in this article and an overview of $\lambda$, $i$, and $\psi$ measurements is shown in  figure~\ref{fig:polar_hist}.\footnote{The data will  be made available via the \href {https://exoplanetarchive.ipac.caltech.edu/}{NASA Exoplanet Archive} and can also be obtained from \href{https://phys.au.dk/exoplanets}{phys.au.dk/exoplanets}.} 

\begin{figure*}
  \begin{center}
     \includegraphics[width=1.0\textwidth]{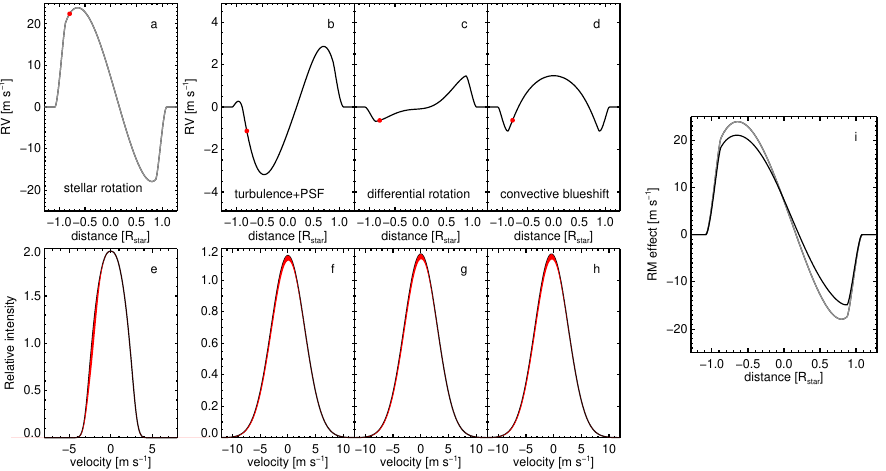}
    \caption {\label{fig:show_lines}
    {\bf Higher-order effects in the anomalous radial velocity,}
    illustrated for a Sun-like star with $v \sin i = 3$~km\,s$^{-1}$
    and a transiting planet with $\lambda=40^{\circ}$,  $r/R=0.12$, and $b=0.2$.
    (a) Limb darkening causes the maxima
    and minima to be rounded, rather than sharp as in
    Figure~\ref{fig:rm}.
    (b) Instrumental broadening (taken to be $2.2$~km\,s$^{-1}$)
    and macroturbulence ($\zeta_{\rm RT} = 3$~km\,s$^{-1}$)
    act oppositely to the rotational effect.
    (c) Differential rotation introduces a
    dependence on the range of stellar latitudes crossed
    by the planet.
    (d) The convective blueshift produces
    an anomalous velocity depending on the planet's distance from the center
    of the stellar disk.
    (e-h) The corresponding models for the stellar absorption lines. The red region identifies the RM distortion at the phase of the transit that is indicated by a red dot in the corresponding upper panel. For the lower panels, the radius of the planet was doubled, to allow improved visibility of the missing velocity components. (i) The combined model including all aforementioned effects. The gray line is the model from panel (a). From \cite{AlbrechtWinnJohnson+2012}. }
  \end{center}
\end{figure*}

\begin{figure*}
  \begin{center}
  \includegraphics[width=0.9\textwidth]{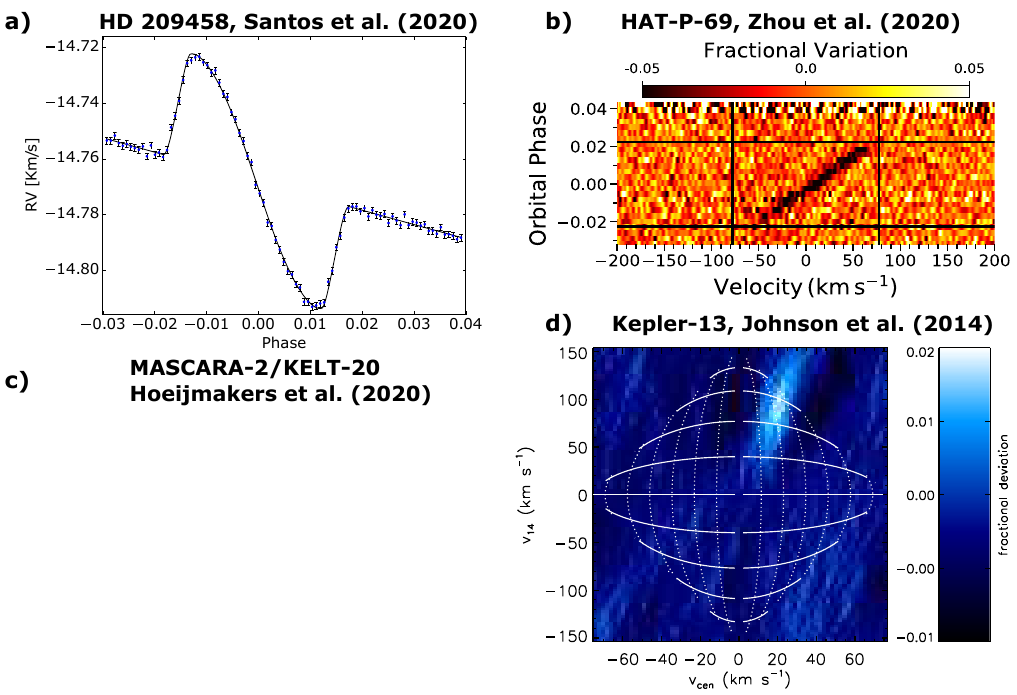} 
    \caption {\label{fig:RM_examples} \textbf{Different visualizations of the RM effect.}
    \textit{Top left}: 
    Time series of anomalous RVs observed during a
    transit of HD\,209458 \citep{Santos+2020}.
    \textit{Top right}:
    Time series of cross-correlation
    functions (CCFs) during a transit of HAT-P-69, after subtracting a model with no RM effect \citep{Zhou+2019}.
    The dark stripe is the planet's Doppler shadow.
    \textit{Lower left}: Time series of the sub-planet
    velocity during a transit of MASCARA-2, inferred
    from the absorption line profiles using the
    ``RM Reloaded'' technique \citep{Hoeijmakers+2020}.
    \textit{Lower right}: Stacked CCF residuals based on
    a spectroscopic time series obtained during 
    a transit of Kepler-13 \citep{Johnson+2014}.
    The residuals
    were shifted and binned for
    various choices of two parameters
    that control the calculated trajectory of the planet.  The first parameter, $v_{14}$, controls the amplitude of the RM effect and the second parameter, $v_{\rm cen}$, controls the asymmetry. Curves of constant $\lambda$ are solid white, and curves of constant impact parameter are dotted white. The bright area is the region of parameter space
    preferred by the data.}
  \end{center}
\end{figure*}

\subsubsection{The RM geometry}
\label{subsec:RM}

Consider a transit of a planet of radius $r$ across a uniformly rotating
star of radius $R$, equatorial rotation velocity $v$, and line-of-sight
inclination $i$.  During the transit,
the stellar absorption lines suffer a fractional loss of light on the order
of $(r/R)^2$ associated with the velocity component
\begin{equation}
\Delta V(t) = (v\sin i)\,x(t),
\end{equation}
which is the rotational radial velocity
of the point on the star directly behind the planet.
Sometimes, this ``sub-planet velocity'' is denoted by $v_{\rm p}$. 
Here, $x(t)$ is the planet's position in units of the stellar radius
along the coordinate axis
running perpendicular to the star's
projected rotation axis, as in Figure~\ref{fig:rm}.

If the radius ratio $r/R$ and transit impact parameter $b$ are
known, then observations of the time series $\Delta V(t)$ can be used
to determine $\lambda$ and $v\sin i$.
In practice, one fits a parameterized model to the time series,
but it is useful to understand
which aspects of the signal provide the information.
Figure~\ref{fig:rm} illustrates the transit geometry and
the corresponding $\Delta V(t)$. The extremes of the signal
occur at ingress ($x_1$) and egress ($x_2$), with amplitudes
\begin{equation}
      \Delta V_1 = (v\sin i)\,x_1,~~\Delta V_2 = (v\sin i)\,x_2.
\end{equation}
Based on the transit geometry,
\begin{eqnarray}
\label{equ:x}
x_{1} & = & \sqrt{1-b^{2}}\cos\lambda - b \sin \lambda,~~{\rm and}\,\,\,\,\,\,\nonumber \\
x_{2} & = & \sqrt{1-b^{2}}\cos\lambda + b \sin \lambda,\,\,\,\,\,\,
\end{eqnarray}
allowing us to write
\begin{align}
\label{equ:RM_amplitude}
\Delta V_2 + \Delta V_1 & = 2 v\sin i \cos\lambda \times \sqrt{1-b^2},  \\
\label{equ:RM_asymmetry}
\Delta V_2 - \Delta V_1 & = 2 v\sin i \sin\lambda \times b.
\end{align}
This system of equations makes clear that 
the {\it amplitude} depends on $\cos\lambda$  (Eq.~\ref{equ:RM_amplitude}), while the {\it asymmetry} of the signal
depends on $\sin\lambda$ (Eq.~\ref{equ:RM_asymmetry}).
It also indicates that
measurements of the amplitude and asymmetry
are sufficient to determine
$v\sin i$ and $\lambda$, as long as $b$ is
not too close to 0 or 1 \citep{AlbrechtWinnJohnson+2011}.

\subsubsection{The Doppler Shadow} 
\label{subsec:line_profile}
 
\begin{figure}
  \begin{center}
  \includegraphics[width=7.5cm]{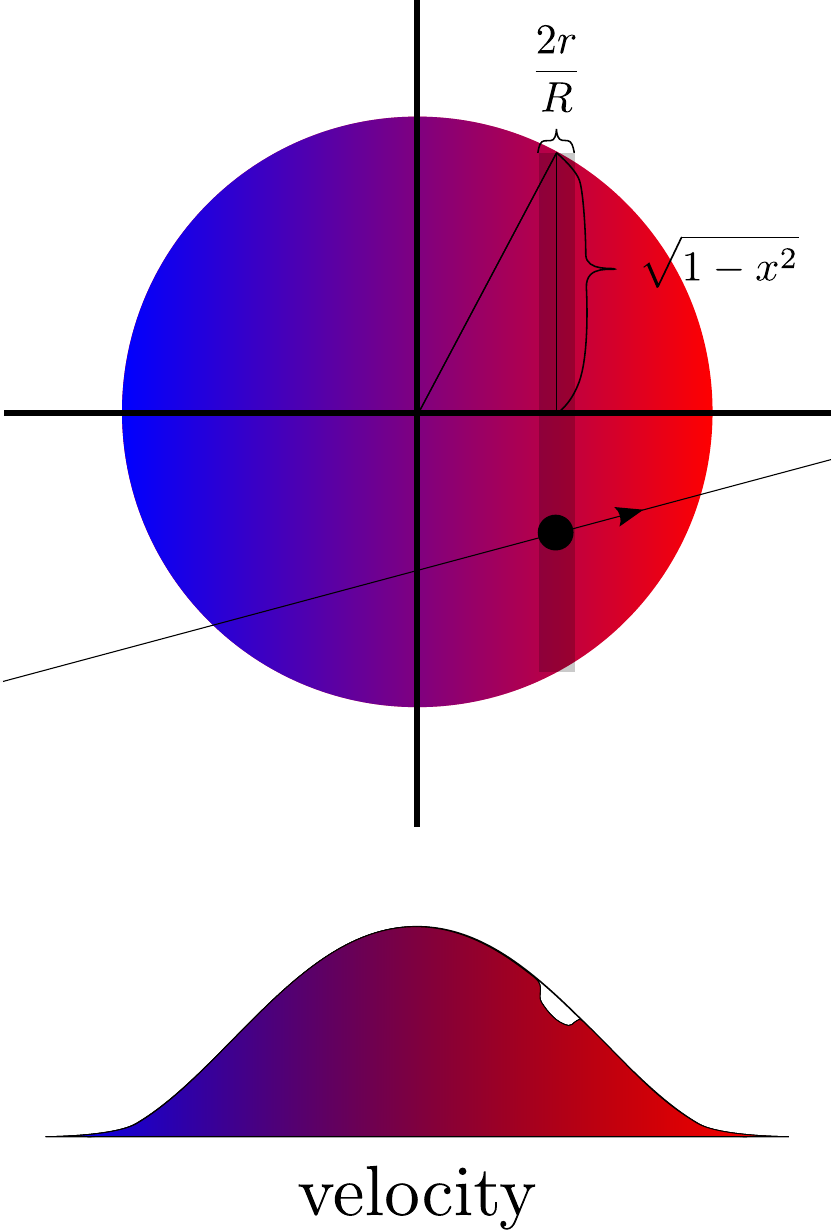} 
    \caption {\label{fig:RM_effect_line} \textbf{Stellar line deformation for a rapidly rotating star}. For this illustration, rotation was assumed to be the dominant broadening mechanism,
     leading to a deformation that is well-localized in wavelength. }
  \end{center}
\end{figure}

The line-profile distortions due to the RM effect can also be
analyzed directly.
For simplicity, consider an idealized spectral line
broadened only by rotation.
When the planet is at position $x_{\rm p}$,
the range of velocity components partially blocked
by the planet is $(v\sin i)\,(x_{\rm p}\pm r/R)$.
Within this velocity range, the fractional loss of light
is equal to the area of the planet's silhouette divided by
the area of the strip of the star within $x_{\rm p} \pm r/R$,
\begin{equation}
\label{equ:distrotion_rm}
\Delta L_{\rm RM}(t) \approx -\frac{\pi}{4} \frac{r}{R} \frac{1}{\sqrt{1-x(t)^2}}.
\end{equation}
This equation gives the intensity contrast of the ``Doppler shadow'' --- the bump
that appears in the line profile.
Because $\Delta L_{\rm RM}$ is proportional to $r$ instead of $r^2$,
this technique is, in principle,
more sensitive to small planets than the anomalous-RV technique.  In practice, though,
unless the star is rotating very rapidly,
the contrast of the bump is reduced by other line-broadening mechanisms,
which at least partially negates this advantage. See Fig.~\ref{fig:RM_effect_line} for an illustration.

\cite{Albrecht+2007} applied this technique to the eclipsing binary V1143\,Cyg.
They created synthetic line profiles
by numerically integrating over a 2-d pixelated stellar disk,
after assigning intensities and velocities to each pixel
due to rotation, limb darkening, velocity fields, etc.
The pixels hidden by the planet were
assigned zero intensity.
\cite{CollierCameron+2010} used a simpler approach in which the stellar line profile
and planetary disturbance were modeled with 1-d functions.
In a variation of this technique dubbed ``RM Reloaded,''
\cite{Cegla+2016} replaced the synthetic line profiles that had been used
in previous studies
with an empirical model based on the spectra obtained outside
of transits. They used a parametric
model only for the portion
of the photosphere covered by the planet, a method developed
further by \citet{Bourrier+2021}.

\subsubsection{The anomalous radial velocity}
\label{subsec:RV_rm}

The effect on a spectral line is a distortion,
not an overall Doppler shift. Nevertheless, a radial-velocity (RV)
extraction algorithm will respond to
the distortion by reporting an anomalous velocity,
\begin{equation}
\label{equ:rv_rm}
RV_{\rm RM}(t) \approx -\left( \frac{r}{R} \right)^2 \Delta V(t).
\end{equation}
A decent approximation for the maximum amplitude of the anomalous RV is
\begin{equation}
    RV_{{\rm RM}({\rm max})} \approx 0.7 \sqrt{1-b^2} \left( \frac{r}{R} \right)^2 v\sin i.
\end{equation}
The factor of 0.7 accounts for limb darkening.
There are other corrections of order unity due to the effects
of turbulent and instrumental
broadening, and the details of the RV-extraction algorithm.
For the case of a cross-correlation
algorithm, a more accurate formula was derived
by \cite{Hirano+2011}, building on 
work by \cite{Ohta+2005}.

\cite{GaudiWinn2007} exposed the information content of the RM signal
in more detail. They derived an approximate formula\footnote{Equations 16 and 17
of \cite{GaudiWinn2007} contain errors; the sine and cosine functions
should be swapped in both cases. The formula given here, as
Equation~\ref{eq:gaudiwinn}, is correct.}
to estimate
the achievable precision in a measurement of $\lambda$,
\begin{equation}
    \label{eq:gaudiwinn}
    \sigma_\lambda = \frac{\sigma_v/\sqrt{N}}{v\sin i} \left(\frac{r}{R}\right)^{\!\!-2}
    \left[ \frac{(1-b^2)\cos^2\lambda + 3b^2\sin^2\lambda}{b^2(1-b^2)} \right]^{1/2},
\end{equation}
assuming the transit is well-sampled
with $N$ uniformly spaced data points, each with an uncertainty
$\sigma_v$ in the radial velocity. The formula was derived
assuming that $\lambda$ and $v\sin i$
are the only two free parameters,
and that limb darkening can be neglected.
The uncertainty diverges as $b\rightarrow 0$,
when the asymmetry vanishes regardless of obliquity (Eqn.~\ref{equ:RM_asymmetry}),
and as $b\rightarrow 1$, when the transit/RM signal itself vanishes (Eqn.~\ref{equ:RM_amplitude}).

Figure~\ref{fig:show_lines} shows some higher-order effects that were neglected
in the preceding discussion. Limb darkening weakens the
RM effect near the ingress and egress phases.
Differential stellar surface rotation causes $\Delta V$ to depend on
both $x$ and $y$, which makes the RM effect sensitive to $i$ in addition to $\lambda$
\citep{GaudiWinn2007,Cegla+2016}.  Turbulence on the stellar
surface also affects $\Delta V$, as does the ``convective blueshift'' --- the 
higher intensity of the hot, upwelling material compared
to the cooler, sinking material \citep{ShporerBrown2011,Cegla+2016}.
Some other effects that are usually neglected, but that may be important
in special cases, are the tidal and rotational deformations
of the star, the saturation or pressure-broadening
of some absorption lines, and the influence of star spots and pulsations.

Whether to analyze the data in terms of the anomalous RV or the
line-profile variations, or both, depends on the instrument and the
system parameters. Roughly speaking, the larger the ratio
\begin{equation}
\label{equ:alpha}
 \alpha = \frac{(v\sin i)(r/R)}{\sqrt{\sigma^2_{\rm inst}+\sigma^2_{\rm mic}+\sigma^2_{\rm mac}}},
\end{equation}
the easier it will be to resolve the planet's Doppler shadow
in the line profiles. Here, $\sigma_{\rm inst}$ is the
instrumental broadening of the spectrograph, and
$\sigma_{\rm mic}$ and $\sigma_{\rm mac}$ are the magnitudes of micro- and macroturbulence \citep{Gray2005}.
Those are the most important terms which
determine the shapes and widths of unsaturated absorption lines, besides rotation.
For rapidly rotating stars, precise RV determination is difficult
but the RM anomalies in the line profiles can reach depths of
several percent of the overall line depth \citep[e.g.][]{Talens+2018},
making them relatively easy to detect.

Fig.~\ref{fig:RM_examples} compares four different representations
of the RM effect drawn from the literature.
The upper left panel shows a time series of the anomalous radial velocity.
The upper right panel
shows a ``Doppler shadow'' as a time series of residual line
profiles, derived from cross-correlation.
Each row represents an observed line profile after
subtracting the best-fitting model of an undisturbed line profile.
As time progresses (upward, on the plot), the negative residual
caused by the planet moves from the blue end to the red end of the line
profile. The lower left panel shows a time series of the
sub-planet velocity inferred with the RM Reloaded technique.
In the lower right panel, the color scale indicates
the strength of the line-profile residuals after shifting and averaging
them as a function of the sub-planet velocity at midtransit ($v_{\rm cen}$)
and the difference in sub-planet velocities at ingress and egress
($v_{14}$) corresponding to Eqn.~\ref{equ:RM_asymmetry} and Eqn.~\ref{equ:RM_amplitude}. Such a data-stacking analysis can be useful in the presence
of correlated noise \citep{Johnson+2014} or a low signal-to-noise ratio
\citep{Hjorth+2021,Bourrier+2021}.  Another approach to detecting
small RM signals, employing Gaussian Processes, was presented by
\cite{Kunovac-Hodzic+2021}.

\subsubsection{The photometric RM effect}
\label{sec:Rotational Doppler boosting}

The intensities of the receding and approaching halves of the stellar disk differ, at least slightly,
due to Doppler beaming as well as the ordinary Doppler shift combined with the finite
bandpass of the observations.  \cite{Groot2012} and \cite{Shporer+2012} evaluated the potential of
using these effects to measure stellar obliquities with precise light curves.
\cite{Shporer+2012} presented the following equation to estimate the maximum amplitude
of the associated
photometric anomaly,
\begin{equation}
    \label{equ:beaming}
    A_{\rm PRM} \approx 10^{-5} \frac{v\sin I}{10 {\rm \ km \ s^{-1}}} \frac{\left(\frac{r}{R} \right)^2}{0.1}\ .
\end{equation}
They concluded that due to the small amplitude of the effect, obliquity measurements will be challenging. The most promising targets are rapidly rotating early-type stars, and possibly white dwarfs.

\begin{figure*}
  \begin{center}
    \includegraphics[width=18cm]{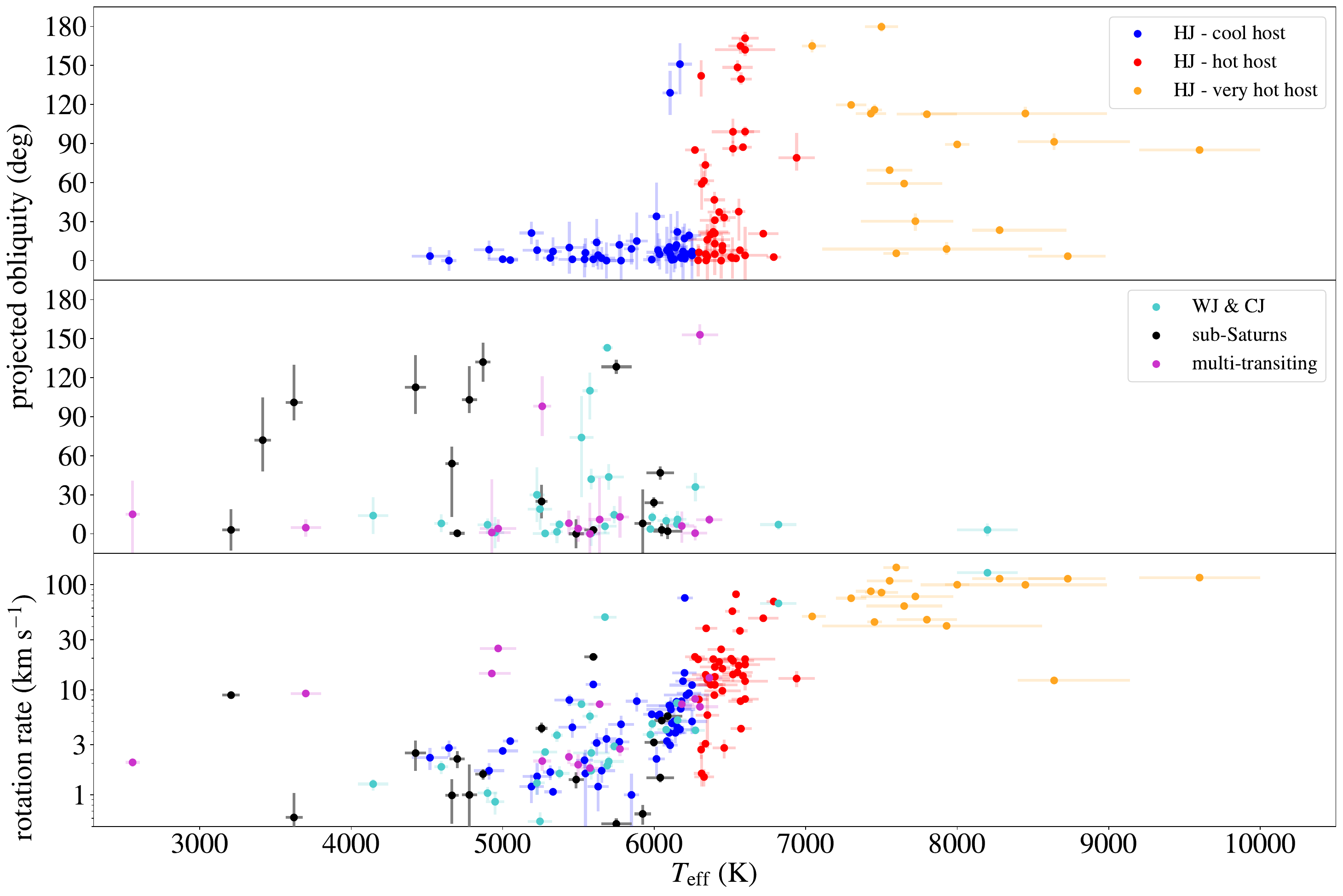} 
    \caption  {\label{fig:teff_projected_obliquity_vsini}
    {\bf Projected obliquity and rotation velocity vs.\ effective temperature.} 
    Data points are color-coded according to the system's characteristics  using the same scheme as in Figure~\ref{fig:polar_hist}.
    Stars are designated as cool, hot, or very hot, using effective temperature
    boundaries of 6250\,K and 7000\,K.
    Planets with masses exceeding $0.3$\,M$_{\rm Jup}$ are designated hot Jupiters (HJ) if $a/R<10$, and warm/cold Jupiters (WJ/CJ)
    if $a/R>10$.
    Planets with masses $\lesssim$\,$0.3~M_{\rm Jup}$ are designated sub-Saturns.
    Systems with at least two transiting planets are designated multi-transiting and are represented
    by only one data point.
    {\it Top:}~For HJs, the obliquity distribution broadens with effective
    temperature, with a relatively sharp transition near 6250\,K. 
     Cool stars with HJs tend to be well aligned, while hotter stars display misalignments more frequently.
    {\bf {\it Middle:}}
    Cooler stars with sub-Saturns or WJ/CJs occasionally have high obliquities.  Also apparent is that almost all the $\lambda$ measurements undertaken for hot stars have involved HJs, as opposed to smaller or wider-orbiting planets.
    {\it Bottom:}~Stellar rotation velocities rise rapidly with effective temperature between about 6000 and 7000\,K, a well-known trend attributed to the magnetic braking
    of cool stars.}
  \end{center}
\end{figure*}

\subsubsection{Hot stars with hot Jupiters have high obliquities}
\label{sec:teff}

The top panel of Figure~\ref{fig:teff_projected_obliquity_vsini} shows the
available measurements of the
projected obliquity as a function of the star's effective temperature.
Focusing attention on stars with hot Jupiters reveals
that cool stars (blue points) tend to have low obliquities, while hot and very-hot stars
(red and orange points) have a broad
range of obliquities.  The transition takes place between approximately 6000 and 6300\,K.
This trend, noted by \cite{Winn+2010} and based on 19 data points,
has persisted even while the sample size has quintupled.
None of the HJ hosts cooler than 6000\,K is known to be misaligned,
where ``misaligned'' is defined (here and elsewhere in this article)
as a reported value of $\lambda$ that exceeds 10$^\circ$ and excludes
$0^\circ$ with $>$3-$\sigma$ confidence.
Between 6250 and 7000\,K, the ratio of misaligned to aligned systems is 1.7 (22 vs.\ 13).
Above 7000\,K, the ratio rises to 4 (12 vs.\ 3), and a Kolmogorov-Smirnov test cannot reject
the hypothesis that the stars are randomly oriented.

\begin{figure*}
  \begin{center}
    \includegraphics[width=\textwidth]{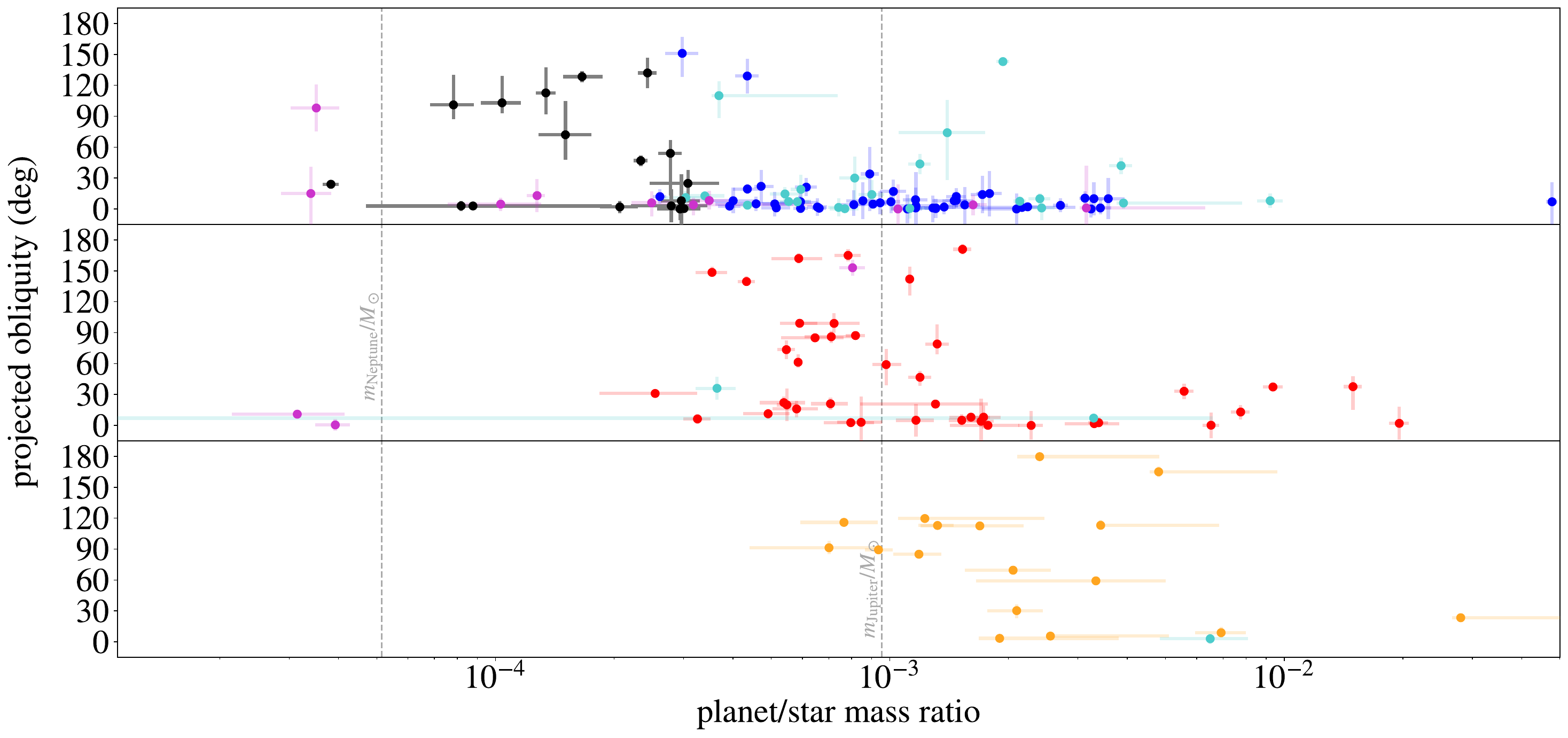} 
    \caption {\label{fig:mass_ratio_projected_obliquity}
    {\bf Projected obliquity vs.\ the planet-to-star mass ratio.}
    Same color scheme as in Figure~\ref{fig:teff_projected_obliquity_vsini}.
    The three panels are for the three different ranges of effective temperature.
    For mass ratios exceeding $0.5\times 10^{-3}$,
    cool host stars have low obliquities, with the exception
    of three WJ hosts.
    Hot and very-hot stars have a broad range of obliquities
    regardless of mass ratio, although hot stars appear to lack retrograde
    systems for mass ratios exceeding $2\times 10^{-3}$.
    }
  \end{center}
\end{figure*}

The range of effective temperatures between 6000 and 6300\,K, where the obliquity distribution
is observed to change, has long been known to be significant in stellar astrophysics.
Observationally, this is the division between the lower and upper main sequences,
where the stellar mass-radius relationship changes
slope and stellar rotation velocities rise sharply with temperature, as shown
in the lower panel of Figure~\ref{fig:teff_projected_obliquity_vsini}.
Theoretically, these changes are explained as consequences of differing
internal structures.  Loosely speaking, a lower main-sequence star has a convective
envelope surrounding a radiative interior, while an upper main-sequence star
has a radiative exterior and may have a convective core (see page 259 of
\citealt{Kippenhahn+2012}).
Lower main-sequence stars rotate more slowly because of ``magnetic braking,'' the steady loss
of angular momentum through a magnetized wind,
which only develops in stars with outer convective zones.
The correct explanation for the
obliquity trend seen in the top panel of Figure~\ref{fig:teff_projected_obliquity_vsini}
is likely to involve these differences in the interior structure of the host stars,
as discussed further in Section~\ref{sec:tides}.

\subsubsection{Stars with especially massive planets have lower obliquities}
\label{sec:oblique_mass}

Figure~\ref{fig:mass_ratio_projected_obliquity} displays the projected obliquity as a function of $m/M$, the planet-to-star mass ratio.
For both cool and hot stars, the obliquities tend to be lower when $m/M$ is larger,
a trend noted by \cite{Hebrard+2011b} using a smaller sample.
All the cool HJ hosts with $m/M<0.5\times 10^{-3}$ are well aligned.
The three misaligned cool stars with $m/M>10^{-3}$ are WASP-8, Kepler-420, and HD\,80606, which all
have $a/R>10$ and are thereby classified as warm Jupiter hosts.
For hot stars, the transition to lower obliquities appears to occur for a higher value
of $m/M$, approximately $2\times 10^{-3}$, above which no retrograde systems are seen.
Very hot stars do not show evidence for any dependence on mass ratio.

\subsubsection{Cool stars with warm Jupiters have high obliquities}
\label{sec:oblique_ar}

Figure~\ref{fig:ar_projected_obliquity} displays the projected obliquity as a function of $a/R$, the orbital
semimajor axis divided by the stellar radius.
As noted by \cite{AlbrechtWinnJohnson+2012},
cool stars orbited by giant planets with $a/R \lesssim 10$ tend to be well aligned,
while those with more distant giant planets have a high obliquity dispersion.
 Out of the 47 HJs orbiting cool stars with $a/R<10$,
36 are well aligned, 3 are misaligned, and 8 have an ambiguous
status according to our definitions.
The misaligned cases are
WASP-60 ($\lambda=19.4^{+4.9}_{-5.1}$~deg; \citealt{Brown+2017}), WASP-94\,A ($151^{+23}_{-16}$~deg; \citealt{Neveu-VanMalle+2014}),
and WASP-60 ($129\pm+17$~deg; \citealt{Mancini+2018}). All three
systems have stars hotter than 6100\,K and orbits with $a/R>7$.
For hot and very-hot stars (the middle and lower panels of Figure~\ref{fig:ar_projected_obliquity}),
there are no obvious trends with $a/R$. However, as shown in \S~\ref{sec:ppp},
the misaligned hot stars with $a/R\lesssim 7$ tend to have $\psi \approx 100$~deg.

\subsubsection{Some cool stars with HJs are very well aligned}
\label{sec:well_aligned}

\begin{figure*}
  \begin{center}
    \includegraphics[width=1\textwidth]{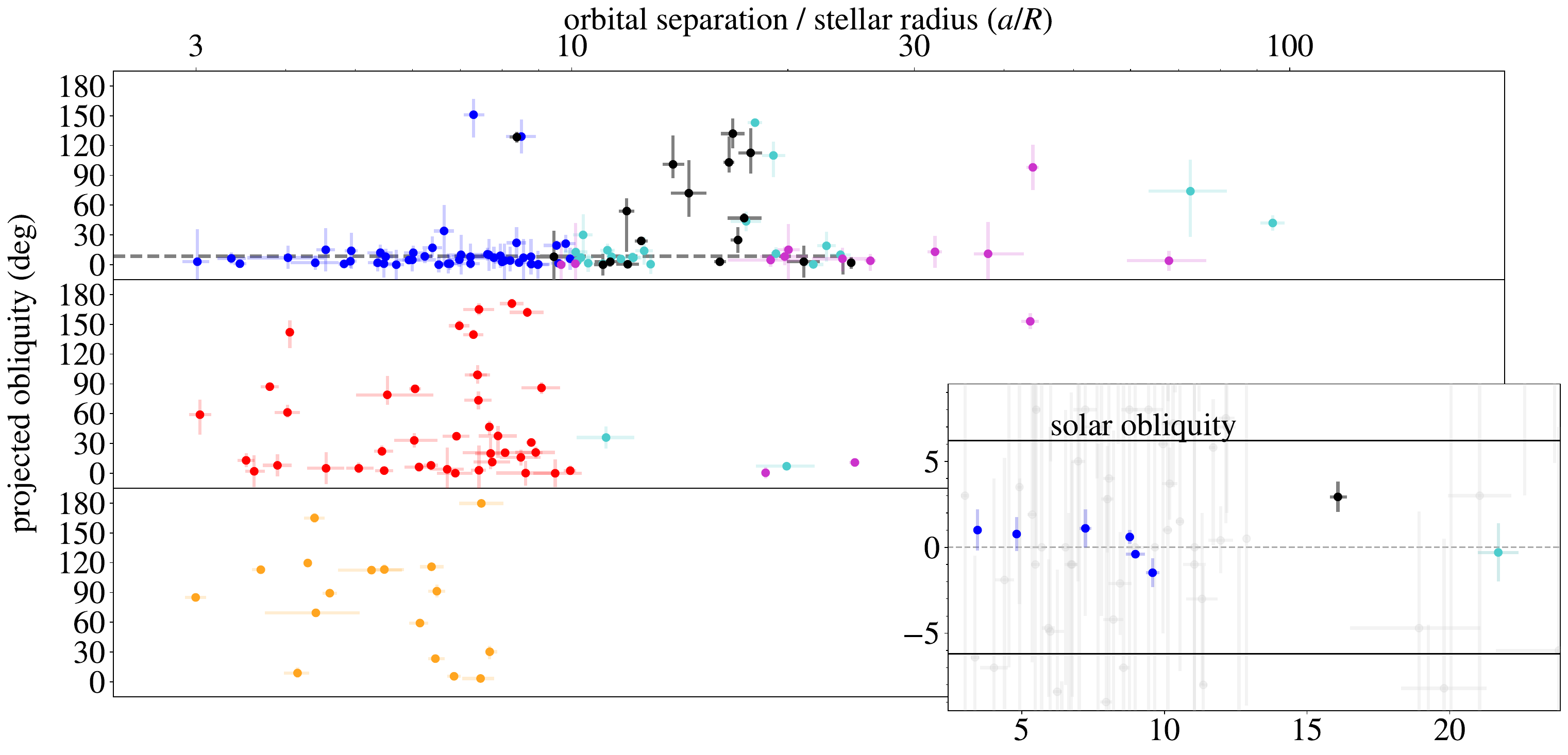} 
    \caption {\label{fig:ar_projected_obliquity}
    {\bf Projected obliquity vs.\ $a/R$.}
    Same color scheme as in Figure~\ref{fig:teff_projected_obliquity_vsini}.
    The obliquity distribution of hot and very-hot stars does not appear to depend on $a/R$.
    In contrast, cool stars tend to be aligned whenever $a/R<10$. 
    The inset panel highlights cool stars for which the measurement uncertainty is below $2^\circ$,
    which appear to have lower obliquities than the Sun.
    Not shown in this plot is the special case of $\beta$~Pic, which has $a/R \approx 10^3$
    and $\lambda = 3\pm 5^\circ$ as determined by spectro-interferometry \citep{Kraus+2020}.}
  \end{center}
\end{figure*} 

For the stars that are compatible with good alignment, it would be interesting
to measure the obliquity dispersion and compare it to the
Sun's obliquity as well as the mutual inclination distribution of the planetary orbits in the
Solar System.
A very low dispersion would suggest that dissipative processes
have damped obliquities, although demonstrating that the dispersion
is very low would require very precise measurements.

The inset panel within Figure~\ref{fig:ar_projected_obliquity}
displays all the projected obliquity measurements for cool stars with
prograde orbits and measurement uncertainties of $2^\circ$ or better.
Positive and negative values of $\lambda$ are plotted to allow
a better look at the region near $0^\circ$.
These stars all have projected obliquities lower than the (unprojected) solar obliquity
of $6.2^\circ$ with respect to the plane defined by the total orbital
angular momentum of the planets.
The standard deviation of the measured values of $\lambda$ for
the prograde cool HJ hosts is $0.91^\circ$, and the
average formal measurement uncertainty is $0.82^\circ$.

Thus, the upper limits on the obliquities of the prograde cool stars with the best measurements
are on the order of a degree, which is several times lower than the solar value.
This might be a hint that at some point during the formation or evolution of
HJs around cool stars, the obliquities were damped by a dissipative process.
The upper limit on the obliquity dispersion
is also comparable to the inferred mutual inclination dispersion of 
compact systems of more than four planets \citep{Zhu+2018}.
It seems worthwhile to expand on the sample of systems with measurement
uncertainties better than 2$^\circ$
and perform a more thorough statistical inference
of the underlying obliquity dispersion.

\subsubsection{Obliquities and stellar age}
\label{sec:age}

\begin{figure}
  \begin{center}
    \includegraphics[width=7.5cm]{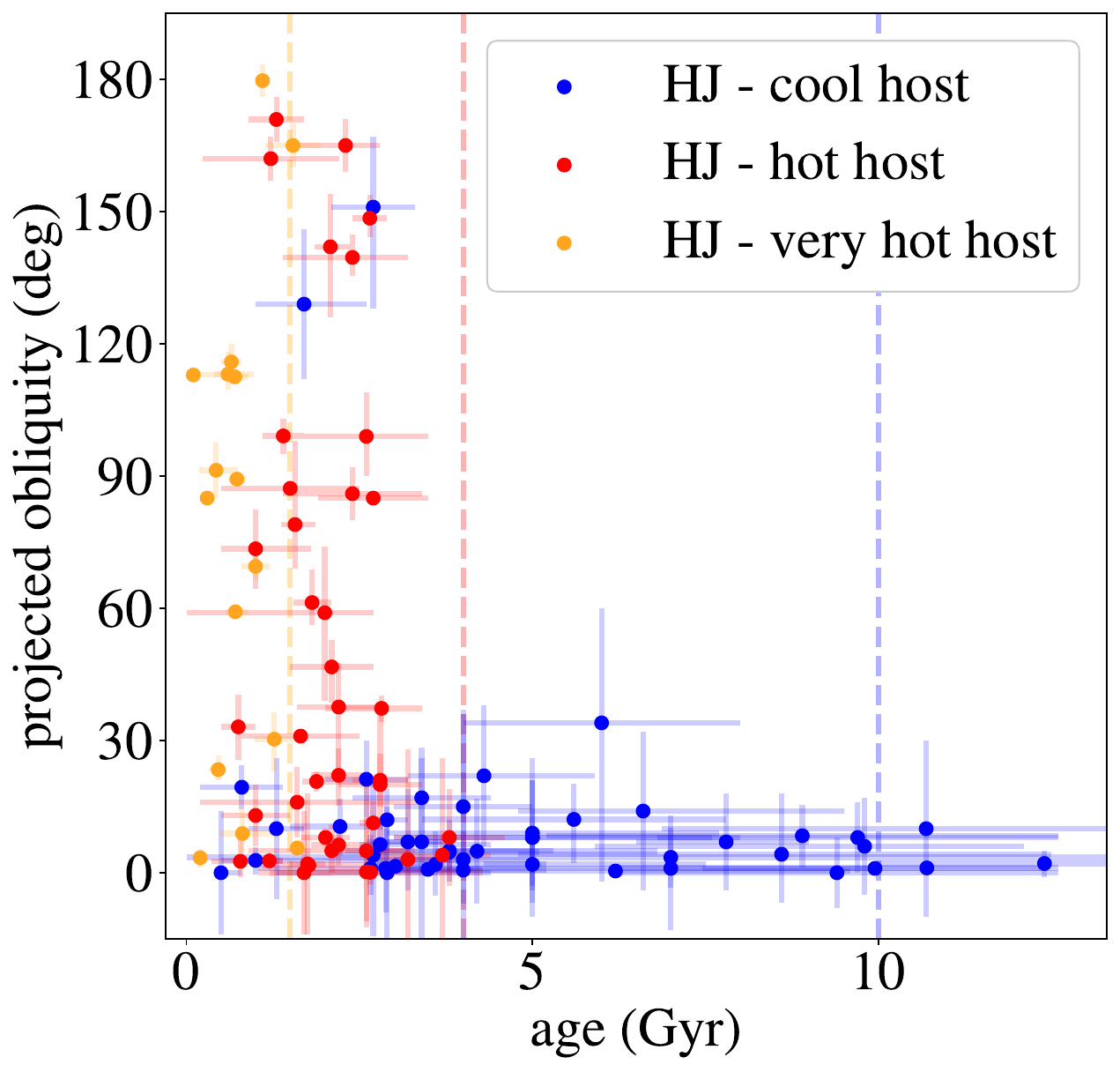} 
    \caption  {\label{fig:age_obli} {\bf Projected obliquity vs.\ age.}
    Same color scheme as in Figure~\ref{fig:teff_projected_obliquity_vsini}.
    Cool stars have a broader age distribution than hot stars
    because cool stars have longer main-sequence lifetimes.}
  \end{center}
\end{figure}

Figure~\ref{fig:age_obli} displays the projected obliquity as function of isochrone age, i.e.,
the age determined by fitting stellar-evolutionary models to the observed stellar properties
such as effective temperature, surface gravity, metallicity, spectral energy distribution,
and luminosity.
The HJ hosts older than about 3\,Gyr tend to be well-aligned,
as noted by \cite{Triaud2011}.
As discussed by \cite{AlbrechtWinnJohnson+2012}, this correlation
is closely related to the previously noted trend involving effective temperature.
Cool stars have a broader range of ages than hot stars, because cool stars have
longer main-sequence lifetimes.  Thus, if misalignments tend to involve
hot stars, the misaligned systems will tend to appear at the young end of the age distribution.
\cite{SafstenDawsonWolfgang2020} used statistical tests to conclude
that the obliquity distribution is more strongly correlated
with effective temperature than age.

\subsubsection{Are very young giant-planet hosts well-aligned?}
\label{sec:newborn}

\begin{figure}
  \begin{center}
    \includegraphics[width=8cm]{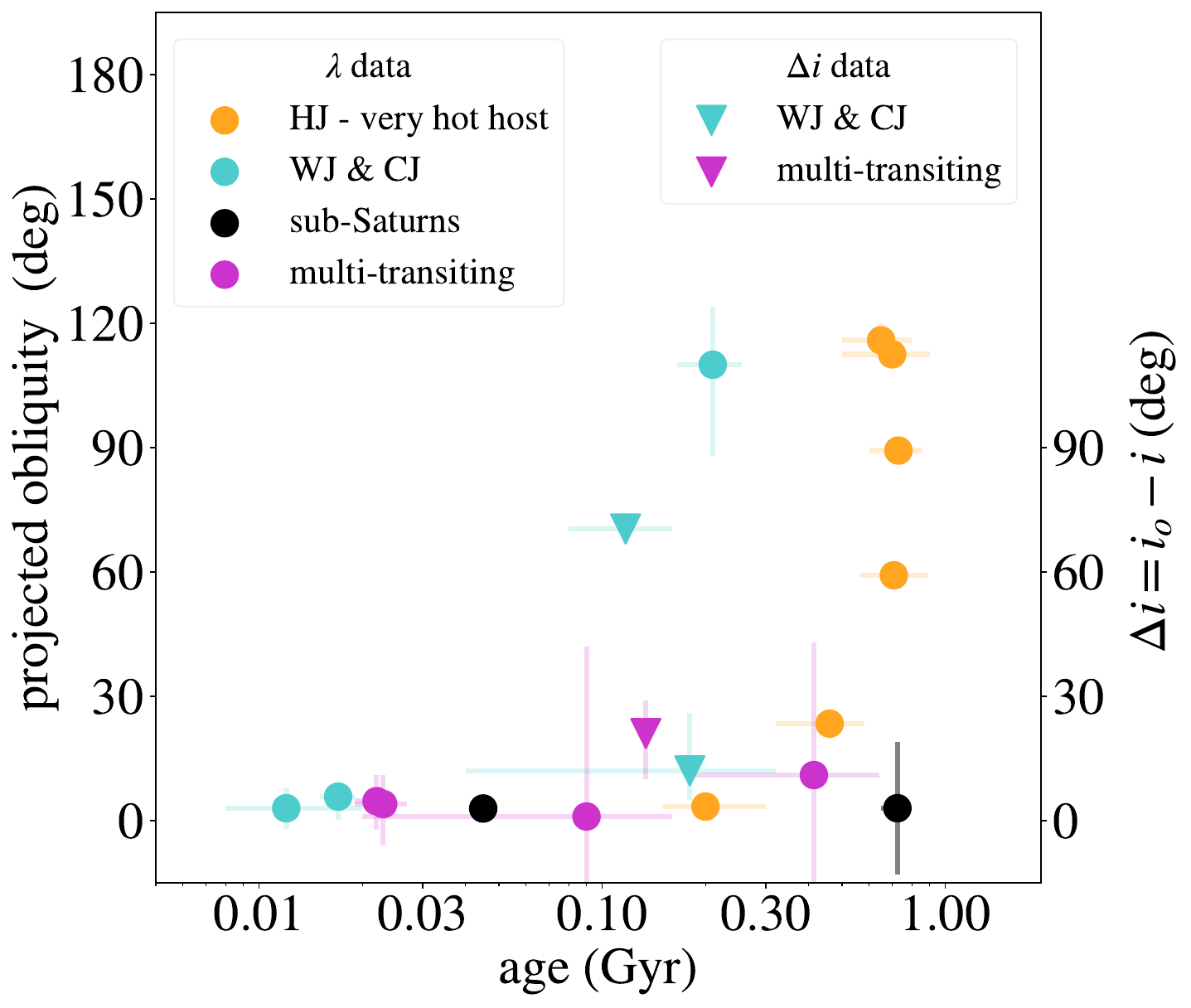} 
    \caption  {\label{fig:newborn} {\bf Spin-orbit alignment for the youngest stars?}
    Shown are the projected obliquity measurements (circles)
    and stellar inclination measurements (triangles) of transit hosts
    younger than 1\,Gyr for which the age uncertainty is smaller than 0.3\,Gyr.}
  \end{center}
\end{figure}

Obliquity measurements for stars younger than 100\,Myr are
scarce, because not many planets have
been detected around such young stars,
and because the intrinsic photometric and spectroscopic variations
of young stars interfere with detailed characterization.
Figure~\ref{fig:newborn} shows the available data
for stars younger than 1\,Gyr that have age uncertainties below 0.25\,Gyr.
The data are based on the RM method
as well as stellar inclination measurements (from the $v\sin i$
and spectro-interferometric methods).  So far, all of the stars
younger than 100\,Myr are consistent with good
alignment.

AU Mic\,b is one of two known transiting planets that orbit
a bright 22-Myr star with an edge-on debris disk \citep{Plavchan+2020}.
Measurements of the inclination angles of the planetary orbit and the debris disk are consistent with alignment,
and RM observations have shown the star to have a low projected obliquity
\citep{Hirano+2020,Palle+2020,Martioli+2020,Addison+2021}.

Another bright and young star with an edge-on debris disk is $\beta$~Pic, which has
two directly-imaged 
giant planets on orbits that are closely aligned with the disk.
Using the spectro-interferometric method (\S~\ref{sec:inter_and_spec}), \cite{Kraus+2020} found
the star to have a projected obliquity of $3\pm 5^\circ$.
This system is quite different from all the others described
in this article because of the
large orbital distances, 4.2 and 10.0 au \citep{Lacour+2021}.

DS\,Tuc\,A is a solar-mass star with an estimated age of 45~Myr,
which is separated by 180\,au from a K3V binary companion, and has
a transiting sub-Saturn planet \citep{Newton+2019}.
RM observations revealed a projected
obliquity of $2.5\pm 1.0^\circ$, and the
$v\sin i$ method also gave results consistent with a low
obliquity.  The $v \sin i$ method (\S~\ref{sec:vsini}) is
well suited to stars such as DS\,Tuc\,A, which rotate rapidly
and undergo large-amplitude light variations that reveal the stellar
rotation period.

The youngest star known to be misaligned is TOI-811, which has an isochrone
age of $117^{+37}_{-43}$~Myr \citep{Carmichael+2020}.
The transiting body in that case has a mass
of $59.9^{+8.6}_{-13} M_{\rm Jup}$ and should probably be considered
a brown dwarf rather than a planet.
The youngest planetary-mass object known to have a misaligned star is Kepler-63\,b
\citep[$210\pm 45$\,Myr,][]{Sanchis-Ojeda+2013}.
Another young and misaligned system is KELT-9, a very hot star
with $\lambda=85.01\pm 0.23^\circ$ and an estimated age of 300\,Myr
\citep{Gaudi+2017}. KELT-9 does not appear in Figure~\ref{fig:newborn} because
the reported age did not include an uncertainty estimate.
Age determinations for rapidly rotating A stars based on
photometry and spectroscopy are
subject to systematic errors due to gravity darkening and rotational oblateness \citep[see, e.g.,][]{Jones+2015}.
The youngest star known to have a well-aligned
planet is HIP\,67522,
with an age of $17\pm2$~Myr \citep{Heitzmann+2021}. The
planet's mass has not been measured; it could be a hot Jupiter or
an inflated planet of lower mass.

While we cannot draw any firm conclusions from the small sample of young stars,
it is noteworthy that the available data for stars with ages $\lesssim$\,100\,Myr
are consistent with low obliquities.
If this trend persists for close-orbiting giant planets, it would 
be an important clue about obliquity excitation.
For example, it might be the case that very young hot Jupiters 
formed {\it in situ} or underwent disk migration, either of which
would preserve the initial alignment, while late-arriving hot Jupiters
formed through orbit-tilting interactions.  We refer the reader
to \cite{DawsonJohnson2018} for more discussion of the possible origins
of hot Jupiters, and to \S~\ref{sec:theory} for more discussion
of the relevance of obliquity measurements.

\subsubsection{Obliquity \& eccentricity}
\label{sec:obliquity_eccentricity}

\begin{figure*}
  \begin{center}
        \includegraphics[width=8cm]{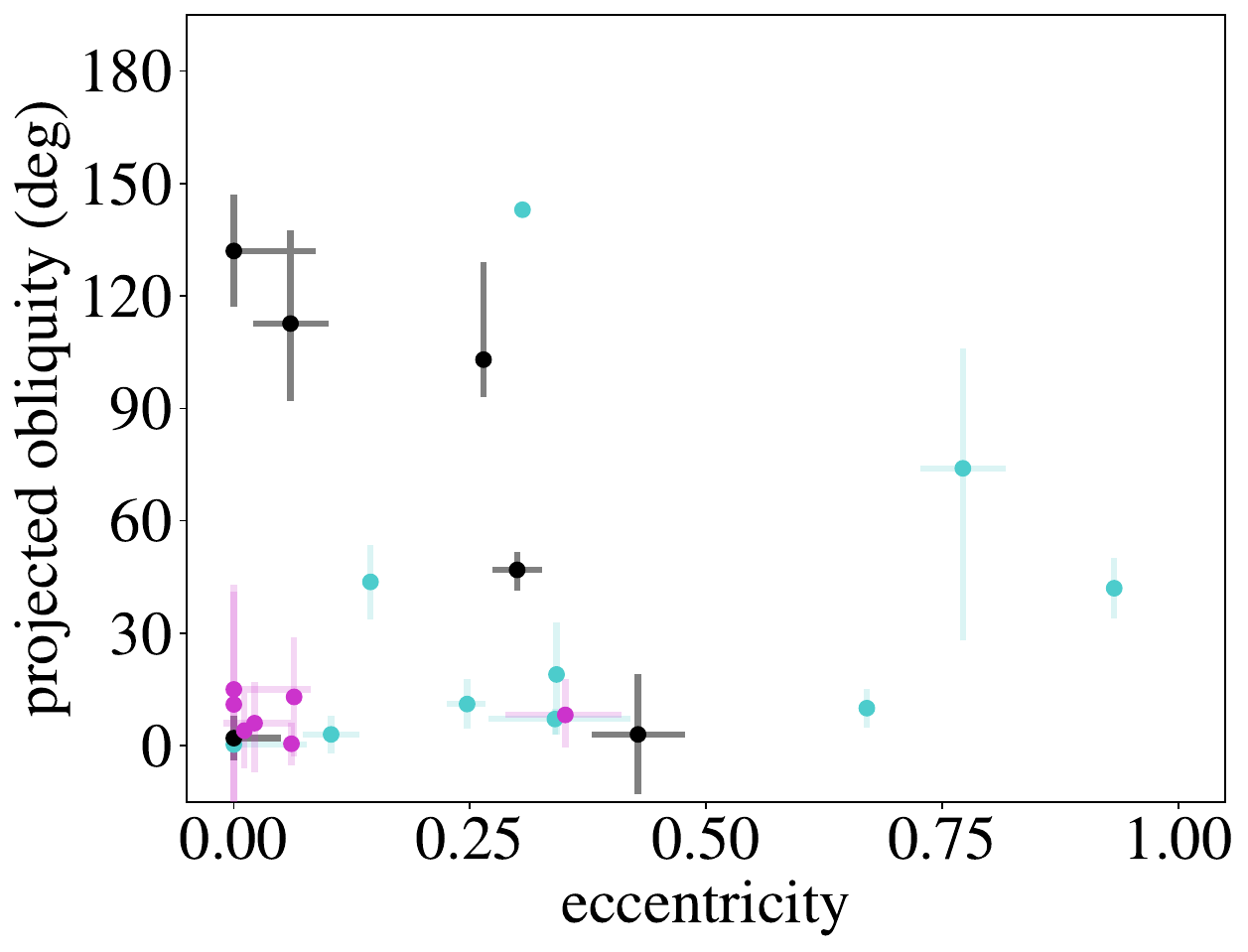} 
        \includegraphics[width=8cm]{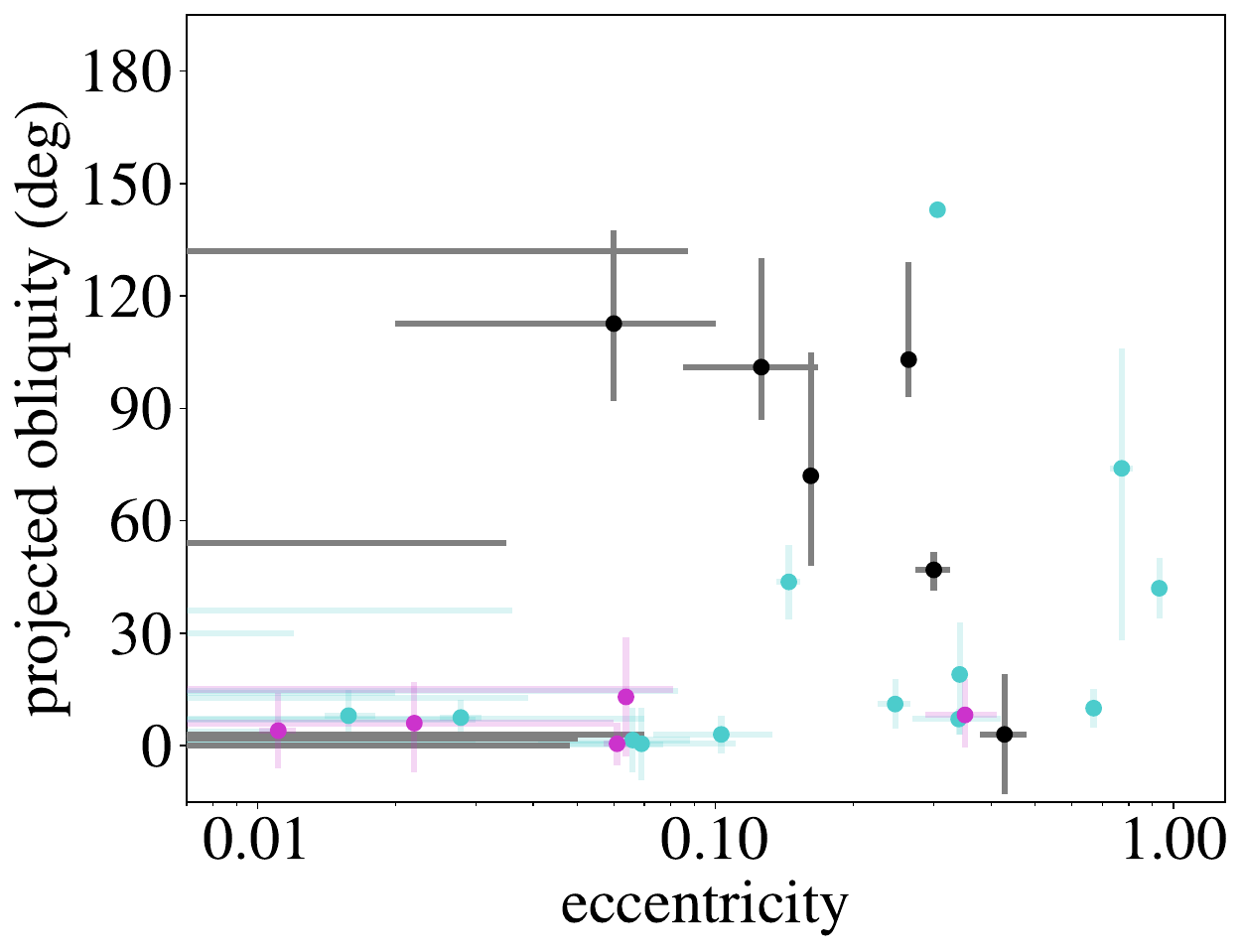} 
    \caption  {\label{fig:eccentricity_projected_obliquity} {\bf Obliquity and eccentricity}, for all systems with $a/R>10$ for which the uncertainty
    in eccentricity is below 0.1.
    The lower limit on $a/R$ is intended to exclude systems
    subject to rapid tidal circularization. The right panel shows
   the same data with a logarithmic scale for the eccentricity axis.
Many systems display nearly circular and aligned architectures, while systems with spin-orbit  misalignments tend to travel on eccentric orbits. We note that our restrictions on distance and eccentricity might bias this sample. In particular, hot host stars are nearly all excluded because eccentricity measurements for such stars are more difficult.
   } 
  \end{center}
\end{figure*}

Dynamical interactions such as planet-planet scattering and Kozai-Lidov cycles, which
are often invoked to explain high obliquities, are also expected
to excite orbital eccentricities (\S~\ref{sec:orbit_plane_change}).  One might therefore
expect an association
between obliquity and eccentricity.
It is difficult to check for such a statistical association,
because there are relatively few systems for which both obliquity
and eccentricity are well determined; often, one or the other
has a large observational uncertainty.
The possible effects of tidal interactions also complicate
the interpretation of the data.
\cite{Wang+2021} noted that cool stars with $a/R\lesssim 10$ tend
to have both low obliquity and low eccentricity.  
This could be due to damping of both obliquity and eccentricity by tides
(\S~\ref{sec:tides}), although according to standard theoretical assumptions,
eccentricity damping is mainly caused by dissipation within planets
and obliquity damping is mainly caused by dissipation within the star \citep[see, e.g.,][]{SchlaufmanWinn2013}.

\cite{RiceWangLaughlin2022} found that cool stars with planets on eccentric orbits tend to have higher obliquities than similar
stars with planets on low-eccentricity orbits.
This is an intriguing finding,
although high obliquity is
associated not only with high eccentricity, but also
lower planet masses and wider orbits (Figures \ref{fig:mass_ratio_projected_obliquity} and \ref{fig:ar_projected_obliquity};
see also Appendix B of \citealt{RiceWangLaughlin2022}),
which complicates the interpretation of the results.

Figure~\ref{fig:eccentricity_projected_obliquity} displays the sample of systems for which
$a/R>10$ and the eccentricity is known with a precision of 0.1 or better. This plot is similar
to the one presented by \cite{Rice+2021}, although they omitted planets on circular orbits. The restriction on orbital separation was imposed to select systems that are probably unaffected by tides. 
(See Tables~\ref{tab:stars} and~\ref{tab:planets} for the relevant parameters and references.)

Two of the systems for which the orbit is consistent with circular are misaligned
at the 2-$\sigma$ level:
HAT-P-12 ($\lambda=54^{+13}_{-41}$~deg) and TrES-1 ($31\pm21$~deg). 
KELT-6 is misaligned with a formal significance of 3-$\sigma$ ($36\pm11$~deg),
but the results are more vulnerable than usual to systematic effects
because no pre-ingress data are available. 
The only clearly misaligned system with $e<0.1$ and $\sigma_e<0.1$
is HAT-P-18, which has a retrograde orbit
and an upper limit of 0.087 on the eccentricity.
HAT-P-18 also appears to be one of the oldest systems
in the sample, although with large uncertainties ($12.4^{+4.4}_{-6.4}$~Gyr).
Only HAT-P-22 appears to be comparably old.
It might be relevant that the systems with large obliquities and small eccentricities have sub-Saturn planets.
Under standard assumptions in tidal theory, lower-mass planets are more likely to
undergo orbital circularization prior to realignment of the host star.
There are also four systems with high eccentricities and low obliquities: HAT-P-17, HD\,17156, Kepler-448, and K2-25.

In summary, we do not think the current data provide a clear answer to the question of
whether misalignment is associated with orbital eccentricity. The relevant
sample is small and heterogeneous, and tidal damping of both
eccentricities and obliquities -- possibly at very
different rates -- complicates the picture.

\startlongtable
\begin{deluxetable*}{l D DD  DDD l}
\tablecaption{Key parameters of systems with multiple transiting planets for which
$\lambda$, $i$, or both were determined. 
For the $\lambda$ measurements, we provide the parameters of the planet
for which the measurement was made. 
In the reference column, the boldface number refers to the work from
which we drew the $\lambda$ measurement. 
The other numbers refer to 
works reporting additional measurements of $\lambda$ or other system parameters not taken from \href{https://www.astro.keele.ac.uk/jkt/tepcat/}{TEPCat}  \citep{Southworth2011}.
\label{tab:multitransits}}
\tablewidth{0pt}
\tablehead{
\colhead{Planet} & \multicolumn2c{$T_{\rm eff}$} 
& \multicolumn2c{$a/R$} & \multicolumn2c{$r$}
& \multicolumn2c{$\lambda$} & \multicolumn2c{$i$} & \multicolumn2c{$\psi$} 
& \colhead{References} \\  
 \colhead{ }  & \multicolumn2c{(K)}
& \multicolumn2c{ }  & \multicolumn2c{(R$_{\rm Jup}$)} 
& \multicolumn2c{($^{\circ})$}& \multicolumn2c{($^{\circ})$} & \multicolumn2c{($^{\circ})$}
& \colhead{} \\
}
\decimalcolnumbers
\startdata
AU Micb & 3700\pm100 & 18.92^{+2.15}_{-2.42} & 0.38\pm0.02 & 4.7^{+6.4}_{-6.8} & 90.0^{+0.0}_{-19.5} & 12.1^{+11.3}_{-7.5} &  {\bf 1},2,3,4,5 \\ 
HD 3167b & 5261\pm60 & 4.08^{+0.99}_{-0.46} & 0.14^{+0.00}_{-0.00} & 6.6^{+7.9}_{-6.6} & \multicolumn2c{$-$} & \multicolumn2c{$-$} &  6,{\bf 7} \\ 
HD 3167c & 5261\pm60 & 43.86^{+0.82}_{-0.86} & 0.27^{+0.04}_{-0.03} & 98.0\pm23.0 & \multicolumn2c{$-$} & \multicolumn2c{$-$} &  {\bf 8} \\ 
HD 63433b & 5640\pm74 & 16.74^{+1.00}_{-1.11} & 0.19\pm0.01 & 8.0^{+33.0}_{-45.0} & 90.0^{+0.0}_{-18.7} & 25.6^{+22.5}_{-15.3} &  {\bf 9},2 \\ 
HD 63433c & 5640\pm74 & 38.00^{+4.60}_{-1.70} & 2.71\pm0.14 & 11.0^{+32.0}_{-35.0} & 90.0^{+0.0}_{-18.7} & 25.6^{+22.5}_{-15.3} &  {\bf 10},2 \\ 
HD 106315c & 6364\pm87 & 24.79^{+0.39}_{-0.43} & 0.39\pm0.01 & 10.9^{+3.8}_{-3.6} & \multicolumn2c{$-$} & \multicolumn2c{$-$} &  {\bf 11} \\ 
K2-290b & 6302\pm120 & 13.15\pm0.69 & 0.27\pm0.01 & 173.0^{+45.0}_{-53.0} & 39.0\pm7.0 & 124.0\pm6.0 &  12,{\bf 13} \\ 
K2-290c & 6302\pm120 & 43.50\pm1.20 & 0.77\pm0.05 & 153.0\pm8.0 & 39.0\pm7.0 & 124.0\pm6.0 &  12,{\bf 13} \\ 
Kepler-9b & 5774\pm60 & 32.05\pm0.74 & 0.74\pm0.01 & 13.0\pm16.0 & 71.9^{+18.1}_{-8.8} & 28.1^{+13.0}_{-13.6} &  {\bf 14},2 \\ 
Kepler-25c & 6270\pm79 & 18.62\pm0.24 & 0.47^{+0.01}_{-0.01} & 0.5\pm5.7 & 90.0^{+0.0}_{-21.3} & 5.7^{+4.2}_{-3.2} &  {\bf 15},16,2,17 \\ 
Kepler-30c & 5498\pm54 & 67.86\pm8.57 & 1.07\pm0.03 & 4.0\pm10.0 & \multicolumn2c{$-$} & \multicolumn2c{$-$} &  {\bf 18} \\ 
Kepler-50b & 6225\pm66 & 10.48^{+1.64}_{-2.73} & 0.15^{+0.00}_{-0.01} & \multicolumn2c{$-$} & 82.0^{+8.0}_{-7.0} & \multicolumn2c{$-$} &  19 \\ 
Kepler-56b & 4840\pm97 & 5.23\pm0.26 & 0.58\pm0.03 & \multicolumn2c{$-$} & 47.0\pm6.0 & \multicolumn2c{$-$} &  20 \\ 
Kepler-65b & 6211\pm66 & 5.24^{+0.32}_{-0.18} & 0.13^{+0.00}_{-0.00} & \multicolumn2c{$-$} & 81.0^{+9.0}_{-16.0} & \multicolumn2c{$-$} &  19 \\ 
Kepler-89d & 6182\pm82 & 23.82\pm2.22 & 1.00\pm0.10 & 6.0^{+11.0}_{-13.0} & \multicolumn2c{$-$} & \multicolumn2c{$-$} &  {\bf 21},15 \\ 
TOI-451b & 5550\pm56 & 6.93^{+0.11}_{-0.16} & 0.17\pm0.01 & \multicolumn2c{$-$} & 69.0^{+11.0}_{-8.0} & \multicolumn2c{$-$} &  22 \\ 
TOI-942b & 4928^{+125}_{-85} & 10.12^{+0.13}_{-0.18} & 0.43^{+0.02}_{-0.00} & 1.0^{+41.0}_{-33.0} & 76.0^{+9.0}_{-11.0} & 2.0^{+27.0}_{-33.0} &  23,{\bf 24} \\ 
TRAPPIST-1b & 2557\pm47 & 20.04^{+0.72}_{-0.69} & 0.10^{+0.00}_{-0.00} & 15.0^{+26.0}_{-30.0} & 90.0^{+0.0}_{-17.4} & 23.3^{+17.0}_{-13.6} &  {\bf 25},2 \\ 
TRAPPIST-1e & 2557\pm47 & 50.82^{+1.83}_{-1.75} & 0.08^{+0.00}_{-0.00} & 9.0^{+45.0}_{-51.0} & 90.0^{+0.0}_{-17.4} & 23.3^{+17.0}_{-13.6} &  {\bf 25},2 \\ 
TRAPPIST-1f & 2557\pm47 & 66.85^{+2.40}_{-2.31} & 0.09^{+0.00}_{-0.00} & 21.0\pm32.0 & 90.0^{+0.0}_{-17.4} & 23.3^{+17.0}_{-13.6} &  {\bf 25},2 \\ 
V1298 Taub & 4970\pm120 & 26.06\pm0.46 & 0.92^{+0.05}_{-0.05} & 4.0^{+7.0}_{-10.0} & 51.0^{+25.0}_{-21.0} & 8.0^{+4.0}_{-7.0} &  26,{\bf 27},28,29 \\ 
V1298 Tauc & 4970\pm120 & 13.19^{+0.15}_{-0.13} & 0.92^{+0.05}_{-0.05} & 5.0\pm15.0 & \multicolumn2c{$-$} & \multicolumn2c{$-$} &  26,{\bf 30},28 \\ 
WASP-47b & 5576\pm67 & 9.67\pm0.15 & 1.12\pm0.01 & 0.0\pm24.0 & \multicolumn2c{$-$} & \multicolumn2c{$-$} &  {\bf 31} \\ 
WASP-148b & 5437\pm21 & 19.80\pm1.50 & 0.80^{+0.02}_{-0.02} & 8.2^{+9.7}_{-8.7} & \multicolumn2c{$-$} & \multicolumn2c{$-$} &  {\bf 32},33 \\ 

\enddata
\end{deluxetable*}
\tablecomments{All data were taken from \href{https://www.astro.keele.ac.uk/jkt/tepcat/}{TEPCat}  \citep{Southworth2011} or the following references: 
 1 \citep{Hirano+2020}, 2 \citep{Albrecht+2021}, 3 \citep{Martioli+2020}, 4 \citep{Palle+2020}, 5 \citep{Addison+2021}, 6 \citep{Christiansen+2017}, 7 \citep{Bourrier+2021}, 8 \citep{Dalal+2019}, 9 \citep{Mann+2020}, 10 \citep{Dai+2020}, 11 \citep{Zhou+2018}, 12 \citep{Hjorth+2019}, 13 \citep{Hjorth+2021}, 14 \citep{Wang+2018}, 15 \citep{AlbrechtWinnMarcy+2013}, 16 \citep{Campante+2016}, 17 \citep{Benomar+2014}, 18 \citep{Sanchis-Ojeda+2012}, 19 \citep{Chaplin+2013}, 20 \citep{Huber+2013}, 21 \citep{Hirano+2012}, 22 \citep{Newton+2021}, 23 \citep{Zhou+2021}, 24 \citep{Wirth+2021}, 25 \citep{Hirano+2020b}, 26 \citep{Davide+2019b}, 27 \citep{Johnson+2022}, 28 \citep{Biddle+2014}, 29 \citep{Gaidos+2022}, 30 \citep{Feinstein+2021}, 31 \citep{Sanchis-Ojeda+2015}, 32 \citep{Wang+2021_wasp148}, 33 \citep{Hebrard+2021}
}

\subsubsection{Interlude: Mutual orbital inclinations}
\label{sec: Mutual orbital inclinations}

Most of the available RM data are for stars with close-orbiting giant planets. Many of the
theories for obliquity excitation involve the same processes that are hypothesized
to allow giant planets to form far from the star and move inward.
On the other hand, there might be reasons for high obliquities that are
unrelated to close-orbiting giant planets, such as `primordial' misalignments
between stars and their protoplanetary disks, i.e., occurring before planet formation. This makes it interesting to
expand the domain of obliquity measurements to smaller and wider-orbiting planets.
  
The {\it Kepler} mission led to the discovery of hundreds of systems of multiple
transiting planets in compact configurations.
The known planets are typically between the Earth
and Uranus in size and have orbital periods shorter than a year.
Through various means, the mutual inclinations between the orbits
have been estimated to be $\lesssim$\,5$^\circ$  \citep{Fabrycky+2014,Xie+2016,Zhu+2018,HermanZhuWu2019}.
Thus, the compact multi-planet systems appear to be at least as flat as the Solar System,
and some are definitely much flatter \citep[see, e.g.,][]{Agol+2021}.
The very innermost planets with periods $\lesssim$\,1~day
tend to have somewhat larger mutual inclinations than more distant planets
\citep{DaiMasudaWinn2018}, but even in those cases the angles are $\lesssim$15$^\circ$.
If we assume the nearly coplanar orbits trace out the plane
of the long-gone protoplanetary disk, then we can test for
primordial misalignments by measuring the stellar obliquity. 
More generally, we can test for differences in
obliquity excitation and evolution between stars with compact multi-planet systems
and those with close-orbiting giant planets.

\subsubsection{Compact multi-transiting systems: aligned with notable exceptions}
\label{sec:multitransits}
The first five obliquity measurements of stars with compact multi-transiting systems
were all consistent with good alignment \citep{AlbrechtWinnMarcy+2013}.
Since then, the sample size has increased to  18 stars.
Of these, 15 have low obliquities.
There are also three stars with high obliquities:
Kepler-56 \citep{Huber+2013},
HD\,3167 \citep{Dalal+2019,Bourrier+2021},
and K2-290\,A \citep{Hjorth+2021}.
Thus, high obliquities are not confined to stars with
close-orbiting giant planets.
In a fourth system, Kepler-129, there is tentative evidence of a high obliquity from
asteroseismology \citep{Zhang+2021}.
As discussed in Section~\ref{sec:theory}, the high obliquities in these systems
might have been caused by gravitational perturbations from
wide-orbiting companions --- a stellar companion
in the case of K2-290\,A, and a planetary-mass companion in the case of Kepler-56.
The properties of the multi-transiting systems are included
in Tables~\ref{tab:stars} and \ref{tab:planets}, along with those
of the single-transiting systems. 
Table~\ref{tab:multitransits} highlights the data
for multi-transiting systems, in particular,
including whether spin-orbit information is available for more than one planet.

\subsection{Asteroseismology}
\label{sec:seismic}

\begin{figure}
  \begin{center}
    \includegraphics[width=8cm,angle=0]{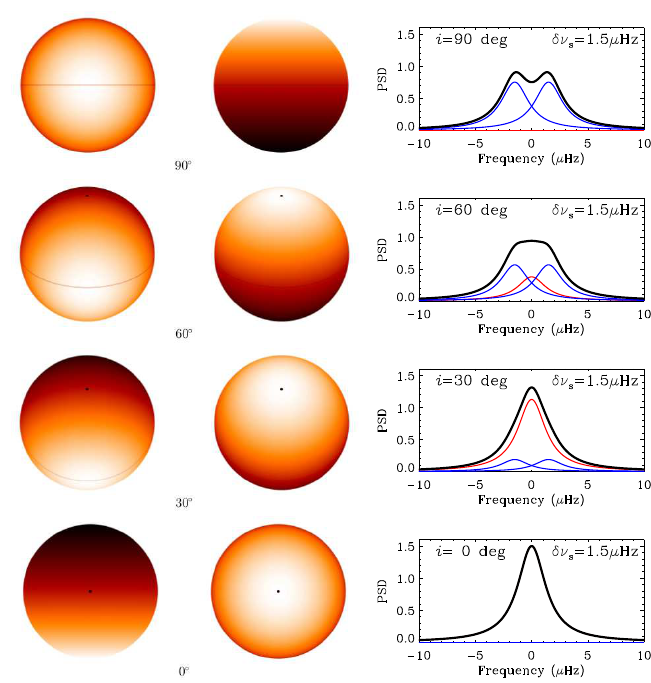} 
    \caption {\label{fig:asteroseismology_principle}
    {\bf Mode visibility for different stellar inclinations.} 
{\it Left and middle columns}: Theoretical intensity perturbations for an $l = 1$ mode with $m=1$ (left) and $m= 0$ (middle). The top row is for a line of sight along the equatorial plane ($i=90$) while the lowest row is for a line of sight along the spin axis ($i=0$). {\it Right column}: Theoretical frequency profiles of the $l=1$, $m=\pm1$ (blue lines) and $m=0$ (red) modes, as well as the combined signal (black line). The  observed amplitude of the $m=\pm1$ modes is reduced
as the inclination is lowered. The reverse is true for the $m=0$ mode. Adapted from
Figures~6 and 7 of \cite{Chaplin+2013}.
    }
  \end{center}
\end{figure}

\begin{figure*}
  \begin{center}
    \includegraphics[width=16cm,angle=0]{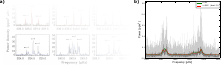} 
    \caption {\label{fig:Asteroseismology}
    {\bf Stellar inclinations from asteroseismology.}
    Shown are the power spectra of two
    transiting-planet hosts, based on {\it Kepler} light curves.
    {\it Left:} Kepler-56, a subgiant star \citep{Huber+2013}.
    The top row shows gravity-dominated modes and the bottom row
    shows pressure-dominated dipole modes. Each mode
    is split into a triplet by rotation.
    All three modes are visible, indicating a moderate inclination.
    Quantitatively, the inclination was found to be $47\pm6^\circ$.
    {\it Right:} Kepler-410, a hotter and less evolved star \citep{VanEylen+2014}.
    The data are shown as gray and dark gray curves (after smoothing).
    The red and green curves are models with $i=90^\circ$ and $45^\circ$.
    Although the triplets are not as well resolved as in Kepler-56,
    the $m=\pm 1$ modes are more prominent than the $m=0$ mode,
    indicating a high inclination.
    }
  \end{center}
\end{figure*}

Just as observing vibrations of the Earth's crust can be used to study
wave propagation in the Earth's mantle and core, observing the vibrations
of stellar surfaces can be used to study the deep interiors of stars.
This technique, asteroseismology, has seen astonishing advances
over the last 15 years due to the availability of photometric
time-series data of long duration, fine time sampling,
and high signal-to-noise ratio \citep{Kurtz+2022}.
Among the information that can be retrieved from
the frequency spectrum of a star's oscillation modes
is the inclination of the star's rotation axis with respect to the
line of sight
\citep{GoughKosovichev1993,GizonSolanki2003,ChaplinMiglio2013}.

Similar to the eigenstates of the hydrogen atom,
the oscillation modes of a star are classified by three integers:
the radial order $n$, the latitudinal degree $l\leq n$, and the azimuthal order $m$
which ranges from $-l$ to $l$.  
Mode patterns with different $m$ but the same $n$ and $l$ are related
by rotation, and therefore, they would have the same frequency
for a stationary spherical star.
Stellar rotation breaks this symmetry.
Each degenerate group of $2l+1$ modes is split into a multiplet with
\begin{equation}
\label{equ:split}
\nu_{nlm} \approx \nu_{nl} + \frac{m\Omega}{2\pi}\,\,\,, 
\end{equation}
where $\Omega$ is the star's average angular rotation rate.
For the Sun, $\Omega/2\pi$ is approximately 0.43\,$\mu$Hz.

For Sun-like stars, the modes are stochastically excited by convection,
making it reasonable to assume that the modes in a given multiplet
are excited to the same intrinsic amplitude.
However, the observed mode amplitudes in a power spectrum
depend on the viewing angle, because they are based on the integrated flux over the visible
stellar hemisphere.
Consider dipole modes ($l=1$).
When $i=90^\circ$, the $m=0$ mode is invisible
because the intensity pattern of the mode is antisymmetric about the stellar equator,
and integrates to zero over the entire disk.  The $m=\pm 1$ modes are seen
at maximum amplitude (Figure~\ref{fig:asteroseismology_principle}).
In contrast, when $i=0^\circ$, the $m=0$ mode is seen with maximum amplitude
and the $m=\pm 1$ modes are invisible.
In general, as shown
by \cite{GizonSolanki2003}, the relative amplitudes are proportional to
\begin{equation}
\begin{aligned}
\mathscr{E}_{l=1,m=0} &= \cos^2 i,\\
\mathscr{E}_{l=1,m=\pm 1} &= \frac{1}{2}\sin^2 i,\\
\mathscr{E}_{l=2,m=0} &= \frac{1}{4}(3\cos^2 i - 1)^2,\\
\mathscr{E}_{l=2,m=\pm 1} &= \frac{3}{8}\sin^2 2i,\\
\mathscr{E}_{l=2,m=\pm 2} &= \frac{3}{8}\sin^4 i.\\
\end{aligned} 
\end{equation}

Therefore, by measuring the relative amplitudes of the modes in each multiplet,
we can determine the inclination. When this technique succeeds for
a star with a transiting
planet, we can compare the orbital and stellar inclinations and thereby
obtain a constraint on the stellar obliquity.
More details on the extraction of the inclination from the photometric
power spectrum are given by
\cite{GizonSolanki2003,BallotGarciaLambert2006,Ballot+2008} and \cite{Kuszlewicz+2019}.
See also \cite{Grundahl+2017}, who discuss
the extraction of the inclination
from radial-velocity data, 
instead of photometric data.

The observational requirements are demanding.
To measure the relative amplitudes of the members of the multiplets,
and thereby constrain the inclination,
it is necessary to resolve the multiplets by obtaining a long-duration time
series.  For uniformly-sampled data, the frequency resolution is $\sim$\,$1/T$
where $T$ is the total duration of a time series.  By this criterion,
resolving a Sun-like splitting
of 0.43\,$\mu$Hz requires a total duration of a month. In most applications, a much longer
time series has been needed to achieve a sufficiently high signal-to-noise ratio.
In addition, the intrinsic frequency widths of the modes need to be
small in comparison to the splittings,
and the mode amplitudes need to be large enough to detect, factors that favor
evolved and rapidly-rotating stars.
Figure~\ref{fig:Asteroseismology} illustrates the method with
two cases: one in which the multiplets are well resolved (left panel),
and another in which they are barely resolved (right panel).

The left panel features the most striking result so far from
the asteroseismic technique for inclination determination: the high obliquity of the
subgiant star Kepler-56 \citep{Huber+2013}.
This star has two transiting planets of radii $6.5$ and $9.8\,R_\oplus$,
which are engaged in a 2:1 mean-motion resonance. The stellar and orbital
inclinations differ by $45^\circ$. This system is discussed further
in \S~\ref{sec:theory}.
Using the same method, \cite{Zhang+2021} found tentative evidence for a misalignment
in the Kepler-129 system,
which has two transiting sub-Neptunes.
\cite{KamikaBenomarSuto+2019} found a misalignment of at least
$42^\circ$ in the Kepler-408 system, which has a super-Earth
on a 2.5-day orbit.  

So far, those are the only misalignments that have been reported
using the asteroseismic method.  \cite{Chaplin+2013} found that the data for the
Kepler-50 and Kepler-65 systems, which have multiple transiting super-Earths,
are consistent with low obliquities. 
\cite{VanEylen+2014} found agreement between $i$ and $i_{\rm o}$
for Kepler-410, a multi-planet system featuring a transiting mini-Neptune on an eccentric orbit  (see the right panel of Figure~\ref{fig:Asteroseismology}).
\cite{Campante+2016} performed a comprehensive study of the 25 Kepler
systems that appeared to have the most favorable properties for the asteroseismic technique.
They confirmed that HAT-P-7 has a nearly polar orbit, as had been found
by \cite{Lund+2014}. Since RM data were also available for this retrograde-rotating
star \citep{Winn+2009,Narita+2009_HAT-P-7,AlbrechtWinnJohnson+2012}, HAT-P-7 was one of the first stars for which
the 3-d obliquity ($\psi$) could be determined.
The data for the rest of the systems analyzed by \cite{Campante+2016} were consistent with good
alignment, although in many cases only weak constraints could be obtained
(e.g., $i \gtrsim 20^\circ$). 

The RM effect would have been difficult to detect
for most of the systems described in this section, because of the small signal sizes.
As noted in \S~\ref{sec:methods results},
the asteroseismic signal depends only on the properties of the star, rather
than the planet.  Once a transiting planet has been detected, the
applicability of the asteroseismic method is independent of the properties
of the planet.  At this point, though, the only successful applications
of this method have been based on data from the Kepler mission.
More detections might come from the ongoing Transiting Exoplanet
Survey Satellite (TESS) mission \citep{Ricker+2015}, 
and especially the future
PLATO mission \citep{PLATO2014}.

\subsection{Spectro-interferometry}
\label{sec:inter_and_spec}

\begin{figure}
  \begin{center}
  \includegraphics[width=7.5cm]{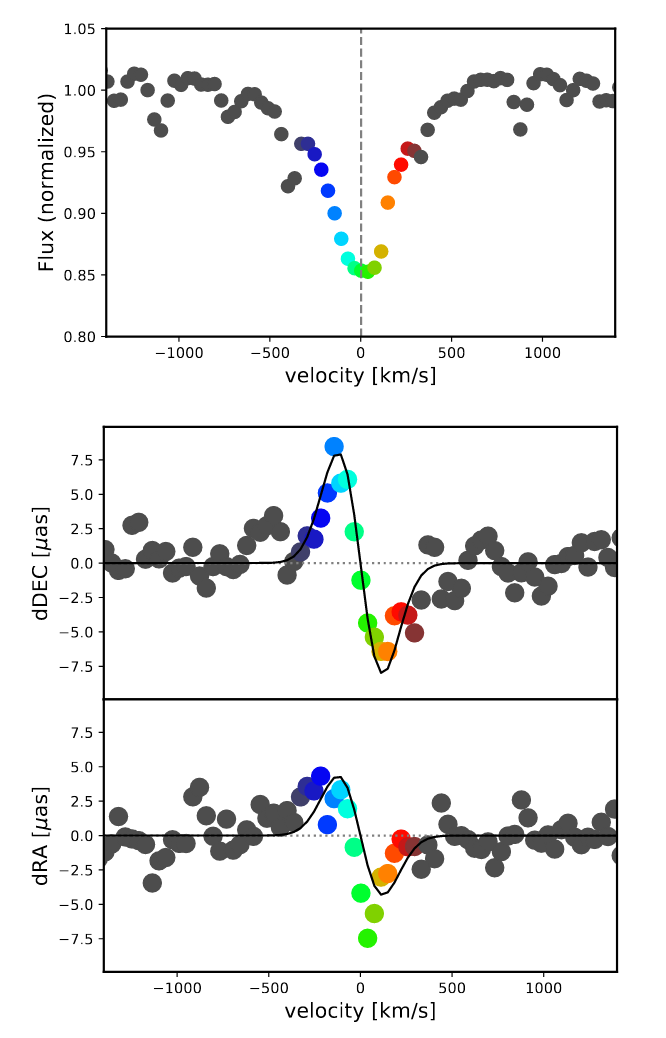} 
  \caption {\label{fig:interferometry} {\bf Spatially resolved Br~$\gamma$ absorption line of $\beta$~Pic.}  From \cite{Kraus+2020}. The top panel displays the spatially integrated Br~$\gamma$ absorption line.
  The two lower panels display the spatial
  offsets of the photocenter relative to the continuum.
  The approaching (blue) side of the star is displaced
  north and east of the receding (red) side of the star.
  }
  \end{center}
\end{figure}

Optical and near-infrared interferometers can
resolve the disks of main-sequence stars in the solar neighborhood.
If equipped with a spectrograph capable of resolving stellar absorption lines, an
interferometer can
measure the orientation of the stellar rotation axis as projected on the
sky plane \citep{Albrecht+2010}.
The approaching and receding halves of the stellar disk have slightly
different locations on the sky. Therefore, the photocenters of the
red and blue wings of a rotationally-broadened absorption line are slightly
displaced.
This displacement is manifested as a phase shift in the fringe pattern recorded
by a pair of telescopes in an interferometer.
For marginally resolved stars, the differential phase between the interferometric
fringes of the two photocenters is (see \citealt{LeBouquin+2009}, and
Eqn.~B.5 of \citealt{Lachaume2003})
\begin{equation}
	\label{equ:interferometry}
    \rho=-2\pi p \frac{B}{\lambda_{\rm w}} {\rm [rad],}
\end{equation}
where $B$ is the baseline length between the telescopes, $p$ is the angular sky separation of the two photocenters projected onto $B$, and $\lambda_{\rm w}$ is the observing wavelength.
More details are given by \cite{Petrov1989} and \cite{ChelliPetrov1995}.

So far, this technique has worked in two cases:
(1) \cite{LeBouquin+2009} showed that Fomalhaut is aligned on the sky
with its debris disk to within a few degrees, and
(2) \cite{Kraus+2020} showed
that $\beta$~Pic is aligned with its
debris disk and with the orbital planes of two directly imaged planets (see Figure~\ref{fig:interferometry}).
Good alignment for $\beta$~Pic had already been suspected, based on
asteroseismology \citep{Zwintz+2019}.
Both of these stars are bright and rapidly rotating,
and in both cases the inclination was derived
from the pressure-broadened Br$\gamma$ line.
It would be advantageous to observe metal absorption lines, because they are
not subject to strong pressure broadening, but no current interferometer
can resolve metal lines in late-type main-sequence stars. 
Preparations are underway for interferometers with higher spectral
resolving power at the CHARA array \citep{Mourard+2018} and the VLT Interferometer  \citep{Kraus2019}. These newer instruments will be able to perform improved fringe tracking,
which will increase the allowable integration time of the spectrograph
and thereby extend the applicability of the technique to fainter stars.

For this method to provide obliquity information,
we must also know the position angle on the sky of the plane of the
surrounding disk, or a planetary orbit.
The position angle of a planetary orbit cannot be obtained from RV or transit observations,
but it can be obtained from direct-imaging or astrometric observations.
Time-series astrometry from the Gaia mission
is expected to lead to the detection of thousands of exoplanets
for which the orbital orientation will be known
\citep{Perryman2014}.
The brightest and nearest stars with Gaia planets might provide a new
pool of targets for interferometric observations of stellar obliquities.

Some other relevant and potentially observable interferometric effects are
worth noting.  In the presence of differential rotation,
it may be possible for interferometry to constrain
the stellar inclination \citep{Souza+2004}.
For the most rapidly-rotating and centrifugally flattened stars,
interferometry can determine the star's sky-projected shape
without the need for spectroscopy \citep[see, e.g.,][]{Souza+2003}.

In addition to studying planetary systems,
interferometers can be used to study the alignment of stars in binary systems,
star-forming regions, and stellar clusters. Such measurements would shed light on
the initial conditions of star and planet formation (as discussed
further in section \ref{sec:theory}).

\subsection{The projected rotation velocity technique}
\label{sec:vsini}

\begin{figure*}
  \begin{center} 
        \includegraphics[width=1.0\textwidth]{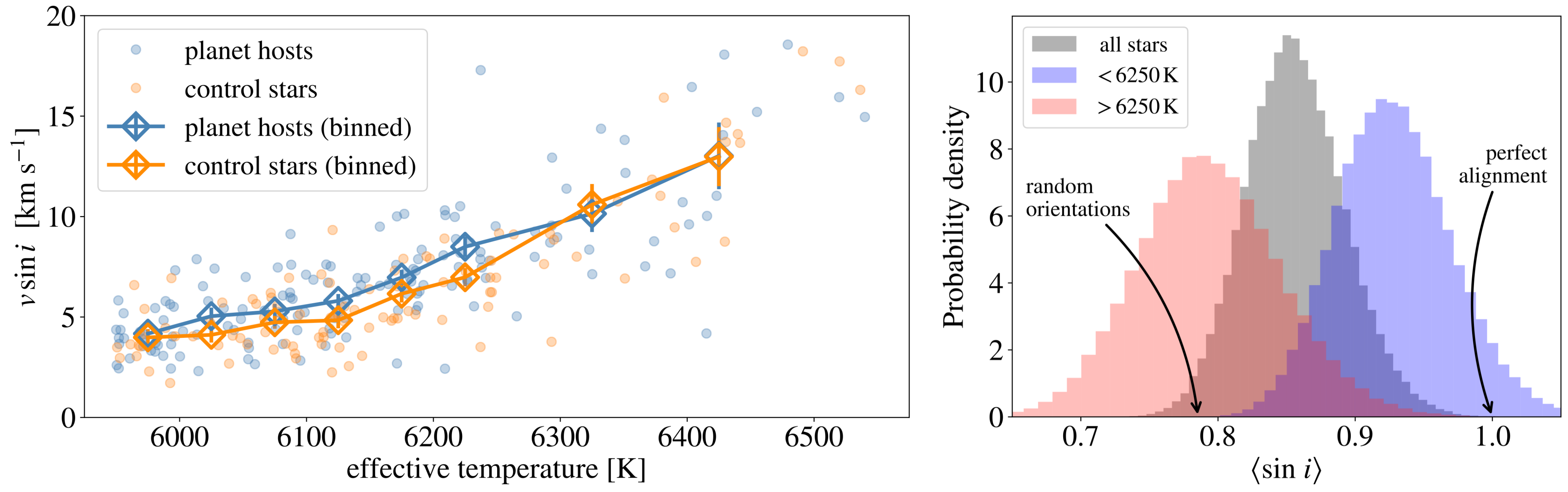}
    \caption { \label{fig:vsini}
    {\bf The projected rotation velocity method applied to Kepler stars,} from \cite{Louden+2021}.
    {\it Left}: Projected rotation velocities of stars with and without detected transiting planets.
    For stars cooler than 6250\,K (the Kraft break), the average value of $v \sin i$ is higher for transit hosts than
    for the non-transit hosts, by approximately the factor one would expect if the transit hosts have
    low obliquities and the other stars are randomly oriented.
    For hotter stars, no difference is detected.
    {\it Right}: Posterior probability distributions for the mean value of $\sin i$ of the
    cool stars (blue), the hot stars (red), and the entire sample (gray).
    }
  \end{center}
\end{figure*}

By combining estimates of a star's radius, $R$, rotation period, $P_{\rm rot}$,
and projected rotation velocity, $v\sin i$, and neglecting any differential
rotation, the inclination can be calculated as
\begin{equation}
\label{equ:vsini_method}
i = \sin^{-1} \left( \frac{v\sin i}{v} \right) = \sin^{-1} \left( \frac{v\sin i}{2\pi R/P_{\rm rot}} \right)\,\,.
\end{equation}
This method is not as straightforward as it might seem.
Measurements of $v\sin i$ are obtained by analyzing a star's spectral
absorption lines or cross-correlation function. The contribution to line
broadening from rotation must be separated from those of other sources, such
as turbulence and the finite instrumental resolution.
To put this into perspective, consider that
the Sun's rotation velocity is 1.9~km\,s$^{-1}$
and the best available spectrographs have a spectral resolution of $\sim$\,$10^5$,
corresponding to a velocity resolution of 3\,km\,s$^{-1}$. 
Thus, data with a very high signal-to-noise ratio are required
to disentangle the sources of broadening for solar-type stars -- and even
then, the results are subject to systematic uncertainties.

The rotation period can be measured by observing the quasiperiodic brightness
variations due to rotating spots and plages.
The advent of the Kepler and TESS missions has made it possible
to measure the rotation periods of thousands of planet-hosting stars
(\citealt{McQuillanMazehAigrain2013,Mazeh+2015method}; see also the review by
\citealt{Maxted2018}).
When the rotation period of a solar-type star
cannot be directly measured (because there are no
suitable data, or because photometric modulation is not detected), one
can guess the rotation period 
based on the star's mass and age. This is because the rotation of
main-sequence solar-type
stars tends to
slow down according to the \cite{Skumanich1972} law,
$P_{\rm rot}\propto t^{1/2}$, with a coefficient depending
on mass \citep[e.g.][]{Barnes2010,EpsteinPinsonneault2014,Meibom+2015,vanSaders+2016}. However, this is not a safe procedure to follow
for stars with hot Jupiters.  Such stars need not obey the usual
pattern because of tidal spin-up
\citep[see, e.g.,][]{TejadaArevalo+2021}.

A subtlety in the statistical inference of $i$ from
Equation~\ref{equ:vsini_method}
is that $v$ and $v\sin i$ are not independent variables.
\cite{MasudaWinn2020} discussed the diagnosis and treatment of this problem.
A relatively simple method to
account for the interdependent variables is
to employ a Monte Carlo Markov Chain method in which the parameters
are $R$, $P_{\rm rot}$, and $\cos i$, with a uniform prior on $\cos i$
and measurement-informed priors on $R$, $P_{\rm rot}$,
and $(2\pi R/P_{\rm rot})\sqrt{1-\cos^2 i}$.

In an important early application of the $v\sin i$ technique,
\cite{Schlaufman2010}
studied a sample of 75 transiting planets for which the stellar rotation
periods had been estimated from an age-mass-rotation relationship.
He found 10 stars with unusually low values of $v\sin i$,
suggesting that they have high obliquities. This was true even
though most of the planets were hot Jupiters and might have been
expected to spin up their stars. All 10 misaligned stars were
in the mass range from 1.2 to 1.5\,$M_\odot$, despite such stars
constituting only $\approx$40\% of the sample.
This suggested that 
misalignments are preferentially found in massive stars, although
this was not necessarily a valid conclusion because 
the $v\sin i$
method is less sensitive to misalignments of lower-mass
stars, due to their slower rotation.
Nevertheless, results from the RM effect and other methods
have indeed shown that misalignments in hot-Jupiter systems are more 
common for massive stars hotter than $\approx$6250~K (see \S~\ref{sec:teff}).

This method was further developed to take advantage of the large sample of
transiting planets and planet candidates supplied by the Kepler mission.
The Kepler data not only led to the detection of thousands of planets, but also
the measurement of the rotation period for about one-third of the
host stars. Therefore, to apply the $v\sin i$ method, the only new observations
that were required were a single good spectrum of each host star.  This made
it possible to study the obliquity distribution of large samples of stars
with non-giant
planets, for the first time. The first such studies,
by \cite{Hirano+2012}, \cite{WalkowiczBasr2013}, and \cite{Hirano+2014},
identified some candidate misaligned systems.  \cite{MortonWinn2014} took a different
approach: Instead of searching for individual cases of misalignment, they 
performed a hierarchical Bayesian analysis
of the obliquity distribution of 70 Kepler stars.
In terms of the vMF model (see \S~\ref{sec:geometry}), they found tentative evidence
($p\approx 0.03$)
that stars with multiple detected transiting planets have a higher $\kappa$
(lower obliquities) than stars with only a single detected transiting planet.

\citet{Winn+2017} performed a larger and more homogeneous study as part of 
the California Kepler Survey \citep[CKS;][]{Johnson+2017,Petigura+2017},
which obtained high-resolution optical spectra of $\approx$\,$10^3$ Kepler stars.
The analysis of this expanded sample did not confirm the previously reported candidate
misalignments, nor 
did it confirm the difference between single and multiple-transiting planets.
\cite{MunozPerets2018} performed a similar study.
In general, both groups found the Kepler stars with non-giant planets
to have obliquities lower than about $30^\circ$.

\begin{figure*}
  \begin{center}
    \includegraphics[width=0.5\textwidth]{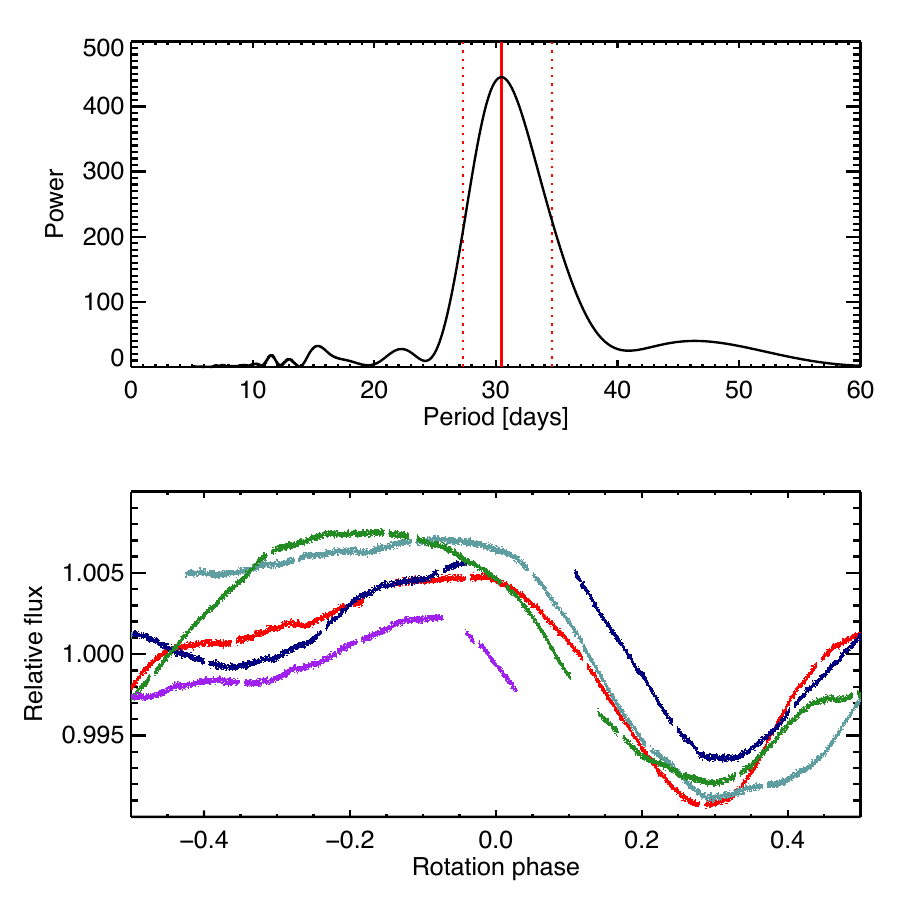}
    \includegraphics[width=0.4\textwidth]{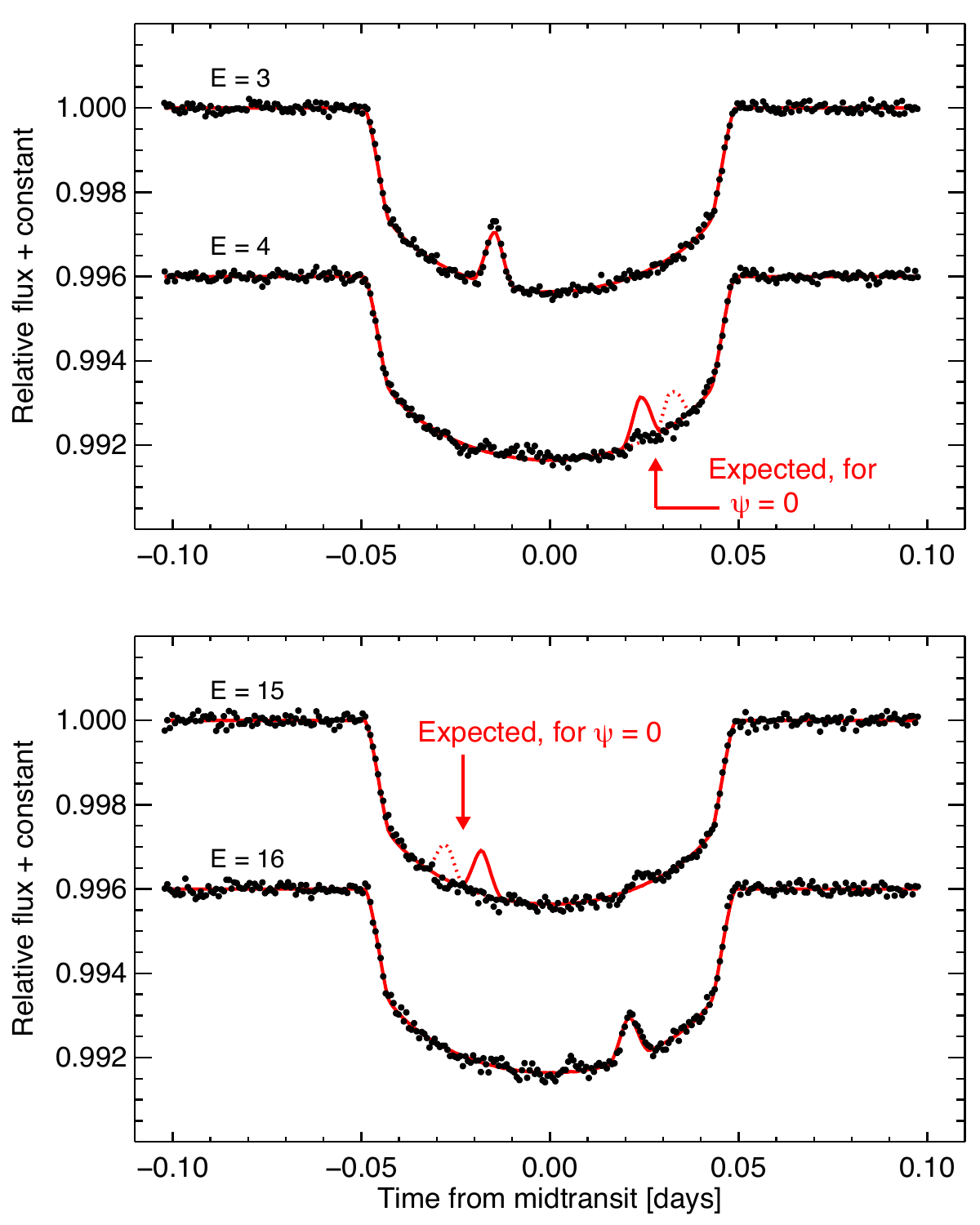}
    \caption {\label{fig:spots} {\bf QPVs and spots crossing transits}.
    From \cite{Sanchis-OjedaWinn2011}.
    {\it Upper left:}~Lomb-Scargle periodogram of {\it Kepler} photometry of HAT-P-11,
    indicating a rotation period of $30.5_{-3.2}^{+3.1}$~days.
    {\it Lower left:} Phase-folded light curve, illustrating
    the quasi-periodic variability.
    {\it Upper right:} Light curves of two consecutive transits.
    Given the orbital and rotation periods, the starspot-crossing anomaly seen in the first
    transit should have appeared in the second transit (red line), if the
    star had a low obliquity.
    {\it Lower right.} A different pair of transits illustrating the
    same effect.
    }
  \end{center}
\end{figure*}

A problem with these population-level analyses
is that the stars with
measurable rotation periods might be biased with respect to age and inclination,
complicating the interpretation of the results. To overcome this problem,
\cite{Louden+2021} returned to the method of \citet{Schlaufman2010}, wherein
the $v\sin i$ distribution of a sample of transiting-planet hosts is
compared with the $v\sin i$ distribution of a sample of control stars.
The control stars were selected to have a distribution of masses, ages, and metallicities
indistinguishable from those of the planet hosts. To the extent that the planet
hosts have low obliquities, their $v\sin i$ distribution should be shifted to higher
values. Using this method, \cite{Louden+2021} found evidence for a difference between the hot and cool
{\it Kepler} stars with transiting planets, see Figure~\ref{fig:vsini}. The stars cooler than 6250\,K 
had $\langle \sin i\rangle = 0.928\pm 0.042$ (1.7-$\sigma$ away from zero obliquity),
while the hotter stars had $\langle \sin i\rangle = 0.794\pm 0.052$ (consistent
with random orientations).  Thus, they suggested that
hot stars with Kepler-like planets
($R\lesssim 4/\,R_\oplus$, $P\lesssim 1$\,year) have a broad obliquity distribution,
as was already known
to be the case for hot Jupiters.
A similar finding was reported earlier by \cite{Mazeh+2015KOIs} using
a technique described below.
These results pertain mainly to stars with compact multi-planetary systems, which
dominate the Kepler sample.  Even the Kepler stars with only a single detected transiting
planet probably are thought to have additional, non-transiting planets most
of the time \citep[see, e.g.,][]{Zhu+2018,Millholland+2021}.

\subsection{Starspots}
\label{sec:starspots}

\subsubsection{Quasi-Periodic Variation}
\label{sec:QPV}

Any inhomogeneities on a star's photosphere (which we will call spots, for simplicity)
will lead to variations in the star's observed brightness
with a period equal to the star's rotation period and its harmonics.
These variations are quasi-periodic, rather than strictly periodic,
because of the evolution of spot properties with time as well as differential rotation.
As described in the preceding section, observing the quasi-periodic variation (QPV)
can be used to measure the rotation period, which can be combined with
$R$ and $v\sin i$ to measure or constrain the stellar inclination.

The QPV amplitude, by itself, also bears some information about inclination.
Sun-like stars tend to have spots appear near the equator. For such stars,
all other things being equal, the observed QPV amplitude is maximized at high
inclination. This is because in
the high-inclination configuration, the spots appear and disappear completely from view.
At low inclination, the spots are limb-darkened and circulate around
the stellar disk, leading to weaker photometric variability.
(The QPVs of stars with polar or nearly-polar spots, such as young and rapidly rotating
stars, might show the opposite trend.)

The amplitude of the QPVs also depends on the spot pattern,
which varies with time and from star to star. \cite{Mazeh+2015KOIs} dealt with this
complication by averaging the QPV amplitudes of large numbers of stars drawn
from the {\it Kepler} survey.  When considering stars cooler than the Sun,
they found the average QPV amplitude of stars without transiting planets (``control stars'')
to be about 0.8 times that of stars with transiting planets.
This is consistent with a picture in which
the QPV amplitude is proportional to $\sin i$,
the transit hosts have low obliquities ($\sin i=1$),
and the control stars are randomly oriented ($\langle \sin i\rangle \approx 0.8$).
Thus, this study supported the idea that the cool Kepler stars tend to have
low obliquities.

For stars between 5800 and 6100\,K, \cite{Mazeh+2015KOIs} found the
transit hosts and the control stars to have similar QPV amplitudes.
This suggests that the pattern found earlier for stars
with close-orbiting giant planets --- hotter stars have
a broader obliquity distribution --- is also shared
by stars with smaller Kepler-type planets.
The transition temperature of 6000~K in the QPV study appears
to be a few hundred degrees lower than that of the RM studies (\S~\ref{sec:teff}),
although it is difficult to compare them directly.

For stars hotter than 6100\,K, \cite{Mazeh+2015KOIs} found the control stars to have {\it higher}
QPV amplitudes than the planet hosts, with an amplitude ratio of about 1.6.
This inverted ratio was surprising.  Interpreted purely in terms of geometry,
and assuming that the latitudinal distribution of starspots of hot stars is similar to that
of cool stars (which might be incorrect),
the results imply that stars tend to rotate perpendicular to the planetary orbits.
At least part of the reason for the inverted ratio is a selection effect: Transiting planets
are easier to detect when the QPV amplitude is low.
However, \cite{Mazeh+2015KOIs} argued that this bias is not likely
to be responsible for the entire observed effect.  There
is also relatively new evidence for 
a tendency toward perpendicularity
based on individual obliquity measurements (\S~\ref{sec:ppp}).
It will be interesting to probe further with these two techniques
and understand the relationship (if any) between the results.

\subsubsection{Starspot-tracking method}
\label{sec:starspot-tracking}

If a transiting planet moves in front of a starspot,
the light curve shows a glitch.
During such a spot-crossing event, the portion of the star blocked by the planet
is not as bright as the surrounding photosphere,
thereby reducing the usual loss of light.
When the spots are dark and localized, the glitches take the
form of brief and partial rebrightenings, as observed for HAT-P-11 \citep[][see Figure~\ref{fig:spots}]{Sanchis-Ojeda+2011}.
When the photosphere is mottled with a more complex pattern,
the light curve becomes jagged, as observed for Kepler-17 \citep{Desert+2011}.

A sequence of transit observations of a star with long-lived spots can sometimes reveal the obliquity,
or at least reveal whether the obliquity is lower than about 10$^\circ$.
When the obliquity is lower than about 10$^\circ$,
the transiting planet's trajectory is aligned well enough
with the trajectories of spots as they move
across the stellar disk to allow for recurrences of spot-crossing events at predictable times.
When the obliquity is high, recurrences require a special coincidence and are rarer.
This type of logic was used to
argue that HAT-P-11, Kepler-63, and WASP-107 have high obliquities
\citep{Sanchis-Ojeda+2011, Sanchis-Ojeda+2013, DaiWinn2017},
and in all three cases the misalignments were confirmed via the RM effect
\citep{Winn+2010_hatp11,Sanchis-Ojeda+2013,Rubenzahl+2021}.
The analysis of starspot crossings was also the technique used in
the first obliquity measurement for a star with multiple transiting
planets \citep[Kepler-30;][]{Sanchis-Ojeda+2012}.
Many other stars have been shown to have low obliquities through this
technique, including 10 hot-Jupiter systems analyzed by
\cite{Dai+2018} using a statistical test for
correlations between the residuals of a sequence of
transit light curves.

\begin{figure}
  \begin{center}
        \includegraphics[width=8cm]{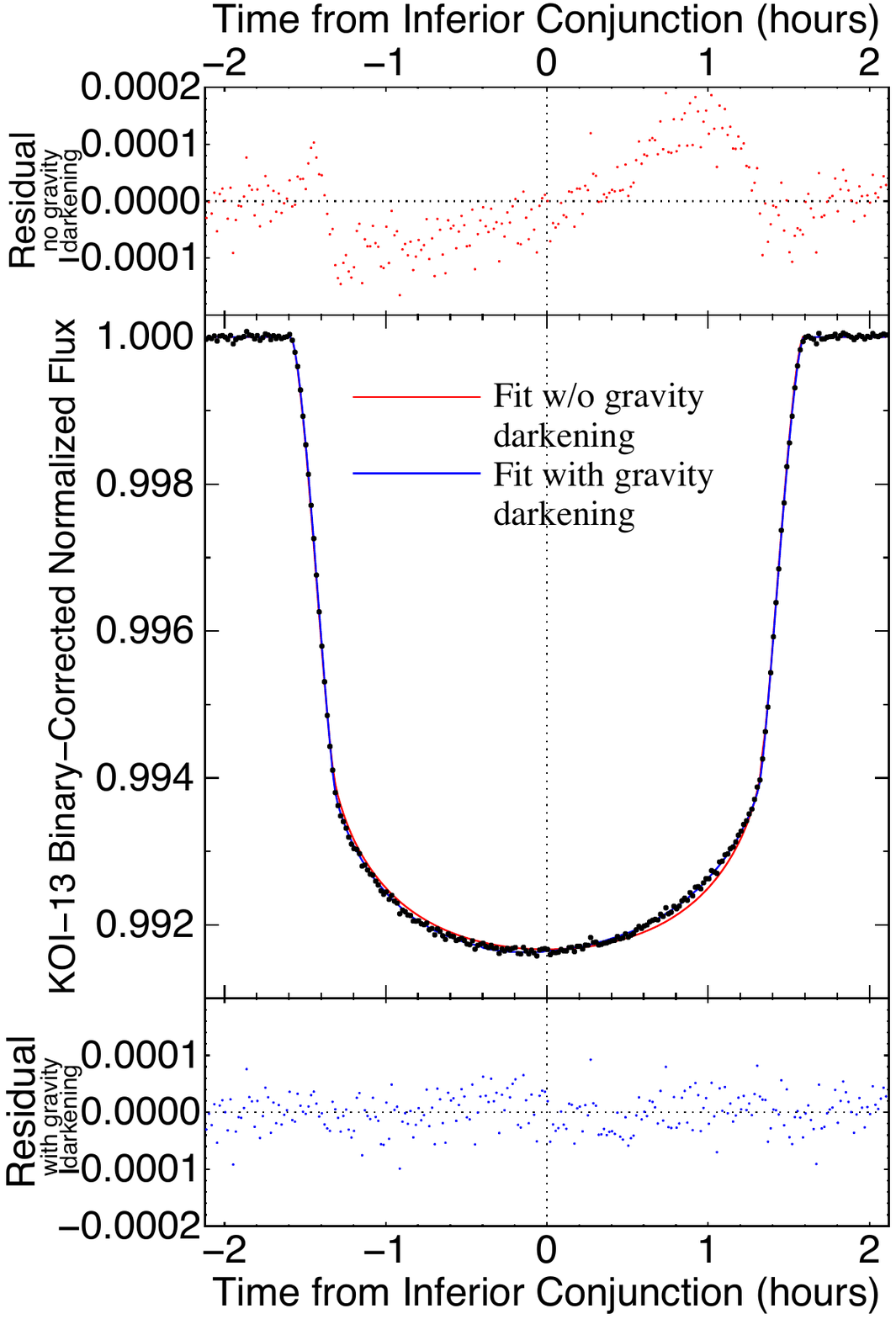}
    \caption { \label{fig:GD} {\bf Gravity darkened transit light curve}. From \cite{BarnesLinscottShporer2011}. {\it Middle}: Phase folded transit light curve of the fast rotating A type star in the Kepler-13 system. {\it Top}: Residuals relative to a light curve model without the effects of gravity darkening. {\it Bottom}: Residuals relative to light curve model including the effects of gravity darkening and a non zero misalignment. }
  \end{center}
\end{figure}

Even when spot-crossing events cannot be well resolved, it is sometimes possible
to tell whether a star is rotating in a prograde or retrograde direction.
Suppose there is a spot on the approaching side of the star, at a latitude
that coincides with the trajectory of a transiting planet.
The stellar flux will be observed to decrease slowly as the spot
rotates toward the meridian.  A transiting planet on a prograde orbit
would encounter the spot in the first half of the transit. If, instead, the planet's orbit
is retrograde, then the spot-crossing anomaly would occur during the
second half of the transit.
Therefore, one can distinguish prograde from retrograde orbits by comparing
the overall asymmetry
of the transit light curve and the trend in the stellar flux outside the transit. Complications
arise when the star has multiple spots. Nevertheless, a few systems
have been shown to have prograde orbits in this fashion
\citep{Nutzman+2011,Mazeh+2015KOIs,Holczer+2015}.

\subsection{Gravity darkening, fast rotators}
\label{sec:gravity}

Another reason for intensity variations across the disk of a star is
gravity darkening.  The equatorial zone of a rapidly rotating star
is centrifugally lifted to higher elevation, causing it to become cooler
and darker than the polar regions.
This pattern is superimposed onto the usual center-to-limb intensity variation
due to limb darkening. Under standard assumptions, the local effective temperature of the photosphere
obeys the von Ziepel theorem \citep[see, e.g., ][and references therein]{Barnes2009}
\begin{equation}
    \label{equ:gravity_darkening}
    T_{\rm eff} \propto g^\beta,
\end{equation}
where $g$ is the local acceleration due to gravity, and 
$\beta=0.25$ for bolometric observations of radiative stars. The exponent varies with the
stellar spectral type and the observing bandpass.

The rotationally-induced intensity variations are manifested as slight
perturbations of the transit light curve.
For the aligned and anti-aligned cases ($\lambda \approx 0^\circ$ or $180^\circ$),
the transit light curve retains its usual symmetry about the midpoint,
making it difficult to use gravity darkening to constrain the obliquity.
Intermediate values of $\lambda$ give rise to transit asymmetries that are
easier to isolate and interpret \citep{Barnes2009}.

The first observation of this effect for an exoplanet system
was for an A type star, Kepler-13 \citep{BarnesLinscottShporer2011,Szabo+2011}. See Figure~\ref{fig:GD}.
The light-curve anomalies in that case were on the order of 0.01\%
and indicated an obliquity of $\approx$60$^\circ$.
Other observations with this technique include HAT-P-7 \citep{Masuda2015}, KOI\,368 \citep{AhlersSeubertBarnes2014} as well as
a less secure detection in KOI\,2138
\citep{barnes2015}.\footnote{Misalignment had been reported for
KOI-89 \citep{AhlersBarnesBarnes2015}, but this result was contested by \cite{MasudaTamayo2020}.}
Gravity darkening has also been used to determine obliquities in a few
TESS systems \citep{Ahlers+2020,Ahlers+2020b}.

Gravity darkening is of particular interest because
the transit light curve depends on both $\lambda$ and $i$,
offering the possibility of determining the 3-d obliquity
from the light curve \citep{Masuda2015,Zhou+2019,Ahlers+2020b}.
Even when the data are not precise 
enough for an unambiguous determination of both $\lambda$ and $i$,
the combination of information from gravity darkening and
the RM effect can provide good constraints on the
3-d obliquity, which can lead to
additional insights, as discussed below.

\subsection{A Preponderance of Perpendicular Planets?}
\label{sec:ppp}

\begin{figure}
  \begin{center}
        \includegraphics[width=0.5\textwidth]{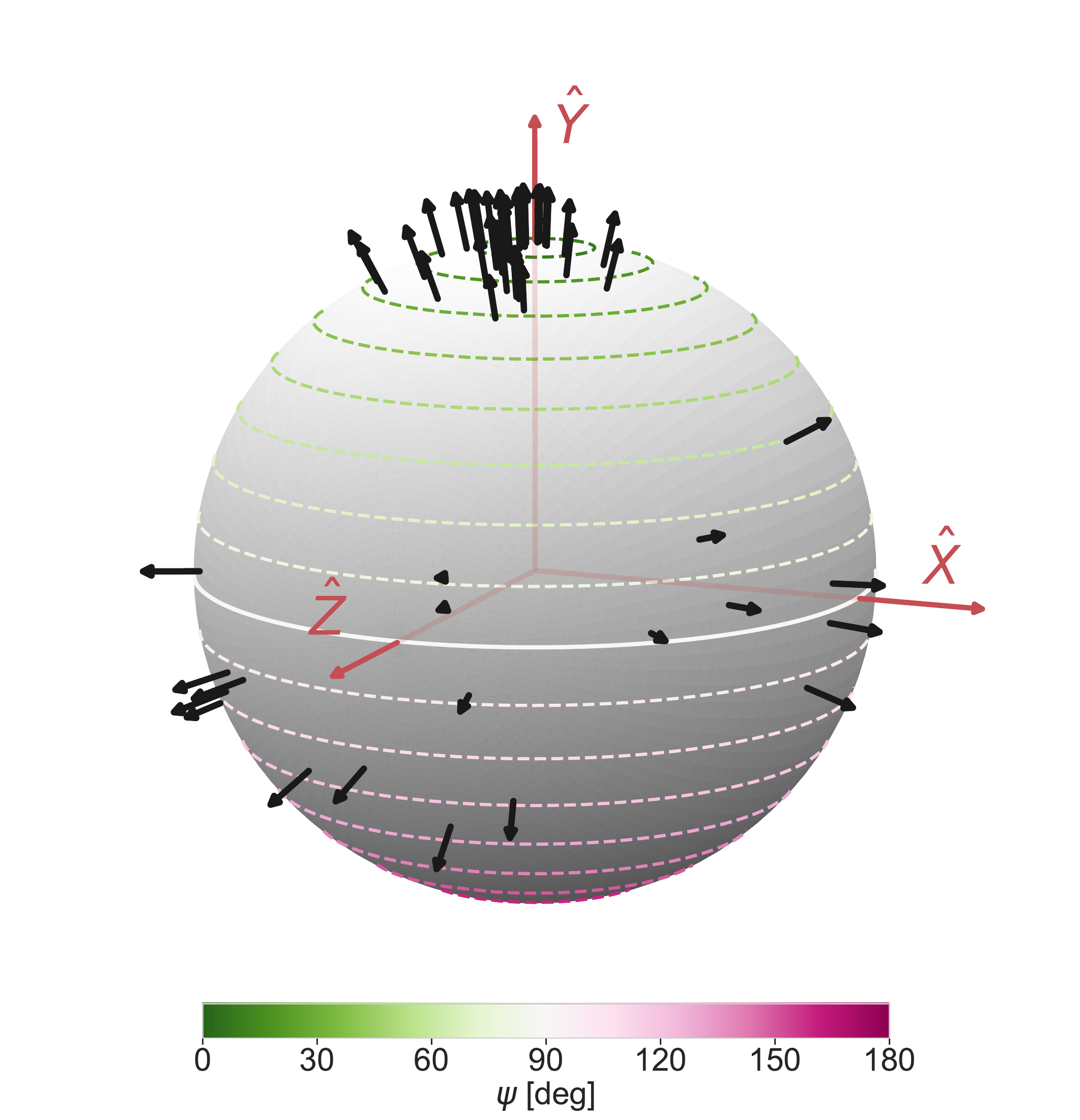}
    \caption { \label{fig:ppp} 
    \textbf{Obliquity distribution in 3-d.} From \cite{Albrecht+2021}. 
Stellar spin axes in the 57 systems analyzed by \cite{Albrecht+2021} are represented by arrows. Here the $z$-axis points along the line of sight, and the $y$-axis points along the orbital axis. 
For plotting purposes $i$ and $i_{\rm o}$ were chosen to be $\leq$\,$90^\circ$ and the signed value of $\lambda$ from the literature was adopted.
}
  \end{center}
\end{figure}

There is a growing number of stars for which both $\lambda$ and $i$ have
been determined, most often through the combination of RM and $v\sin i$ methods.
\cite{Albrecht+2021} analyzed 57 such cases and found a surprising feature in the resulting
distribution of $\psi$, the 3-d obliquity.
Even though the results for $\lambda$ span the full range from $0^\circ$ to $180^\circ$,
the range of $\psi$ appears to be limited to a maximum of about $120^\circ$, with
apparent preferences for values near $0^\circ$ and $100^\circ$ (Figure~\ref{fig:ppp}).
There were 38 well-aligned stars, most of which are consistent with zero obliquity,
and 19 misaligned stars, of which 18 have values of $\psi$ between $80^\circ$ and $125^\circ$.
The sole exception was Kepler-13, with $\psi \approx 60^\circ$.
Notably, the planets with nearly polar orbits have a wide range
of masses (0.1 to 3 Jupiter masses; see also Fig.~\ref{fig:mass_ratio_projected_obliquity}),
orbital separations ($a/R$ from 3 to 30), 
and stellar effective temperatures
(3000 to 8000\,K).

It will be interesting to see if this pattern becomes more
statistically significant after collecting additional data.
If so, the pattern would seem to be an important clue about
the processes sculpting the obliquity distribution. 
Since the study by \cite{Albrecht+2021},
at least two additional $\psi$ measurements of misaligned systems have been published: MASCARA-1\,b ($\psi=72.1\pm2.5$~deg) by \cite{Hooton+2021}, and GJ\,3470\,b ($\psi=97^{+16}_{-11}$~deg) by \cite{Stefansson+2022}. 

Looking further ahead, highly precise light curves from TESS, CHEOPS \citep{CHEOPS2021}, and eventually PLATO may provide
additional opportunities to measure 3-d obliquities
using the gravity darkening method.
PLATO may also enable additional asteroseismic
inclination measurements to complement $\lambda$
measurements (\S~\ref{sec:seismic}).

\subsection{Interlude: Planetary obliquities}
\label{sec: Planetary obliquities}

Although this article is a about stellar obliquities, there have been some
interesting developments regarding {\it planetary} obliquities.
Some of the proposed techniques for exoplanet obliquity measurements
are related to the techniques developed for stars. 

The planets of the Solar System show a wide range of obliquities. Earth's obliquity
is famously 23.5~deg, Venus rotates very slowly in the retrograde direction, and
Uranus is tipped over on its side. The root-mean-squared obliquity of the other five planets is 20.7~deg.
The planetary obliquities are not considered to be primordial \cite{LaskarRobutel1993}; instead,
planetary obliquities are thought to have been influenced by a variety of processes.
Collisions can lead to large obliquities
\citep{DonesTremaine1993,LiLai2020}, as can earlier interactions with
the protoplanetary or external perturbers \citep{Tremaine1991,JenningsChiang2021}.
Nonzero planetary obliquities in exoplanet systems might also lead to
important tidal effects (\citealt{HellerLeconteBarnes2011,MillhollandLaughlin2018,MillhollandLaughlin2019,SuLai2022};
see also \citealt{WinnHolman2005} and \citealt{Fabrycky+2007}). 
A planet's obliquity evolution can be coupled to the properties of
its moons, if it has any \citep{AtobeIda2007}. Moons
can stabilize the planet's
spin-axis orientation, as appears to be the case for the Earth \citep{LaskarJoutelRobutel1993}.
Moons can also drive large obliquity variations \citep{SaillenfestLariBoue2021}. Finally,
planetary obliquities
have been discussed as possible factors in the habitability of planets \citep{Deitrick+2018}.

An early idea for measuring planetary obliquities was
based on the slight difference between the transit
light curve of a spherical planet and an oblate planet
with the same cross-sectional area
\citep{SeagerHui2002,BarnesFortney2003}, which was used by
\cite{CarterWinn2010} to set joint upper limits on the oblateness and
obliquity of HD\,189733\,b.
Soon after, \cite{CarterWinn2010_TdepthV} proposed a related detection
method based on the transit depth variations induced by
the spin precession of an oblique planet,
which was put into practice by
\cite[][although without any secure detections]{BierstekerSchlichting2017}.

More recently, with the advent of high-resolution spectroscopy of
directly imaged planets, it became possible to use the $v\sin i$ method (\S~\ref{sec:vsini}) to constrain
planetary obliquities.
The broadening of the planet's spectral lines is used to
measure $v\sin i$, the planet's photometric variations reveal
the rotation period, and the planet's radius
is estimated from its spectrum
and thermal evolution models.
Using this technique,
\cite{Bryan+2020,Bryan+2021} found evidence for large obliquities
of two different planets, a major step forward.
With current and forthcoming high-resolution infrared spectrographs,
it may be possible to apply this method to a larger sample
\citep{Snellen+2014}.

Further in the future, it may become possible to detect
moons around directly imaged planets. Spectroscopic observations
of the transit of a moon would allow the planet's obliquity to
be constrained via the RM effect \citep{HellerAlbrecht2014}.

%%%%%%%%%%%%%%%%%%%%%%%%%%%%%%%%%%%%%%%%%%%%%%%%%%%%%%%%%%%%%%%%%%%

\begin{figure*}
  \begin{center}
        \includegraphics[width=1\textwidth]{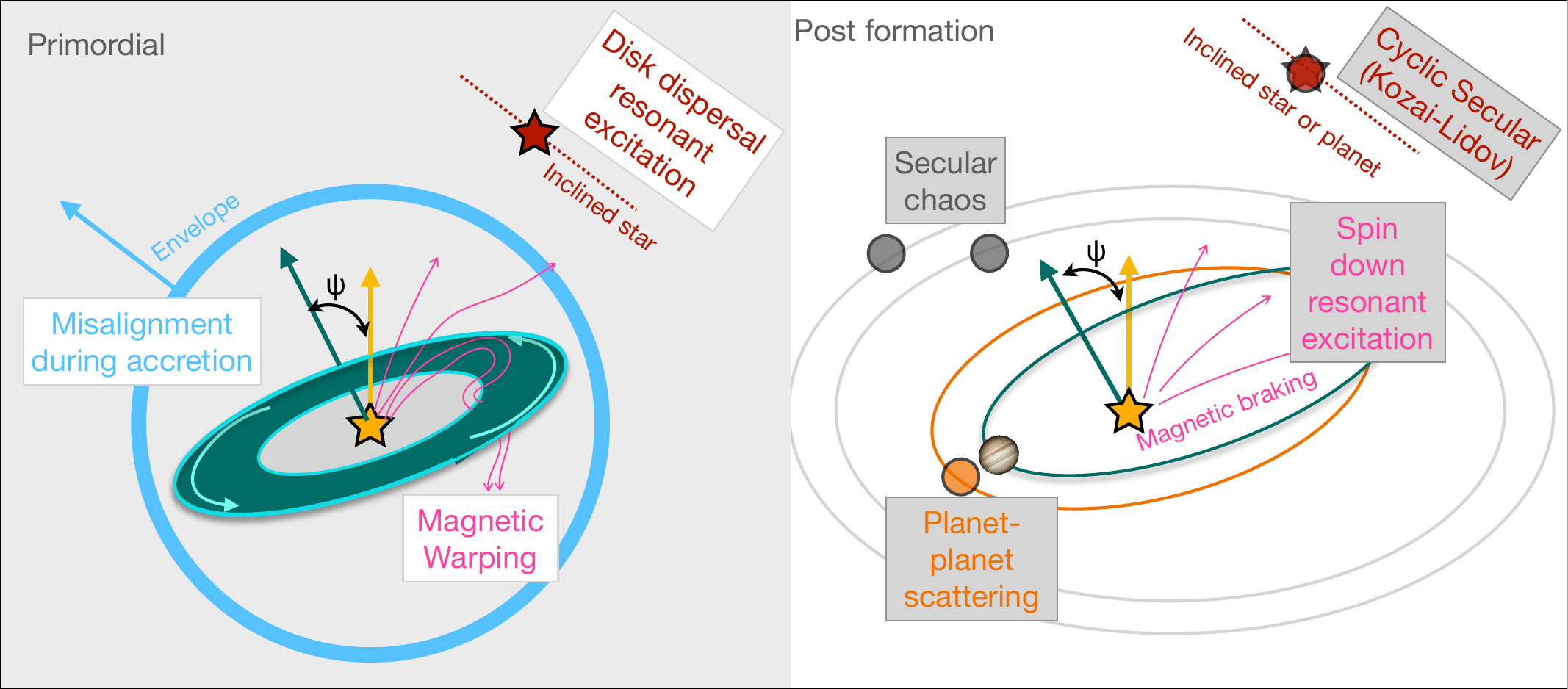}
    \caption { \label{fig:misalign_cartoon} Processes that create spin-orbit misalignments before (left) or after (right) planet formation. }
  \end{center}
\end{figure*}

\section{Processes that influence obliquities}
\label{sec:theory}

In this section, we review the physical mechanisms that might be responsible
for establishing and altering a star's obliquity.
We compare the predictions of some specific models with the available data.
We also suggest future measurements that would help to
clarify which mechanisms are operative, both for individual systems and in general.

We treat separately the subjects of obliquity {\it excitation} and {\it damping}.
There are observational clues for obliquity damping
due to tidal dissipation in systems with low-mass stars and hot Jupiters (\S~\ref{sec:tides}).
In contrast, the observations do not appear to single
out any particular mechanism for obliquity excitation.
A similar situation occurs in the interpretation of orbital
eccentricities: It is widely accepted that tidal
circularization occurs for at least some hot Jupiters,
while the mechanisms for eccentricity excitation of giant
planets are still debated. The proposed mechanisms for obliquity excitation fall into three groups:
primordial misalignment, i.e., occurring before the formation of the planet (\S~\ref{sec:primordial}), post-formation misalignment (\S~\ref{sec:orbit_plane_change}), and changes in the stellar spin vector that
are not associated with planet formation (\S~\ref{sec:stellar_spin_changes}). The predicted obliquity trends for the different categories of processes are summarized in Table~\ref{tab:trends} and
Figure~\ref{fig:misalign_cartoon} is a graphical summary of this categorization. 

\newpage

 \begin{table*}%[h!]
    \begin{center}
    \begin{tabular}{c}
\includegraphics[width=0.99\textwidth]{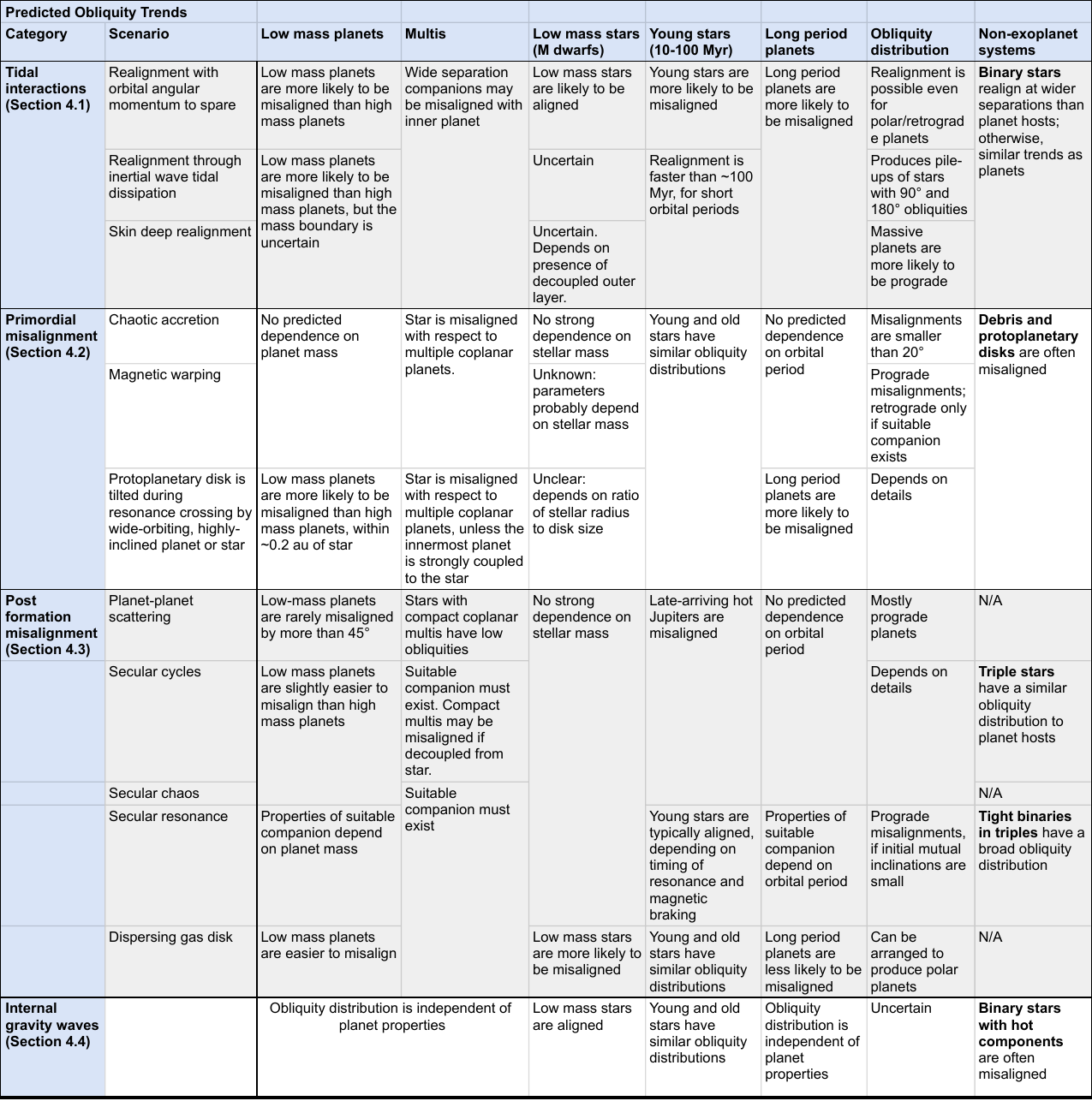}
     \end{tabular}
 \end{center}
 \caption{{\label{tab:trends}} This table summarizes what might be observed in different populations of systems if any or some of the various proposed processes are shaping spin-orbit angles of exoplanet host stars.}
\end{table*}

\newpage

\subsection{Tidal realignment}
\label{sec:tides}

Tidal dissipation in any binary system typically tends to circularize the orbit,
align the rotational and orbital axes, and synchronize the rotational and orbital periods. 
\footnote{For binary systems in general, tidal interactions can temporarily excite instead of damp eccentricities and inclinations, depending on the starting ratio between rotational \& orbital angular momenta, the rotational and orbit frequencies, and the orbital eccentricity. However, the eccentricity and obliquity are damped for typical parameters of exoplanet host stars with close-in planets. Many observed short period planets have insufficient angular momentum to synchronize their stars (Section \ref{subsec:equil}). See \cite{Hut1981,BarkerOgilvie2009,Ogilvie2014}.}
It is straightforward to demonstrate these facts based on simple considerations of energy
and angular momentum \citep[see, e.g.,][]{Hut1980}, but it is much more
difficult to model the specific processes by which angular momentum is
exchanged and energy is dissipated.
For reviews of 
tidal phenomena in binary-star and exoplanet systems,
we refer the reader to
\cite{Zahn2008}, \cite{Mazeh2008}, and \cite{Ogilvie2014}.

Obliquity damping --- if it occurs --- must happen after obliquity
excitation. Nevertheless, we discuss obliquity damping first, because the interpretation of the
data seems clearer and because obliquity damping must be kept in mind
when evaluating the evidence for obliquity excitation mechanisms.
We summarize here the evidence for tidal obliquity damping:
\begin{enumerate}
 
\item The narrowing of the obliquity distribution of hot Jupiter hosts
as the effective temperature becomes lower than about 6250\,K (\S~\ref{sec:teff}).
The transition temperature appears to be coincident with the 
``Kraft break,'' below which cool stars have deep convective envelopes and undergo magnetic braking.
The different internal structure and slower rotation
of cooler stars may hasten the
effects of tidal dissipation.

\item The exceptions to the preceding statement --- cool stars
with misaligned planets --- are generally the systems with lower planet
masses and wider orbital separations than the others (\S~\ref{sec:oblique_mass} and \S~\ref{sec:oblique_ar}).
Tidal dissipation is expected to be weaker in such cases;
thus, the exceptions seem to prove the rule.

\item The most precise measurements of $\lambda$ involving low-mass
stars with hot Jupiters are consistent with zero within about a degree (\S~\ref{sec:well_aligned}),
which is smaller than the Sun's obliquity and smaller than
the typical mutual
inclinations of the known multi-planet systems.

\item Low-mass stars
with hot Jupiters tend to rotate more rapidly than similar stars
without hot Jupiters \citep{Brown+2014,PoppenhaegerWolk2014,MaxtedSerenelliSouthworth2015,TejadaArevalo+2021}.
This is evidence that hot Jupiters tidally affect the spin rates
of their host stars, which in turn suggests that hot Jupiters can also tidally affect the stars' obliquities.

\end{enumerate}
Although this evidence points toward tidal dissipation in general, the issue of
tidal dissipation has raised interesting theoretical questions related
to which types of tidal oscillations are relevant, and how long it takes
for obliquity damping to occur. Before getting to these details, though,
we present below
a simplified tidal model that is consistent with the patterns
seen in the data.

\subsubsection{Simplified tidal friction model}
\label{subsec:equil}

In the theory of the {\it equilibrium tide}, fluid elements of the star closer to the planet feel a stronger gravitational force than those farther away, stretching the star into a slightly ellipsoidal shape. The near
and far ends of the ellipsoid are the ``tidal bulges.'' Unless the planet's orbit is aligned and synchronized
with the star's spin, the tidal bulges are not stationary in the rotating frame of the star.
If the response of the stellar fluid is delayed by viscosity, the tidal bulges point toward a previous position of the planet instead of the instantaneous position. This offset leads to a torque on the star that transfers angular momentum and also a steady loss of energy from the system. When the orbital period $P_{\rm orb}$ is shorter than the star's rotation period $P_{\rm rot}$, the planet drags the bulges in the prograde direction, spinning up the star. Conversely, when $P_{\rm orb} > P_{\rm rot}$, the planet drags the bulges in the retrograde direction, slowing down the star's rotation. When there is a spin-orbit misalignment, the tidal bulges oscillate in latitude in addition to being dragged around the star's circumference. 

Stars also experience {\it dynamical tides}. The star can be modeled
as a fluid oscillator that is being driven by the planet's periodic tidal perturbations. The oscillations take different forms, depending on the characteristics of the perturbations and the structure of the star.
For example, in the radiative zone of a star, tidal perturbations can generate gravity waves (in which the restoring force is gravity), and in the convective zone, tidal perturbations that can generate inertial waves (in which the restoring force is the Coriolis force) that reflect off the radiative core.
Any dissipation of the energy associated with the star's oscillations ultimately comes at the expense of the kinetic and potential energy associated with orbital motion and rotation.

\begin{figure*}
  \begin{center}
    \includegraphics[width=1\textwidth]{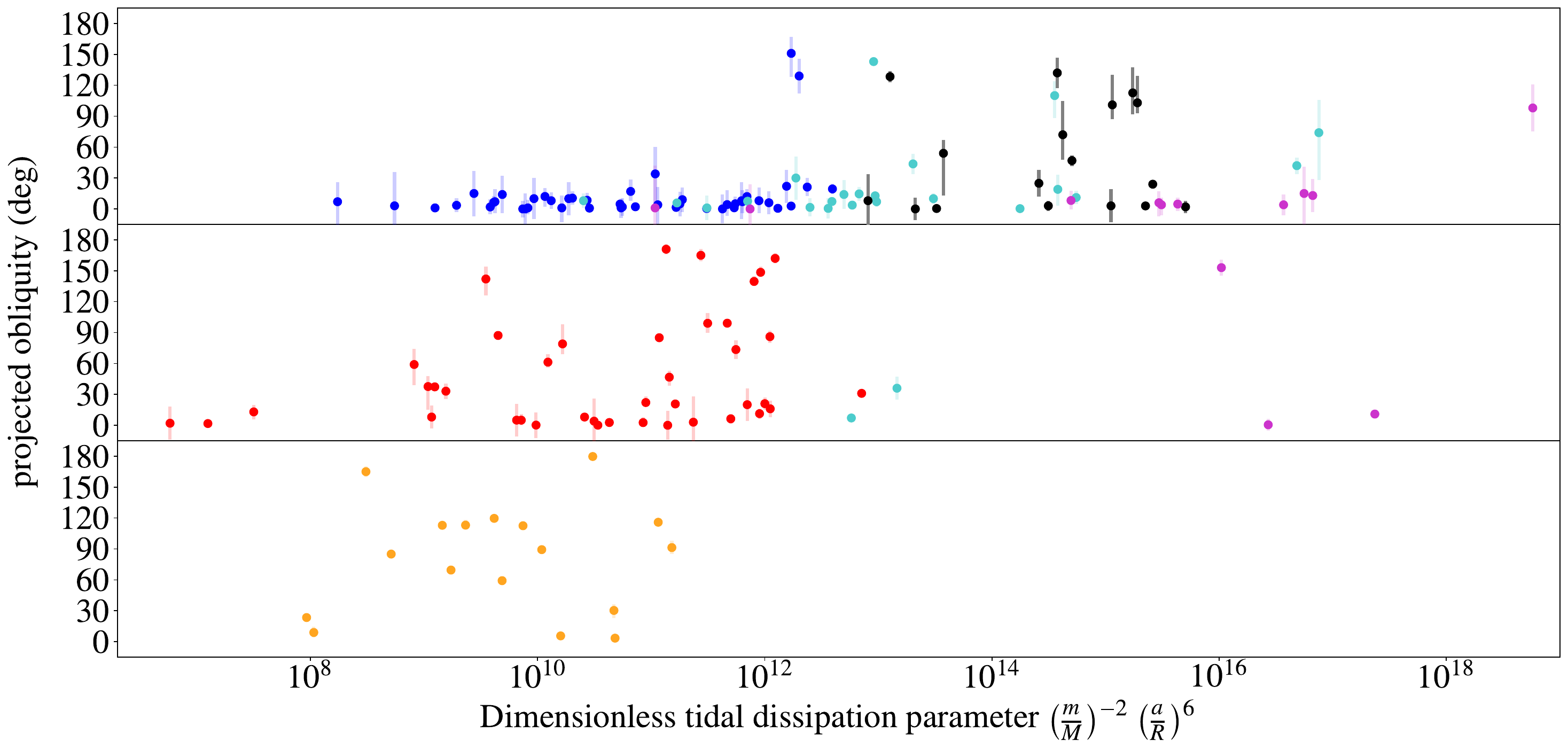} 
    \caption  {\label{fig:tau_obli} {\bf Projected obliquities of
       exoplanet systems as a function of expected tidal dissipation
       timescale}. Plotted on the $x$-axis is a dimensionless parameter $(m/M)^{-2} (a/R)^6$ that appears in a simple theory
       for tidal dissipation. 
       For cool host stars, the systems with the shortest expected
       tidal dissipation timescales tend to have low projected obliquities. For hotter hosts, any such a trend is less obvious or absent.
       }
  \end{center}
\end{figure*}

Figure \ref{fig:tau_obli} shows the measurements of $\lambda$ as a function of an approximate tidal scaling factor $(m/M)^{-2} (a/R_\star)^6$ 
taken from the equilibrium-tide theory of \citep{Zahn1977}. 
For cool host stars, the systems with the lowest values of the tidal scaling factor
tend to have low obliquities. The systems with larger values of the tidal scaling factor
show a much broader range of obliquities, while also including
a concentration of well-aligned systems. Less of a trend is seen for the hotter host stars, and no trend is seen for stars with $T_{\rm eff}> 7000$~K. 
Overall, the data appear consistent with the hypothesis of tidal
damping. This type of argument was first made
by \cite{AlbrechtWinnJohnson+2012}, and the trend still holds true
after the sample has more than tripled in size.

Despite the agreement between the data and the simple scaling
relation drawn from equilibrium-tide theory, we must acknowledge
that dissipation of the equilibrium tide is not a realistic model
for obliquity damping in star-planet systems.
In that theory, the timescales for realignment and orbital decay
should be comparable, and therefore, there is only a fleeting time interval
when the star has been realigned and the planet still exists.
The orbit decays because a short-period planet (even a giant planet)
generally does not have enough orbital angular momentum to synchronize the star and thereby halt the dissipation.
Tidal dissipation ultimately causes the planet to transfer all its angular momentum
to the star and become engulfed or tidally disrupted.
Observing hot Jupiters around cool stars that have been tidally
realigned is most likely
when the realignment timescale is much shorter than the orbital decay timescale. In the context of the equilibrium-tide theory,
this condition, in turn, implies
that the angular momentum ratio $\frac{L_{\rm orb}}{L_\star}$ should
be much smaller than one. However, the actual ratios are on the order of unity:
\begin{multline}
\label{eqn:amrat}
    \frac{L_{\rm orb}}{L_\star} = \frac{m a^2 (2\pi/P_{\rm orb})}{k_\star M R^2 (2\pi/P_{\rm rot})} \\ 
    \sim 2.5 \left(\frac{0.1}{k_\star}\right)\left(\frac{m/M}{0.001}\right) \left(\frac{a/R}{5}\right)^2 \left(\frac{P_{\rm rot}/P_{\rm orb}}{10}\right).
\end{multline}

\smallskip

\subsubsection{More realistic tidal models}
\label{subsec:complextides}

Many possible answers have been offered to the question of
how to tidally damp a star's obliquity without destroying the planet,
and how to account for the difference in the obliquity
distributions of hot and cool stars.
\begin{itemize}
\item{\bf Planets with orbital angular momentum to spare:} For systems
with unusually massive planets, wide orbits, or
slowly rotating stars, the ratio $\frac{L_{\rm orb}}{L_\star}$ can
exceed 10, and the realignment timescale becomes shorter than the orbital decay timescale
\citep[see, e.g.,][]{Hansen2012,ValsecchiRasio2014}.
These planets may be able to realign their stars without undergoing much orbital decay over the star's lifetime.
\item {\bf Inertial wave dissipation:} Tidal perturbations can launch inertial waves
within the convective zone of a cool star. As noted above, the restoring force is
the Coriolis force. Their dispersion relation is:
$\omega = 2\vec{k}\cdot \vec{\Omega_\star}$, where $\omega$ is the angular frequency,
$\hat{k}$ is the direction of the wavevector and phase velocity, and $\Omega_\star \equiv 2\pi/P_{\rm rot}$.
Thus, inertial waves always have frequencies lower than $2\Omega_\star$,
and cannot be excited by perturbations with higher frequencies.
For a short-period planet, most of the components of the tidal perturbation
are too fast to excite inertial waves --- but when the star is misaligned,
there is a component that oscillates at the frequency $\Omega_\star$.
This component affects the spin direction, but not the orbital distance
\citep[see, e.g.,][]{ Lai2012,LinOgilvie2017,DamianiMathis2018}.
By itself, inertial wave tidal dissipation would drive
the star toward one of three equilibrium orientations: $\psi=0$, 90, or $180^\circ$.
The destination depends on the initial condition, the ratio of orbital to spin angular momentum (smaller ratios lead to more $\psi=90^\circ$ planets), and the ratio between the rates of inertial-wave tidal dissipation and equilibrium tidal dissipation (larger ratios lead to more $\psi=90^\circ$ planets). For
more details, see \cite{ Xue+2014} and \cite{LiWinn2016}.
For an attempt to use this theory to constrain the timing of
hot Jupiter formation, see \cite{SpaldingWinn2022}.

\item {\bf Steeply frequency-dependent tidal dissipation:} The tidal dissipation rate
is likely to depend strongly on the forcing period.
If the rate of tidal dissipation drops sharply with decreasing period,
then a hot Jupiter can realign the star while its orbit decays by a moderate
amount, before tidal evolution slows to a crawl as its orbital
period decreases.  There is some empirical evidence for such a sharp frequency
dependence, based on the observed degree of tidal spin-up as a function
of the system parameters \citep{Penev+2018,AndersonWinnPenev2021}. Although resonance locking to a tidally excited stellar gravity mode can generate such a frequency dependence, but the waves break at the core, preventing effective  dissipation \citep{barker2010}. Thus, it is thought to only operate for very massive stars with convective cores or for Neptune-mass and lower-mass planets around Sun-like stars \citep{MaFuller2021}.
In this scenario, the differences in the obliquity distributions of
hot and cool stars might arise because hot stars have a lower tidal dissipation efficiency,
a different frequency dependence, or both.

\item {\bf Realignment is only skin deep:} The planet might realign only the outer convective
zone of the star, if this outer zone could somehow remain decoupled from the interior (e.g., \citealt{DobbsdixonLinMardling2004,Winn+2010}).
This would lower the amount of orbital angular momentum required for realignment.
Stars lacking an outer convective zone would, naturally, remain misaligned.
Moderately hot stars could also remain misaligned because they rotate quickly (due to a lack
of magnetic braking; \citealt{Dawson2014}),
because their convective outer layers couple strongly to the interior, or
because they have less efficient tidal dissipation.

\end{itemize}
\citet{Cebron+2013} proposed that, rather than damping the obliquity of cool stars, tides could excite the obliquities of hot stars. However, later global simulations found that the mechanism would tend to damp obliquities \citep{barker2016} and that obliquity excitation would require tides to generate rather than dissipate energy \citep{Ogilvie2014}.

Figure \ref{fig:model} shows the results of a toy population-synthesis model, which is
described in detail in Appendix \ref{app:sims}. 
The first column shows the observed distributions of $\lambda$, $T_{\rm eff}$, and $v\sin i$
for stars having planets with $a/R < 10$ and $m > 0.5 M_{\rm Jup}$.\footnote{\bf We decided not to show predictions for
low-mass planets because the mass cut-off for alignment is very sensitive to uncertain tidal dissipation parameters.}
The second column shows a hypothetical initial distribution of stellar properties, before
any tidal effects.
The next four columns show the
final distributions according to 
four of the proposed solutions to the realignment problem described above.

In the equilibrium-tide theory (third column),
the most massive planets manage to realign their stars, but lower-mass hot Jupiters remain misaligned
even around cool stars, contradicting the data.

With inertial waves excited by dynamical tides (fourth column),
cool stars are realigned.  However, some simulated systems end up stalled
at $\psi=180^\circ$, even though the effects of equilibrium
tides are also being taken into account \citep{Xue+2014,LiWinn2016}.
We have not observed a population of anti-aligned stars,
neither in exoplanet systems nor binary stars.
We think it would be difficult
to resolve the disagreement by altering the initial obliquity distribution;
even stars that start with obliquities near $90^\circ$ can evolve
to 180$^\circ$ by the end of the simulation.

Theories in which tidal dissipation declines sharply with frequency (fifth column)
and in which tidal realignment is only skin deep (sixth column)
successfully reproduce the observed broadening of the obliquity distribution
at the Kraft break.
However, these models are {\it ad hoc}. Further work needs to be done
to try connecting them to more physically grounded
theories of stellar oscillations.
In particular, skin-deep realignment would require the outer and inner zones
of the star to rotate with different rates and in different directions for
billions of years, which seems at odds with the observation that
the Sun's convective zone rotates at the same rate as its radiative interior.
On the other hand, the Sun does have a thin near-surface shear outer layer (starting
at around 0.95 $R_\odot$) which rotates at a different rate from the rest of the convective zone (see, e.g., \citealt{Thompson+1996}).

\begin{figure*}
  \begin{center}
    \includegraphics[width=1\textwidth]{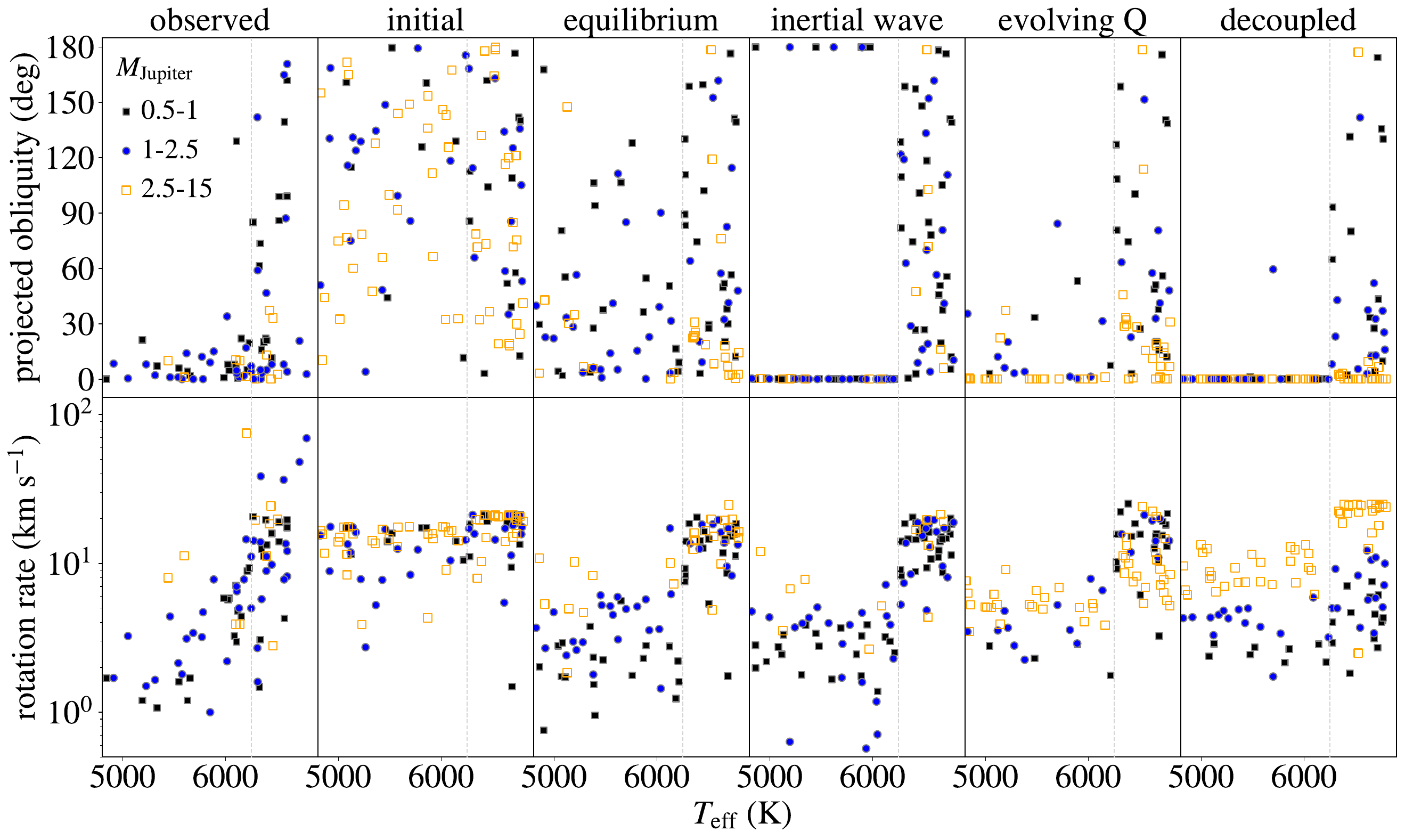} 
    \caption  {\label{fig:model} {\bf Results of illustrative population-synthesis models for
    tidal realignment.}  Column 1 shows real data. Column 2 shows the initial properties
    of the simulated distribution. Columns 3-6 show the final distributions.
    See the text, and Appendix, for more details.}
    \end{center}
\end{figure*}

In summary, tidal alignment appears to play an important role in shaping
the obliquity distribution of stars with close-orbiting giant planets.
However, there is no completely satisfactory theory
that specifies the nature of the tidal perturbations and how they are damped.

\subsection{Primordial misalignment}
\label{sec:primordial}

One might expect a star to be well-aligned with its protoplanetary disk,
because the star and the disk both inherit their angular momentum
from the same part of a collapsing molecular cloud.
Furthermore, gas funnels onto the young star through the disk, a process that would
help maintain good alignment.
Nevertheless, three processes have been proposed to generate a ``primordial misalignment'':
chaotic accretion, magnetic warping, and tilting by a companion star.
Figure~\ref{fig:misalign_cartoon} illustrates these possibilities.

{\bf Chaotic accretion} refers to the fact that stars form in the dense and chaotic environment of
a gravitationally collapsing and fragmenting molecular cloud.
Interactions between a protostar and neighboring protostars or clumps of gas
might cause the early-accreting gas
to arrive from a different direction than the gas that accretes later.
The late oblique infall of material might warp or tilt the disk away from
the star \citep{BateLodatoPringle2010,Thies+2011,Fielding+2015,Bate2018,Kuffmeier+2021}.
This possibility was studied further by \cite{TakaishiTsukamotoSuto2020},
who found that subsequent accretion from the disk onto the star tends to eliminate any temporary misalignments.
According to this work, by the time planets form, the disk and star are probably aligned to within 20$^\circ$.

{\bf Magnetic warping} occurs when differential rotation between a young star and the ionized inner disk twists the magnetic field lines that link them, generating a toroidal magnetic field. The toroidal magnetic field,
in turn, generates an electrical current in the inner disk. The Lorentz force on this current generates a misaligning torque that would amplify
any initially small misalignment and warp the inner disk
(\citealt{FoucartLai2011,LaiFoucartLin2011}; see also
\citealt{Romanova+2013,Romanova+2021}, for three-dimensional magnetohydrodynamical simulations).
Modest misalignments can be generated by this torque if 
the toroidal field is sufficiently strong and the realigning torques are sufficiently weak. Realigning torques are caused by accretion,
magnetic braking, disk winds (which transport angular momentum from the star to the disk), and viscosity (which couples the misaligned inner disk to the aligned outer disk). A realigning torque can also
be created if the magnetic field becomes wrapped around the stellar rotational axis instead of around the disk axis \citep{Romanova+2021}.
A broad distribution of obliquities (including retrograde stars)
can be achieved if the warping torques are accompanied
by an external disturbance to the outer disk,
perhaps generated by a stellar companion \citep{FoucartLai2011}.

{\bf Inclined stellar or planetary companions} can tilt disks (see, e.g., \citealt{Borderies_Goldreich_Tremaine1984,Ludow_Ogilvie2000,Batygin2012,matsakos2017}). Although the disk is coupled to the star through accretion, a misalignment can be generated if the
system encounters a resonance between the spin precession period of the star around the disk axis,
and the nodal precession period of the disk around the binary axis \citep{BatyginAdams2013,lai2014}.
Such a ``secular resonance crossing'' might be encountered as a disk's precession period gradually changes in response to mass loss \citep[see, e.g.][]{spalding2014}.
However, hot Jupiters are tightly coupled by gravitational 
forces to the spins of their host stars.
\cite{ZanazziLai2018} found that this coupling generally
prevents misalignments from occurring due to secular
resonance crossings, even if the planet formed at a large orbital
distance and underwent disk-driven migration after the resonance crossing.
Therefore, this mechanism is unlikely to be responsible for the observed high obliquities of stars with hot Jupiters. 

Among the three primordial misalignment mechanisms, magnetic warping (under certain disk conditions) remains viable as an explanation for the observed obliquity distribution. The incidence of primordial misalignments might increase with stellar mass.
For example, \citet{spalding2015,spalding2016} proposed that young stars
with masses lower than about 1.2\,$M_{\odot}$ are able to realign their disks due to their stronger magnetic fields. However, in that case, misalignment would more strongly correlate with the initial main sequence effective temperature than with the present-day effective temperature. Because primordial misalignment occurs prior to planet formation,
the observed correlations
between misalignment and planet mass and orbital separation
(\S~\ref{sec:oblique_mass}) and~\ref{sec:oblique_ar})
would need to have a different explanation, such as tidal realignment. 

What about stars with other kinds of planets, besides hot Jupiters?
A star that was primordially misaligned might be expected
to have multiple coplanar planets that are all misaligned with the star by the same angle.
As noted in \S~\ref{sec:vsini} and \S~\ref{sec:starspots}, there is indeed evidence for a broad obliquity
distribution among hot stars hosting
Kepler-type planets, many of which are in flat multi-planet systems.
However, this is not true of cool stars, and it is hard to understand why primordial misaligning mechanisms
would leave the cool stars alone. Planet-star tides are weak and probably
irrelevant for almost all the Kepler systems.

Regarding individual systems, the sample of 17 compact multi-transiting systems with obliquity measurements presented in \S~\ref{sec:multitransits} contains 14 well-aligned systems of super-Earths and mini-Neptunes .
There are three misaligned systems:
\begin{itemize}
\item HD\,3167 is a K dwarf with mass $0.88\,M_\odot$, two transiting planets, and a third non-transiting planet orbiting in between the
transiting planets \citep{Christiansen+2017}. The currently available data do not show clear evidence for any
wider-orbiting companion, whether planet or star.
The orbit of the innermost planet is misaligned on the sky by $-97\pm 23^\circ$ relative to the star \citep{Dalal+2019}. The orbit of the outer transiting planet was found to be aligned with the star \citep{Bourrier+2021},
implying the two planetary orbits are nearly perpendicular, an
extraordinary architecture, although the evidence for this claim
relies on a signal that was only barely detected. Reobservation of this remarkable
system is warranted.

\item Kepler-56 is a 1.3\,$M_\odot$ star with two transiting planets,
and is misaligned by at least 45$^\circ$ with respect to both of their orbits \citep{Huber+2013}.
Kepler-129 is a 1.2\,$M_\odot$ star that also has two transiting planets,
for which a $\approx$40$^\circ$ misalignment was tentatively detected \citep{Zhang+2021}.
%RID: move to footnote
Both stars have wide-orbiting massive planets with properties that
make them good candidates for tilting the orbital plane of the inner
planets, either before or after these planets have formed \citep{GratiaFabrycky2017,Zhang+2021}.

\item K2-290 is a 1.2\,$M_\odot$ star with two
transiting planets --- a warm Jupiter and an inner Neptune-sized planet --- and has an obliquity of $124\pm 6^\circ$ \citep{Hjorth+2019,Hjorth+2021}.
A stellar companion, K2-290~B (projected separation $\approx$110~au), was also detected and appears capable of tilting the protoplanetary disk but not capable of tilting the system today. This system is probably the best known candidate for
a primordial misalignment, although a recent study by \cite{BestPetrovich2022} suggested an alternative possibility: The third known star in that system, K2-290~C (projected separation $\approx$2500~au) might have been responsible for the misalignment.

\end{itemize}

To make further progress, it would be helpful to measure the obliquities of stars that
still have protoplanetary disks.
The problem is that the disks and the surrounding material often blocks the stellar photosphere.
It may be relevant that misalignments have been found
between the planes of the inner and outer parts
of a protoplanetary disk
(see, e.g., \citealt{MarinoPerezCasassus2015,Sakai+2019,Ginski+2021}; or \citealt{Casassus+2016} for a review).
However, the occurrence rate of such misalignments is not known.
These internally misaligned or ``broken'' disks might form planets with large mutual inclinations,
setting the starting conditions for some of the obliquity-excitation processes discussed
in the next section. The obliquities of stars with resolved debris disks
have been investigated, usually using the projected rotation velocity method,
and have generally been found to be lower than $\approx$30$^\circ$
\citep{Watson+2011,Greaves+2014,Davis2019}.

To complete the discussion of stars and disks, we note that there
are a few wide binary systems in which the two protoplanetary disks are misaligned
with respect to each other and with respect to the orbital plane.
These findings are based on polarization observations of disk jets \citep[][and references therein]{Monin+2007} and interferometric imaging of protoplanetary disks
(e.g., HK~Tauri, \citealt{JensenAkeson2014}, IRS 43 \citealt{Brinch+2016}).
Misaligned circumbinary debris disks have also been found, such as the disk
surrounding and eclipsing the two stars of KH\,15D \citep{Winn+2004,ChiangMurray-Clay2004,PoonZanazziZhu2021}. 

The $v \sin i$ method has been used to test for alignment in binaries with separations of several au \citep[e.g.][]{Weis1974,Hale1994,GlebockiStawikowski1997,HoweClarke2009}. However, \cite{JustesenAlbrecht2020} argued that the available data do not allow for strong
conclusions, contradicting previous results.

\subsection{Post-formation misalignment}
\label{sec:orbit_plane_change}

After formation, gravitational interactions between a planet and other bodies could alter the planet's orbital plane, leading to misalignment with the host star's spin.
These gravitational interactions might also initiate high-eccentricity tidal migration,
in which a giant planet forms on a distant orbit and acquires a high orbital eccentricity, bringing
it close enough to the star for tidal dissipation to shrink and circularize the orbit.\footnote{In this
case, tidal evolution can be driven by dissipation within both the star and the planet.
The planetary contribution is thought to be dominant \citep[see, e.g.,][]{Mazeh2008}.}
An appealing aspect of this scenario is that it has the potential to explain three things at once:
hot Jupiters, eccentric orbits, and misaligned stars
(see \citealt{DawsonJohnson2018}, for a review of theories for the origins of hot Jupiters).

Since the discovery of misaligned hot Jupiters,
it was hoped that stellar obliquities would provide clues about the
dynamical histories of hot Jupiters. Specifically, many authors
have interpreted a low obliquity as a sign that a hot Jupiter formed through disk-driven migration or {\it in situ} formation,
and a high obliquity as an indicator of
high-eccentricity migration \citep[see, e.g.,][]{FabryckyWinn2009}.
The picture has become more complicated, partly because
of the evidence for tidal sculpting of the obliquity distribution (Section \ref{sec:tides}).
If tides can damp obliquities, then a low obliquity cannot necessarily
be counted as evidence
for disk-driven migration or {\it in situ} formation.
The initial obliquity distribution could have been very broad (Figure~\ref{fig:misalign_cartoon}).
Some of the proposed primordial theories
for obliquity excitation, discussed in Section \ref{sec:primordial},
can generate a broad initial obliquity distribution 
and also account for the statistical evidence for misalignments
of hot stars hosting compact coplanar systems (Sections~\ref{sec:vsini} and \ref{sec:starspots}).

{\bf Planet-planet scattering} can tilt orbits with respect to each other and the host
star, on timescales as short as thousands of years.
Close encounters between planets disturb their orbits,
causing eccentricities and mutual inclinations to undergo random walks.
Planet-planet scattering can take place shortly after the dissipation of the gas disk,
when it becomes possible for large eccentricities to develop.
Scattering might also occur when planets are brought together
by longer-timescale chaotic evolution (see below)
or the influence of a passing star
\citep{MalmbergDaviesHeggie2011}.

\begin{figure}
  \begin{center}
    \includegraphics[width=8.5cm]{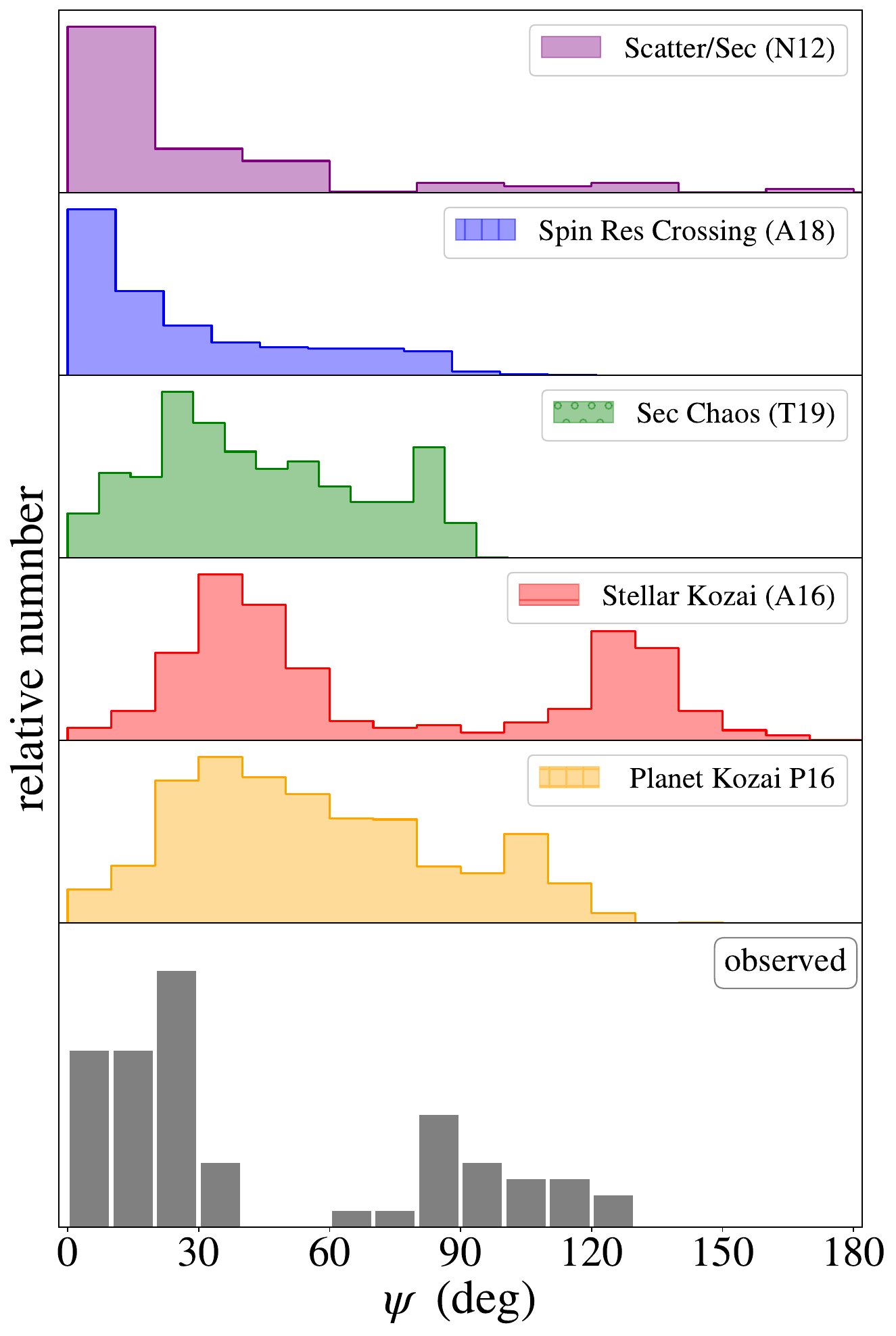} 
    \caption  {\label{fig:misalign} {\bf 
    Population-synthesis predictions for the 3-d obliquity distributions} based on different misalignment mechanisms: 
    planet-planet scattering with secular cycles (\citealt{BeaugeNesvorny2012}; see their Fig.~16 for hot Jupiters from systems with four planets at 3 Gyr), resonance crossing for hot Jupiters (Fig.~9 of \citealt{AndersonLai2018};
    secular chaos (Fig.~11 of \citealt{TeyssandierLaiVick2019}),
    star-planet Kozai-Lidov cycles for a 1 $M_{\rm Jup}$ HJ orbiting an F star (Fig.~14 of \citealt{AndersonStorchLai2016}), and planet-planet Kozai-Lidov cycles (Fig.~10 of \citealt{PetrovichTremaine2016}). Bottom: Observed distribution of $\psi$.}
  \end{center}
\end{figure}

When the planets have low masses or small orbits (i.e., when the escape velocity from the surface of the planet is less than the planet's orbital velocity), 
encounters tend to lead to collisions rather than scattering.
In such cases, planet-planet scattering can only produce small mutual
inclinations \citep[see, e.g.,][]{GoldreichLithwickSari2004}.
For giant planets on wide orbits --- the types of planets
that might become hot Jupiters --- the expected distribution of mutual inclinations
ranges up to about 60$^\circ$ and does not include retrograde systems
\citep{Chatterjee+2008}. This is a problem when trying to match the data.
Possibly, planet-planet scattering establishes the initial
conditions for subsequent secular interactions that further
broaden the obliquity distribution (see, e.g., \citealt{NagasawaIdaBessho2008,NagasawaIda2011,BeaugeNesvorny2012}).
In Figure~\ref{fig:misalign}, which compares the obliquity distributions predicted by
various excitation theories, the top purple histogram is for planet-planet scattering \citep[taken from][]{BeaugeNesvorny2012}.

A related mechanism that can produce a more isotropic distribution
is the direct disturbance of a giant planet's orbit through a hyperbolic encounter with another star.
This is very improbable in general, but it might occur in a very dense cluster environment, such as the center of a globular cluster where stars are moving in random directions \citep{HamersTremaine2017}.
Unfortunately, we do not yet know of any hot Jupiters in globular clusters
\citep{Gilliland+2000,MasudaWinn2017}.

{\bf Secular cyclic interactions} refers to processes that allow
planets and stars to gradually and periodically
exchange angular momentum, over timescales
of thousands of orbits or more.
Eccentricities and mutual inclinations oscillate as the bodies
torque each other.
When the oscillations of a close binary are driven by a distant tertiary
object on a highly inclined or highly eccentric orbit,
these interactions are known as
{\bf Kozai-Lidov} cycles \citep{Kozai1962,Lidov1962}.
They can be driven by either a stellar or planetary companion.
Many authors have investigated Kozai-Lidov cycles as a possible
route to the formation of hot Jupiters; see, e.g., \citealt{WuMurray2003,FabryckyTremaine2007,Naoz+2011}, or \citealt{Naoz2016} for a review.
The period of the secular oscillations
depends on the separation and mass of the perturbing companion, with typical timescales on the order of millions of years.
Most calculations of secular interactions have assumed
that the protoplanetary disk has already disappeared.
However, mutual inclinations can also be excited
through secular interactions between a planet, a protoplanetary
disk, and a companion
\citep{PicognaMarzari2015,LudowMartin2016,FranchiniRebeccaLubow2020}. 

{\bf Secular resonant excitation of the stellar obliquity} can occur if
a resonance is encountered between a secular frequency
and some other frequency in the system.
In a triple system, as the primary star spins down
due to magnetic braking, the star's rotational precession around
the planet's orbital axis slows down, possibly leading to a resonance between
the stellar spin precession frequency and the nodal precession of the inner planet's orbital axis about the outer companion's orbital axis.
The result would be a large misalignment \citep{AndersonLai2018}. 

Because of the reliance on magnetic braking, we would expect this mechanism to
work only for cool stars. Thus, this mechanism might struggle to explain
the high obliquities of hot and rapidly rotating stars.
It also has trouble producing retrograde obliquities,
as illustrated in the second histogram (blue) in the top panel of Figure~\ref{fig:misalign} when initial mutual inclinations are small. Larger initial mutual inclinations -- for example, if the third body is a binary, or if other interactions tilts its plane -- would produce more retrograde orbits.

In a system with an outer planetary companion and a dispersing gas disk, the outer planet's precession frequency (due to the inner planet and to the disk) might come to coincide with the inner planet's precession frequency (due to the outer planet and to the star's rotation).
Such a resonance crossing is capable of generating a large mutual inclination between the inner and outer planetary orbits. With the aid of general relativistic precession, the stellar obliquity can be driven to $90^\circ$ \citep{Petrovich+2020}. This mechanism is most effective for close-in Neptune-mass planets with outer Jupiter-mass companions, such as HAT-P-11.

{\bf Secular chaos} refers to the stochastic growth of eccentricities and
mutual inclinations due to the overlap of secular frequencies in multi-planet systems
\citep{Laskar2008,WuLithwick2011,HamersAntoniniLithwick+2017,TeyssandierLaiVick2019} or multiple-star systems \citep{Hamers2017,GrishinLaiPerets2018}.
The growth takes place over many secular timescales, typically
hundreds of millions of years or longer.
The resulting obliquity distribution depends on the initial architecture.
Producing planets on retrograde orbits requires initially
large eccentricities and inclinations \citep[see, e.g.,][]{LithwickWu2014},
which could have been established by planet-planet scattering \citep{BeaugeNesvorny2012}
or stellar flybys \citep[see, e.g.,][]{HaoKouwenhovenSpurzem2013}.
\citet{TeyssandierLaiVick2019} argued that secular chaos has an
insurmountable problem in producing retrograde planets, as illustrated
with the green histogram in Figure~\ref{fig:misalign}.
This is because the planet's orbit tends to circularize and decouple from the companion before the obliquity grows very large.

Kozai-Lidov cycles with tidal friction tend
to produce a bimodal obliquity distribution
\citep{FabryckyTremaine2007}, as illustrated with the red histogram in Figure~\ref{fig:misalign}. The location of the peaks and the degree of bimodality depend on the properties of the star and the perturber, including the star's oblateness and spin evolution \citep{DamianiLanza2015,Petrovich2015,AndersonStorchLai2016}. The spin evolution of the star can also
lead to chaotic obliquity variations \citep{StorchAndersonLai2014}.
The expected obliquity distribution can be broadened (and the fraction of retrograde obliquities can be increased)
by placing the companion on an eccentric or nearby orbit
\citep{Naoz+2011,Teyssandier+2013,Li+2014,Li+2014c,PetrovichTremaine2016}.
Such companions might result from planet-planet scattering.
The case of an eccentric or nearby planet perturber is illustrated with the yellow histogram in
Figure~\ref{fig:misalign}. Intriguingly, there is evidence for a peak in the observed
obliquity distribution near 100$^\circ$ (\S~\ref{sec:ppp})
which might be related to Kozai-Lidov cycles. More
work needs to be done to see if theory and observations
can be brought into better agreement.

For Kozai-Lidov cycles to significantly raise the mutual inclination, the orbital precession induced by the companion must be faster than
precession from other sources, such as stellar oblateness, tides, and general relativity. In compact systems where planets are more tightly coupled to each other than to an exterior companion, the exterior companion can misalign the entire interior system relative to the host star's equatorial plane, as observed for Kepler-56 \cite[see, e.g.,][]{Innanen+1997,Takeda+2008,Kaib+2011,BoueFabrycky2014,Li+2014b,GratiaFabrycky2017}.
This explanation does not work for K2-290 (a system highlighted in \S~\ref{sec:primordial}) because oblateness-induced precession is too fast \citep{Hjorth+2021}.

In summary, none of the proposed mechanisms, by itself,
seem capable of producing a broad obliquity distribution with
plenty of retrograde planets.
However, more complex and multistep dynamical histories, such as planet-planet scattering followed by secular cycles, are at least qualitatively consistent with the hot Jupiter data (Fig.~\ref{fig:misalign}).
The low probability of retrograde outcomes in these models
might not be a fatal problem. 
Although a fair fraction of hot stars
have retrograde hot Jupiters, the measured 3-d obliquities are
almost all below 130$^\circ$ (\S~\ref{sec:ppp}).
In addition, a low initial fraction of retrograde systems might remove 
one of the objections to the idea that obliquities are damped
due to inertial-wave dissipation,
by reducing the probability of
systems evolving into the anti-aligned state (see \S~\ref{sec:tides}).

The possible influence of tidal damping --- which might play a
role even for hot stars --- 
complicates the effort to compare the theoretical and observed obliquity distributions and tease out the relative contributions of different mechanisms \citep[see, e.g.,][]{MortonJohnson2011,NaozFarrRasio2012}.
Achieving an isotropic distribution for small, compact, coplanar planets orbiting hot stars (see \S~\ref{sec:vsini}) may be even more challenging; it has not yet been attempted, to our knowledge.

As noted above, the basic premise
of all the theories described in this section is
that a hot Jupiter forms far away from the star,
undergoes interactions that raise its orbital
eccentricity and inclination,
undergoes high eccentricity tidal migration,
and arrives in its final orbit in a misaligned state.
None of the obliquity excitation mechanisms would
work well after the giant planet is already in a tiny orbit.
Planet-planet scattering generally fails to
generate large misalignments very close to the star.
Secular mechanisms would require very nearby planets
to overcome the coupling between the hot Jupiter
and the star, and hot Jupiters only rarely have such companions
(see below).

Can post-formation misalignment scenarios be tested with stellar ages? Like the primordial misalignment mechanisms (Section \ref{sec:primordial}), the disk-companion secular mechanism and disk dispersal mechanism predict that misaligned planets are in place as soon as the gas disk disperses. In contrast, the magnetic braking resonance crossing mechanism generally
needs billions of years to generate misaligned planets. For the mechanisms that act in concert with high-eccentricity tidal migration, the migration time (rather than the eccentricity excitation timescale) typically dominates the time to deliver a misaligned hot Jupiter. Depending on the planet's periastron distance after eccentricity excitation,
the planet can achieve a short orbital period on timescales ranging from less than a Myr to several Gyr.

The current sample of obliquity measurements for young stars ($\lesssim$100\,Myr) contains only well-aligned systems (\S~\ref{sec:newborn}), but it
is not yet possible to use this sample to rule out any of the proposed
post-formation misalignment mechanisms. This is partly because
the sample is small. In addition, the planets in the sample
are not of the types that are commonly
found to be misaligned around older stars.
To derive constraints on the timing of
misalignments, it would be more useful to obtain measurements for
young stars with close-orbiting giant planets,
which are often seen to be misaligned around older stars.

If a larger sample of young systems, including hot Jupiters, still
shows very few misalignments,
then the primordial misalignment scenario would be cast into doubt.
Such a result would be consistent with the magnetic braking resonance crossing mechanism. It would also be consistent
with the hypothesis of dual origins for hot Jupiters: a population of initially aligned orbits from disk-driven migration,
along with late-arriving hot Jupiters from high-eccentricity
migration. If the sample could be enlarged to include
stars with accurately determined ages spanning a wide
range, we may be able to test other predictions as well. For example, \citet{BeaugeNesvorny2012} predicted that retrograde orbits generated by planet-planet scattering and secular interactions should go hand in hand with small periastron distances, and are thus more likely to experience tidal decay, leading to a dearth of retrograde orbits around older stars.

One way to test the secular cycle hypothesis, in particular, is to search for companions capable of driving Kozai-Lidov cycles. A prime example is the HD\,80606 double star system with its highly eccentric warm Jupiter \citep{Naef+2001} on an oblique orbit \citep{Winn+2009_HD80606,Hebrard+2010}. This planet may be in the midst of high-eccentricity tidal migration
\cite{WuMurray2003}. The Friends of Hot Jupiters
survey \citep{Knutson+2014,Ngo+2015,Piskorz+2015,Bryan+2016,Ngo+2016} found that most hot Jupiters lack a stellar companion capable of driving
Kozai-Lidov oscillations --- but also that many hot Jupiters
do have potentially suitable planetary companions.
Time-series astrometry from the Gaia mission will probe whether these companions have sufficiently high mutual inclinations. 
Using the currently available Gaia data, several groups have found correlations between the orbital orientation of a planet and the orbital orientation of a stellar companion \citep{Dupuy2022,BehmardDaiHoward2022,Christian+2022}. Future Gaia data releases are expected to significantly improve our understanding of companionship and mutual orbital inclinations.

\begin{figure}
  \begin{center}
    \includegraphics[width=8.5cm]{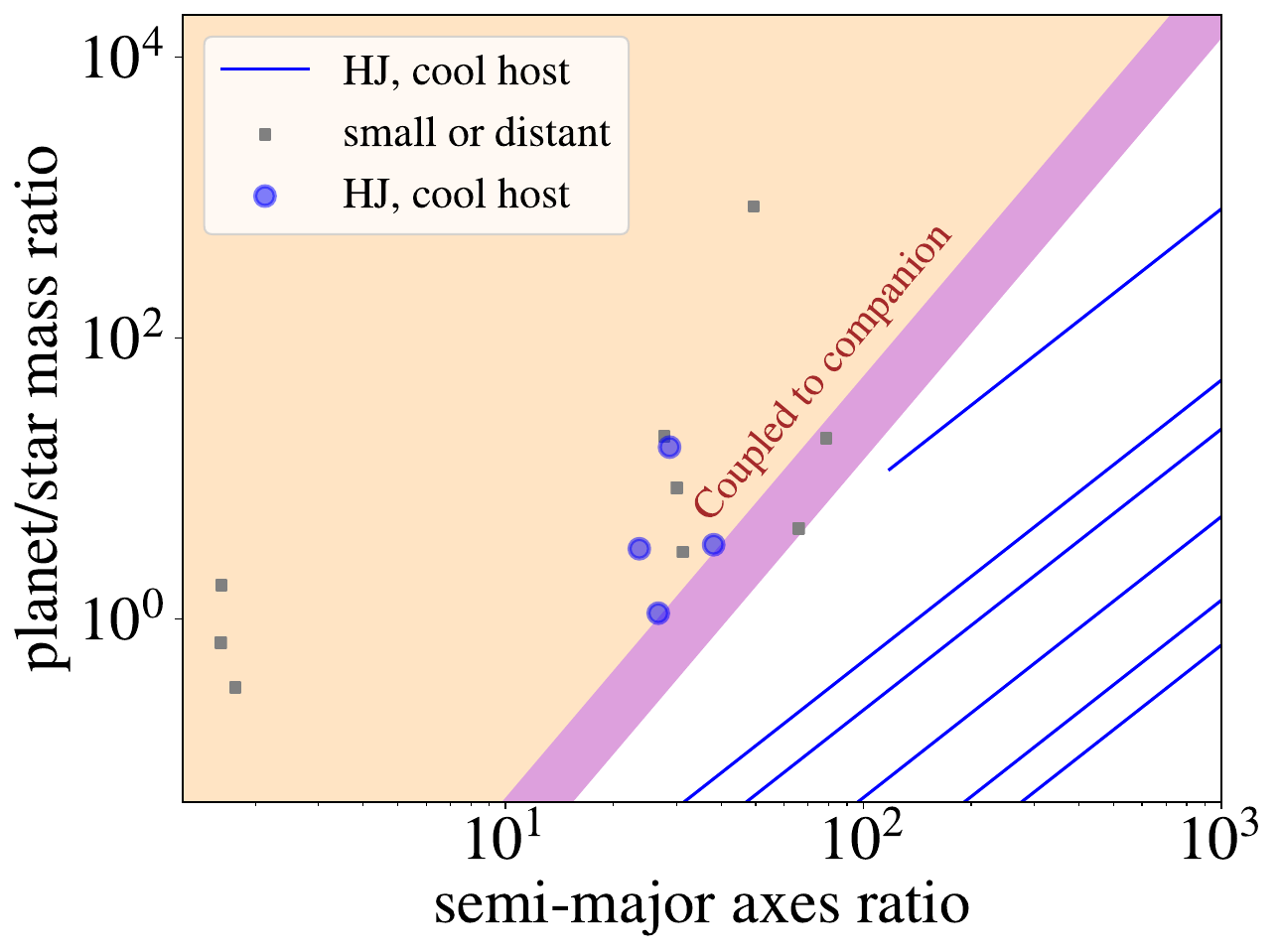} 
    \caption  {\label{fig:hotjupiter_coupling} {\bf Planet-planet coupling. } A handful of cool stars with HJs and low obliquities (blue symbols in the orange region) are strongly coupled to a nearby companion, preventing tidal realignment. Lines represent companions that are detected as radial velocity trends (for which the companion
    mass and semi-major axis cannot be independently determined).     
    }
  \end{center}
\end{figure}

Searching for wide-orbiting companions to hot Jupiters can also shed light on the tidal realignment hypothesis. A mutually inclined companion can continue to drive secular cycles \citep{Becker+2017} or, with sufficient tidal dissipation, drive the hot Jupiter to a mutually inclined equilibrium known as a Cassini state \citep{correira2015}. However, the mutually inclined companion must be nearby and massive enough to compete with the spin-orbit coupling arising from stellar oblateness \citep{LaiAndersonPu2018}, i.e., a giant planet companion interior to $\sim$1~au. The majority of known companions are too far away or
too low in mass to compete with the stellar oblateness coupling  (Figure~\ref{fig:hotjupiter_coupling}). Qatar-2, HAT-P-13, WASP-41, and WASP-47 are currently known systems with an aligned hot Jupiter and a nearby, massive companion; the mutual inclinations of the companions for these systems must be small.

\subsection{Altering the stellar spin vector}
\label{sec:stellar_spin_changes}

The processes discussed so far involve tilting a planet's orbital plane
or reorienting a star with an external torque.
Some theorists have proposed that a star is capable of
reorienting itself.
Based on 2-d hydrodynamical simulations, \cite{RogersLinLau2012,RogersLinMcElwaine2013} argued that stars with convective cores and radiative envelopes
have photospheres that undergo random tumbling.
This is because internal gravity waves (IGW), generated at the
radiative/convective boundary, propagate upward within the radiative zone and crash near the stellar surface, depositing angular momentum
stochastically. The imperfect cancellation of prograde and retrograde waves leads to a self-reinforcing effect that causes the net orientation of the photosphere to wander randomly,
potentially by large angles. The timescale of the changes would be on the order of $10^4$ rotation periods or shorter. 

Because upward-propagating IGWs are only expected in hot stars,
this hypothesis can explain
why hot stars have a broader obliquity distribution than cool stars
(\S~\ref{sec:teff}).
The IGW hypothesis is also compatible with the evidence that hot stars
are misaligned in general, even when they host Kepler-type planets
instead of hot Jupiters (\S~\ref{sec:vsini} and \S~\ref{sec:starspots}).
However, this mechanism does not provide an explanation for
the high obliquities that are observed for some cool
stars with low-mass or wide-orbiting giant planets (\S~\ref{sec:oblique_mass}, \ref{sec:oblique_ar}).

The IGW theory might be tested by seeking evidence
for time variations of $\lambda$ and $v \sin i$ for hot stars.
The interpretation of any observed changes
would be complicated by the possibility of
precession caused by other effects \citep[see, e.g.,][]{Szabo+2012,Johnson+2015,Masuda2015}.
Another signature of IGWs would be the asteroseismic detection of
radial differential rotation within the host stars \citep{Christensen-DalsgaardThompson2011}.

We might also test the IGW theory using binaries with separations beyond the reach of tides that contain both a cool star and a hot star.
Assuming initial spin-orbit alignment, IGWs would tend to cause
the hot star to become misaligned with the binary orbit, while leaving the cool star
alone.
Obliquities have been measured in a few double-star systems \citep[see][for a listing]{MarcussenAlbrecht2021}
but none of them fulfill all these requirements.
There are two systems, DI\,Her and CV\,Vel, in which both components have radiative envelopes and measured obliquities --- but in those systems, the time variations in $\lambda$ and $v \sin i$
are caused by the precession of the misaligned stars around the total angular momentum vector \citep{ReisenbergerGuinan1989,Albrecht+2009,PhilippovRafikov2013,Albrecht+2014,LiangWinnAlbrecht2022}.

\section{Summary and discussion} \label{sec:summary}

The textbook picture of a planetary
system features a star rotating around the axis defined by the plane of its protoplanetary disk or the
orbits of its planets. Experience has shown that this
picture is missing some aspects
of planet formation or the long-term evolution of
planetary systems.
Here, we summarize the general trends in the data that were discussed in more
detail in Section~\ref{sec:methods results}. Table~\ref{tab:obs_trends} provides an
even more concise summary.

\begin{itemize}

\item The orbital planes of hot Jupiters orbiting G and K dwarfs are almost always found to be aligned with the stellar equatorial plane. However, misalignments are
at least occasionally found in all the other classes of systems that have
been investigated (Figure~\ref{fig:obli_techniques}).

\item The risk factors for a star with a close-orbiting giant planet
to develop a high obliquity are: a stellar mass exceeding
about 1.2\,$M_\odot$ or an effective temperature exceeding 6250\,K
(\S~\ref{sec:teff}),
a relatively low planet mass (Neptunian instead of Jovian; \S~\ref{sec:oblique_mass}),
and a relatively wide orbit ($a/R\gtrsim 10$; \S~\ref{sec:oblique_ar}).

\item The most precise measurements so far show that giant planets traveling on short-period orbits around stars with deep convective envelopes tend to be aligned to within
about 1~degree (\S~\ref{sec:well_aligned}), i.e., more closely aligned than the Sun is with respect to the orbits of the planets of the Solar System.

\item The two preceding observations can be interpreted as evidence for tidally-driven spin-orbit alignment, although we lack a good theoretical understanding of the process (\S~\ref{sec:tides}).

\item The occurrence of misalignments for the youngest systems that have been investigated ($\lesssim 100$~Myr) appears to be lower than for their older siblings, although this is not a firm conclusion because of the small sample size (\S~\ref{sec:newborn}).

\item One might expect high obliquities and high eccentricities to be correlated.
Such a correlation might already be present in the current data, though the interpretation is complicated by the possibility of tidal alignment and tidal circularization, by the difficulty of measuring both eccentricity and obliquity in the same system, and by the inhomogeneity of the sample (\S~\ref{sec:obliquity_eccentricity}).

\item Coplanarity of orbits does not guarantee alignment between the planetary orbital plane and the stellar equator. Most multi-transiting systems for which a component of the obliquity vector has been measured tend to be consistent with alignment, but exceptions do exist (\S~\ref{sec:multitransits}). Statistical studies also suggest that alignment is not universal among multi-transiting systems, at least for hot stars (\S~\ref{sec:QPV} and \ref{sec:vsini}).

\item Planets comparable in size to Neptune, or smaller, have been found both on well-aligned
orbits and on misaligned orbits (\S~\ref{sec:rm} and \S~\ref{sec:seismic}). Statistical studies point toward widespread misalignments up to
about 30$^\circ$ (\S~\ref{sec:vsini} and \S~\ref{sec:QPV}).

\item Studies of planets with orbital distance larger than $\sim$1~au are only just becoming possible (\S~\ref{sec:inter_and_spec}).

\item Almost all the 3-d obliquities that have been determined fall into two groups: Most are lower than 40$^\circ$, and the rest are between about 80 and 125$^\circ$ (\S~\ref{sec:ppp}). The misaligned systems do not seem to have much in common: they include
planets ranging from mini-Neptunes to Jupiter-sized, and stars ranging from spectral type M to A.

\end{itemize}

The observational evidence does not single out a unique explanation
for spin-orbit misalignments.
Many plausible theoretical mechanisms have been proposed.
The multiplicity of mechanisms is not too surprising. Undoubtedly, there is more than one way
to perturb stars and planetary orbits. To paraphrase Tolstoy:
circular and coplanar systems are all alike;
each eccentric and misaligned system is eccentric and misaligned
in its own way.
We do not yet know which of the proposed misalignment mechanisms actually occurs, and if so, how often.
Nevertheless, there is evidence for at least two pathways to spin-orbit misalignment: primordial misalignment between a star
and its protoplanetary disk, and gravitational dynamical processes taking place after planet formation. 

Evidence for {\bf primordial misalignment} comes from the small collection of misaligned stars with multiple coplanar planets (Kepler-56, K2-290\,A,
and possibly Kepler-129 and HD\,3167).  In particular, K2-290\,A is known
to have a companion star with properties suitable for misaligning
the protoplanetary disk.  
In addition, the $v\sin i$ and photometric-variability statistical
techniques have provided evidence that hot Kepler stars have a broad distribution of obliquities.
This suggests that misalignment operates independently of the size or orbital period of the planets. Primordial misalignment is such a mechanism.

Furthermore, observations of jets and disks in young stellar systems show indirect evidence for occasional misalignments between a star and the
surrounding protoplanetary disk. Some young binary systems observed with ALMA have circumstellar disks tilted with respect to the orbital plane and each other, such as HK~Tauri. 
However, based on theoretical work, observations of resolved debris disks, obliquity measurements in a few young exoplanet systems, and obliquity measurements in compact transiting multi planet systems with cool host stars suggest primordial misalignments do not seem to
be widespread --- or perhaps they do not persist throughout the entire planet formation epoch.
The evidence for {\bf post-formation gravitational dynamical interactions} comes
from systems with close-orbiting giant planets. Observational support comes from the higher incidence of misalignments for close-orbiting giant planets compared to compact systems of multiple coplanar planets. The observations of alignment in the (admittedly small) number of systems younger than 100~Myr systems with close-orbiting giant planets suggests that misalignments occur later, as expected from
long-term dynamical interactions that stimulate high-eccentricity migration.

There are a few ``poster child'' systems for Kozai-Lidov cycles caused by stellar companion, such as HD\,80606. Of all the post-formation misalignment scenarios, Kozai-Lidov cycles can most easily generate retrograde orbits. However, surveys have generally not found evidence for stellar companions
with suitable properties. This still leaves open the possibility of Kozai-Lidov
cycles induced by wide-orbiting massive planets, if these companions
somehow acquire large mutual inclinations.
Different post-formation mechanism scenarios lead to different predictions for the obliquity distribution, which can in principle be used as an observational test.  The possibility of tidal realignment, and our lack of a quantitative understanding of tidal timescales,
complicates such tests.

{\bf Tidal interactions between the star and planet} appear to be important in the subset of systems with short-period hot Jupiters and stars with thick convective envelopes.
While tides are not fully understood, there are several indications that the problem of tidal realignment without catastrophic orbital decay
might not be as severe as originally feared. Tides are successful in explaining observational trends with stellar structure, orbital separation, and planetary mass. Tides also explain
why the hot Jupiters with cool host stars and the most precisely measured obliquities are very well aligned, with an inclination
dispersion on the order of 1$^\circ$.

The {\bf current sample of systems with obliquity measurements is biased}, and especially favors close-in giant planets orbiting main sequence FGK stars. 
{\bf New projects and space missions} have the potential to end this preoccupation with a small subset of planetary systems.
Bright and well-characterized TESS systems allow for more precise RM and $v \sin i$ measurements over a more diverse range of system properties.
For example, it will be interesting to see if the obliquity distribution
of M-dwarf planet hosts differs from that of G and K dwarfs.
PLATO will follow TESS in this regard, while also providing longer time baselines for seismic and starspot-based measurements.
Gaia will dramatically improve our understanding of companionship and measure orbital inclinations in some systems. This in turn will motivate and enable obliquity measurements for such systems via some of the methods discussed above. Together with the availability of future spectrographs and interferometers,
there will be many ways to improve our knowledge, including:
\begin{itemize} 

\item Bright and well-studied TESS systems harboring close-in giant planets will allow for precise RM measurements ($\lesssim 2^{\circ}$) employing new ground-based spectrographs. Such observations would be useful in
testing whether tidal obliquity damping occurs.
 
\item Theories of primordial misalignment can be tested by observing young systems, planets with large orbital separations, systems with multiple transiting planets with and without wide-orbiting
companions, and resolved disks with stars amenable to the
$v\sin i$ technique. 
Also relevant are observations of misalignments and warps between the inner and outer portions of disks, and the angles between disks around the members of wide double stars. 

\item The increasing number of transiting planets
found around bright stars
will allow for more precise obliquity and eccentricity measurements in systems with lower stellar masses, wider orbits, and smaller planets. It also allows for a more complete characterization of companionship. 
Combined with a better understanding of tidal alignment (first point above), it may be possible to make meaningful comparisons between the measured obliquity distribution and predictions of post-formation misalignment mechanisms.  
 
\item Obliquity measurements in multiple-star systems over a wide
range of separations may help to determine whether primordial disk misalignment mechanisms and stellar KL-cycles are important. These samples could also serve to better test the role of internal gravity waves.

\end{itemize} 
Measurements of the stellar obliquity have opened a new window through which we can observe the 3-d structure of exoplanetary systems. However, the currently available measurements are mainly confined to stars with close-orbiting giant planets or Kepler-type compact multi-planet systems. Other classes of exoplanets have been left out, so far. Two related types of information, mutual orbital inclinations and planetary obliquity, have proven more difficult to obtain. Not before long, all of these quantities will be more accessible for a more diverse exoplanet population. This will be thanks to time-series astrometric data from Gaia, advances in spectroscopic and interferometric instrumentation, and newly discovered exoplanet systems with nearby and bright host stars. The combined information on the orientation of the various components of angular momentum in a planetary system will have an impact on our understanding of star and planet formation, exoplanet orbital dynamics, and theories of tidal interactions between stars and planets. The drama of planet formation and evolution --- initially portrayed on a two-dimensional, planar stage --- is gradually becoming a three-dimensional spectacle.

\acknowledgments
S.A.\ acknowledges the support from the Danish Council for Independent Research through the DFF Sapere Aude Starting Grant No.~4181-00487B, as well as a DFF Research Project 1 grant No.2032-00230B, and the Stellar Astrophysics Centre whose funding is provided by The Danish National Research Foundation (Grant agreement no.: DNRF106).
R.I.D.\ acknowledges the support from grant NNX16AB50G awarded by the NASA Exoplanets Research Program and the Alfred P.\ Sloan Foundation's Sloan Research Fellowship. The Center for Exoplanets and Habitable Worlds is supported by Pennsylvania State University, the Eberly College of Science, and the Pennsylvania Space Grant Consortium. 
J.N.W.\ gratefully acknowledges the hospitality and support of the Institute for Advanced Study.

We thank J.J.\ Zanazzi and Hans Kjeldsen for insightful comments
on a manuscript draft.
We are grateful for feedback and helpful suggestions from Adrian Baker, Subo Dong, Dan Fabrycky, Ren\'{e} Heller, Emil Knudstrup, David Latham, Marcus Marcussen, Gordon Ogilvie, Malena Rice and Scott Tremaine.  We thank the anonymous reviewer for helpful comments that improved the review.
We thank John Southworth for curating  the \href{http://www.astro.keele.ac.uk/jkt/tepcat/tepcat.html}{TEPCat} catalog.

\newpage

\appendix

\section{Population synthesis simulations}
\label{app:sims}

The simulations are based on work by \citet{Dawson2014}, with updates to incorporate options for inertial wave tidal dissipation and a frequency dependent tidal dissipation efficiency. We numerically integrate the planet's specific orbital angular momentum vector $\vec{h}$ and the host star's spin angular frequency vector, assuming a circular orbit. The equations here correspond to \citet{BarkerOgilvie2009}, Eqns.~A7 and A12 with the eccentricity vector $\vec{e}=0$.

\begin{eqnarray}
\label{eqn:h}
\left(\dot{\vec{h}}\right)_{eq} =-\frac{1}{\tau_{eq}}\vec{h}+\frac{1}{\tau_{eq}}\frac{\Omega_\star}{2n}\left(\frac{\vec{\Omega_\star}\cdot\vec{h}}{\Omega_\star~h}\cdot\vec{h}+\frac{h}{\Omega_\star}\vec{\Omega_\star}\right)\nonumber \nonumber\\
\left(\dot{\vec{\Omega_\star}}\right)_{eq,\alpha}=-\frac{m}{k_{\star,\rm eff} M R^2}\dot{\vec{h_{eq}}}-\alpha_{\rm~brake}\Omega_\star^2\vec{\Omega_\star}, \nonumber \\
\end{eqnarray}
\noindent for which 
\begin{eqnarray}
\label{eqn:tau}
\tau_{eq}=\frac{Q}{6k_L}\frac{M}{R^5(M+m)^8G^7}\frac{M}{m}h^{13} \nonumber \\=\tau_{eq, 0}\left(\frac{h}{h_0}\right)^{13}\frac{0.5M_{\rm~Jup}}{m}
\end{eqnarray} is an orbital decay timescale, $k_L$ is the Love number, $Q$ is the tidal quality factor,  $k_{\star,\rm eff}$ is the effective constant of the stellar moment of inertia participating in the tidal realignment, $\alpha_{\rm~brake}$ is a braking constant, and $h_0 = \sqrt{a_0 G (M+m)}$ is the initial specific angular momentum. By default, we use $k_{\star,\rm eff} M R^2 = 0.08 M_\odot R_\odot^2$ for cool stars, $k_{\star,\rm eff} M R^2 = 0.08 (1.2 M_\odot) (1.4 R_\odot)^2$ for hot stars, and $\Omega_{s,0} = 800$ AU$^{2}$yr$^{-1}$. We use $\alpha = 3 \times 10^{-16}$ for hot stars, $\alpha = 1.4 \times 10^{-14}$ for cool stars, $\tau_{eq,0} $ = 500 Gyr, and $h_0 = 1.33$ au$^{2}$yr$^{-1}$. For the pure equilibrium tides simulation, we use $h_0 = 1.68$ au$^{2}$yr$^{-1}$. For the frequency-dependent Q simulations, we use $\tau_{eq,0} $ = 20 Gyr, and $h_0 = 2$ au$^{2}$yr$^{-1}$. For inertial wave tidal dissipation, we use $\tau_{eq,0} $ = 50 Gyr for cold stars, $\tau_{eq,0} = 5 \times 10^6$ Gyr for hot stars, and $h_0 = 1.25$ au$^{2}$yr$^{-1}$. For the decoupled outer envelope simulations, we use $\tau_{eq,0} $ = 100 Gyr, $\alpha =  \times 10^{-13}$ for cool stars, $k_{\star,\rm eff} M R^2 = 0.0008 M_\odot R_\odot^2$ for cool stars, and $k_{\star,\rm eff} M R^2 = 0.0004 (1.2 M_\odot) (1.4 R_\odot)^2$ for hot stars.

For inertial wave tidal dissipation, the tidal forcing component that excites inertial waves exerts a torque. Here, we follow \citet{ Lai2012} to compute the resulting effects on $\vec{h}$ and $\vec{\Omega_\star}$. One component is parallel to the stellar spin, i.e., in the $\vec{\Omega_\star}$ direction. A second component is perpendicular to both $\vec{h}$ and $\vec{\Omega_\star}$, i.e., in the  $\vec{\Omega_s} \times \vec{h}$ direction, and is ignored in our calculations because it does not affect the alignment. The third component is perpendicular to the other two, and thus we compute its unit vector as:
\begin{equation}
    \hat{x} = (\vec{h}\times\vec{\Omega_s}) \times \frac{\vec{\Omega_s}  }{ \Omega_\star^2~h \sin \psi}
\end{equation}
where
\begin{eqnarray}
\cos \psi = \frac{\vec{\Omega_\star}\cdot\vec{h}}{\Omega_\star~h} \nonumber\\
\sin \psi = \frac{|\vec{\Omega_\star}\times\vec{h}|}{\Omega_\star~h}\nonumber\\
\end{eqnarray}
We add the following terms to Eqn.~\ref{eqn:h}.
\begin{eqnarray}
    \left(\dot{\vec{\Omega_\star}}\right)_{dy} &=& -\frac{1}{\tau_{dy}}\left(1-\frac{\tau_{0,dy}}{\tau_{0,eq}} \right)  \nonumber \\ 
    &&\nonumber \left [ \left(\sin \psi \cos \psi\right)^2 \vec{\Omega_\star} - \sin \psi \cos \psi^3 \Omega_\star \hat{x} \right ] \nonumber \\
 \left(\dot{\vec{h}}\right)_{dy}  &=& -\frac{k_{\star,\rm eff} M R^2}{m} \left(\dot{\vec{\Omega_\star}}\right)_{dy} \nonumber \\
\end{eqnarray}
where
\begin{equation}
    \tau_{dy} = \frac{\tau_{0,dy}}{\tau_{0,eq}} \frac{\Omega_\star}{\Omega_{\star,0}}\frac{h_0}{h} \tau_{eq}.
\end{equation}
We set $\frac{\tau_{0,dy}}{\tau_{0,eq}} = 10^{-5}$ for Fig. \ref{fig:model}.

For the frequency-dependent tidal dissipation efficiency model \citep{Penev+2018}, we use Eqn.~\ref{eqn:h} with a modified value of $t_{\rm eq}$:
\begin{equation}
    t_{\rm eq,f} = t_{\rm eq} \frac{\rm max \{10^6/P_{\rm tide}^{3.1}, 10^5\}}{\rm max \{10^6/P_{\rm tide,0}^{3.1}, 10^5\}}
\end{equation}
where $P_{\rm tide} = \pi/(n-\Omega_s)$ is in units of days.

To generate the populations for Fig.~\ref{fig:model}, we select an effective temperature drawn randomly from a uniform distribution
between 4800 and 6800\,K, a planet mass drawn 
from a log-uniform probability distribution
between 0.5 and 15 Jupiter masses, an initial obliquity
$\psi$ drawn from an isotropic distribution,
an evolution time drawn from a uniform
distribution between 0 and 10\,Gyr for cool stars ($T_{\rm eff} < 6250$\,K)
and between 0 and 4\,Gyr for hot stars, and a longitude of ascending node
drawn from a uniform distribution between 0 and 360$^\circ$.
We integrate the momentum equations given above, between $t=0$ and $t=t_\star$. We compute
$\sin~i=\sqrt{1-(\sin\psi\cos\Omega)^2}$, $v\sin~\left(i\right)/R=\Omega_\star \sin~i$,
and
$\lambda=\tan^{-1}\left(\tan\psi\sin\Omega\right)$ (\citealt{FabryckyWinn2009}, Eqn.~11; Column 2 of our Fig.~\ref{fig:model} shows the initial distribution of $\lambda$).

\section{Histogram plot}
\label{app:histrograms}
 
\begin{figure*}[h!]
  \begin{center}
    \includegraphics[width=1.0\textwidth]{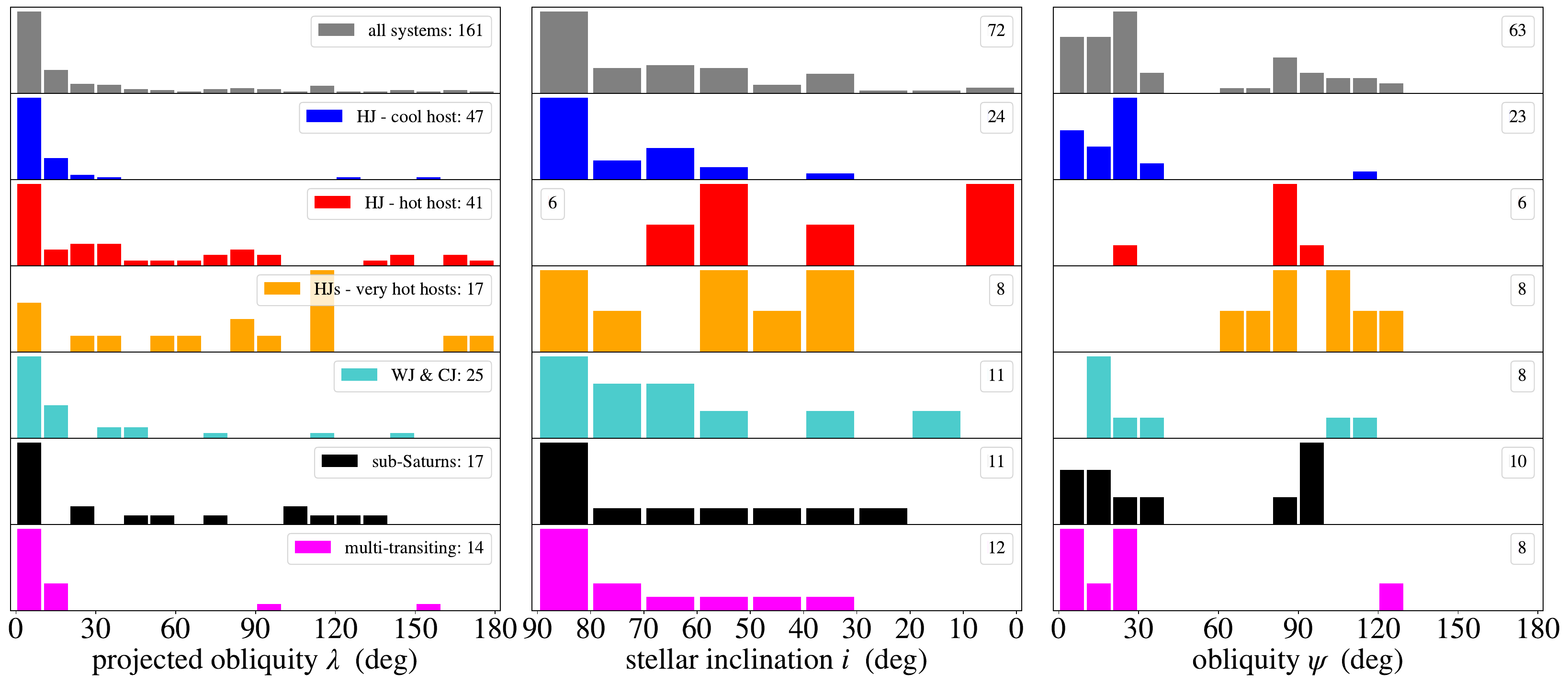}
    \caption {\label{fig:histograms}
    {\bf Overview: Histograms of host star obliquities.} {\it Left}:~Projected obliquities $\lambda$ folded onto the range $[0,180^\circ]$. 
    {\it Middle}:~Stellar inclination measurements folded onto $[0,90^\circ]$.  
    {\it Right:}~Three-dimensional obliquities ($\psi$), for the cases in which
    $\lambda$ and $i$ have both been measured. 
    Histograms are color-coded according to the system's characteristics.
    Stars are designated as cool, hot, or very hot, using effective temperature
    boundaries of 6250\,K and 7000\,K.
    Planets with masses exceeding 0.3\,M$_{\rm Jup}$ are designated hot Jupiters (HJ) if $a/R<10$, and warm/cold Jupiters (WJ/CJ)
    if $a/R>10$.
    Planets with masses $\lesssim$\,$0.3~M_{\rm Jup}$ are designated sub-Saturns.
    }
  \end{center}
\end{figure*}

\section{Systems}
\label{app:data}
 
Here we describe the sources of the system parameters and our vetting process that led to the sample of obliquity data that forms
the basis of many of the figures in this article.
On 5~January~2021, we downloaded the data from the TEPCat catalog,
which is curated by John Southworth and is available here: \href{http://www.astro.keele.ac.uk/jkt/tepcat/tepcat.html}{TEPCat} \cite{Southworth2011}.

We added the following obliquity measurements
which were not in TEPCat at the time of download:
$\beta$~Pic \citep{Kraus+2020},
HD\,332231 \citep{KnudstrupAlbrecht2022},
K2-290 \citep{Hjorth+2021},
a measurement of the second planet in HD\,63433 \citep{Dai+2020},
TOI-942 \citep{Wirth+2021},
HIP 67522 \citep{Heitzmann+2021},
GJ\,3470 \citep{Stefansson+2022},
V1298 Tau c \citep{Feinstein+2021} \& V1298 Tau b \citep{Johnson+2022}, K2-140 \citep{Rice+2021},
K2-232 \citep{Wang+2021},
TOI-1518 \citep{Cabot+2021},
WASP-148 \citep{Wang+2021_wasp148},
HD\,3167 b \citep{Bourrier+2021},
TOI-2109 \cite{Wong+2021},
KElT-25 and KELT-26 \cite{Rodriguez+2020},
and TOI-1268 \cite{Dong+2022}.

We also included some stellar inclination measurements obtained with the method described by \cite{MasudaWinn2020}. These were:
TOI-251 \& TOI-942 \citep{Zhou+2021},
TOI-451 \citep{Newton+2021},
TOI-811 \& TOI-852 \citep{Carmichael+2020},
and TOI-1333 \citep{Rodriguez+2021}.

For some systems, more than one measurement of the stellar inclination or projected obliquity have been reported. TEPCat indicates a ``preferred''
measurement, which we adopted 
except for the following cases:
for HAT-P-7, we chose ``Solution 1'' of \cite{Masuda2015};
for HAT-P-16, we chose \cite{Moutou+2011};
for Kepler-25, we chose \citep{AlbrechtWinnMarcy+2013};
for MASCARA-4, we chose \cite{Dorval+2020};
for WASP-18 \& WASP-31, we chose \cite{AlbrechtWinnJohnson+2012};
and for WASP-33, we chose the 2014 data of \cite{Johnson+2015}.
We do not think that different choices would have led to significantly different conclusions.

We folded all the measurements of projected obliquity onto a half circle ranging from $0^\circ$ to $180^\circ$. (The only exception is one panel in Fig.~\ref{fig:ar_projected_obliquity}.)

We obtained orbital eccentricity data either from the papers reporting
on radial-velocity data of individual objects, or, whenever possible, from the comprehensive work by \cite{Bonomo+2017}. For Kepler-448, we used the eccentricity determined by \cite{Masuda2017}. We obtained information on companionship from the ``Friends of hot Jupiters'' series of papers
\citep{Knutson+2014,Piskorz+2015,Ngo+2016}. 

We excluded systems with uncertainties in the projected obliquity larger than $50$~deg, specifically HAT-P-27 \citep{Brown+2012} and WASP-49 \citep{Wyttenbach+2017}. We also excluded the following specific systems,
for which we think the true uncertainties are larger than the formal
uncertainties would suggest.
For the hot Jupiter system CoRoT-1, two RM datasets have been
reported, one indicating good alignment \citep{Bouchy+2009}, and one indicating strong misalignment \citep{Pont+2010}.
For CoRoT-19, \cite{Guenther+2012} found a projected obliquity of $\lambda=-52_{-22}^{+27}$~deg. However, no post-egress data were obtained, and the RM effect was only detected with 2.3-$\sigma$ confidence. 
For HATS-14, \cite{Zhou+2015} reported a misaligned orbit ($|\lambda|=76_{-5}^{+4}$~deg). However, there are no post-egress data and, as highlighted by the authors, making different assumptions about the orbital semi-amplitude leads to different conclusions about the obliquity.  
For similar reasons, the WASP-134 results by \citep{Anderson+2018} were excluded.
WASP-23 has a low transit impact parameter and a low $v \sin i$, which prevented \cite{Triaud+2011} from drawing any conclusions other than that the orbit is probably prograde. 
We excluded the WASP-1 and WASP-2 systems \citep{Triaud+2010,Simpson+2011} for reasons that were discussed
in detail by \cite{AlbrechtWinnJohnson+2011} and \cite{Triaud2018}. 
\cite{BourrierHebrard2014} presented evidence for a significant misalignment in the 55~Cnc system,
but \cite{Lopez-Morales+2014} presented other data suggesting that the detection
of misalignment had been spurious.
There was also a tentative detection of misalignment in KOI-89 \citep{AhlersBarnesBarnes2015}, but a recent reanalysis of the Kepler data showed that the obliquity is unconstrained by the data \citep{MasudaTamayo2020}. 

Most of the $v \sin i$ measurements adopted for our sample
were obtained by modeling the RM effect. However, for some cases in which the RM data have a low signal-to-noise ratio \citep[e.g., Qatar-2][]{Esposito+2017}, we opted to report the value of $v\sin i$
based on the observed line broadening. It is also worth noting that
in general, the $v \sin i$ value derived from line broadening is based on the disk-integrated stellar spectrum, whereas the value obtained by
modeling the RM effect is more closely connected to the spectrum of the portion of the star hidden by the planet.

\startlongtable
\begin{deluxetable*}{l DD DDD l}
\tabletypesize{\scriptsize}
\tablewidth{50pt}
\tablecaption{\label{tab:stars} Key properties of the systems considered in this review. Additional planetary parameters are given in Table~\ref{tab:planets}.}
\tablewidth{0pt}
\tablehead{
\colhead{System} & \multicolumn2c{T$_{\rm eff}$} & \multicolumn2c{M}
& \multicolumn2c{R} & \multicolumn2c{age} & \multicolumn2c{$v \sin i$}  
& \colhead{References} \\
\colhead{ } & \multicolumn2c{(K)}  & \multicolumn2c{(M$_\odot$)}
& \multicolumn2c{(R$_\odot$)} & \multicolumn2c{(Gyr)} &  \multicolumn2c{(km\,s$^{-1}$)}
& \colhead{} \\
}
\decimalcolnumbers
\startdata
AU Mic & 3700\pm100 & 0.50\pm0.03 & 0.75\pm0.03 & 0.02\pm0.00 & 9.23^{+0.79}_{-0.31} &  1,2 \\ 
Beta PIC & 8200\pm200 & 1.75^{+0.03}_{-0.02} & 1.87\pm0.03 & 0.01^{+0.01}_{-0.00} & 130.00\pm10.00 &  3,4,5,6 \\ 
CoRoT-2 & 5598\pm50 & 1.00\pm0.05 & 0.90\pm0.02 & 2.66\pm1.62 & 11.25\pm0.45 &  7,8 \\ 
CoRoT-3 & 6558\pm44 & 1.40\pm0.06 & 1.58\pm0.09 & 2.20\pm0.60 & 17.00\pm1.00 &  9,10 \\ 
CoRoT-11 & 6343\pm72 & 1.26\pm0.14 & 1.37\pm0.06 & 2.66\pm1.62 & 38.47\pm0.07 &  11,12 \\ 
CoRoT-18 & 5440\pm100 & 0.88\pm0.07 & 0.88^{+0.03}_{-0.03} & 10.69\pm3.82 & 8.00\pm1.00 &  7,13 \\ 
DS Tuc & 5598^{+28}_{-59} & 0.96\pm0.03 & 0.87\pm0.03 & 0.04\pm0.00 & 20.58^{+0.31}_{-0.24} &  14,15 \\ 
EPIC 246851721 & 6202^{+52}_{-50} & 1.32\pm0.04 & 1.62^{+0.04}_{-0.04} & 3.02^{+0.44}_{-0.46} & 74.92^{+0.62}_{-0.60} &  16 \\ 
GJ 3470  & 3622^{+58}_{-55} & 0.53^{+0.02}_{-0.03} & 0.50^{+0.00}_{-0.02} & 1.65\pm1.35 & 0.61^{+0.43}_{-0.37} &  17,18,19 \\ 
GJ 436 & 3416\pm54 & 0.51^{+0.07}_{-0.06} & 0.46\pm0.02 & 6.00\pm2.00 & 0.33^{+0.09}_{-0.07} &  20 \\ 
HAT-P-1 & 5975\pm50 & 1.15\pm0.02 & 1.17^{+0.03}_{-0.03} & 1.90\pm0.60 & 3.75\pm0.58 &  21,22 \\ 
HAT-P-2 & 6290\pm60 & 1.28\pm0.05 & 1.68\pm0.15 & 2.60\pm0.50 & 19.50\pm1.40 &  23,24 \\ 
HAT-P-3 & 5190\pm80 & 0.93\pm0.04 & 0.85\pm0.02 & 2.60\pm0.60 & 1.20\pm0.36 &  25 \\ 
HAT-P-4 & 6036\pm46 & 1.27^{+0.12}_{-0.07} & 1.60^{+0.12}_{-0.04} & 4.20^{+2.60}_{-0.60} & 5.83\pm0.30 &  26,27 \\ 
HAT-P-6 & 6570\pm80 & 1.29\pm0.07 & 1.52\pm0.07 & 2.30^{+0.50}_{-0.70} & 7.80\pm0.60 &  28,29 \\ 
HAT-P-7 & 6310\pm15 & 1.59\pm0.03 & 2.02\pm0.01 & 2.08^{+0.28}_{-0.23} & 2.70\pm0.50 &  30,29 \\ 
HAT-P-8 & 6200\pm80 & 1.19\pm0.07 & 1.48\pm0.03 & 3.40\pm1.00 & 14.50\pm0.80 &  31,32 \\ 
HAT-P-9 & 6350\pm150 & 1.28\pm0.10 & 1.34\pm0.08 & 1.60^{+1.80}_{-1.40} & 12.50\pm1.80 &  33,32 \\ 
HAT-P-11 & 4780\pm50 & 0.80\pm0.03 & 0.68\pm0.01 & 6.50^{+5.90}_{-4.10} & 1.00^{+0.95}_{-0.56} &  34,35 \\ 
HAT-P-12 & 4665\pm45 & 0.69\pm0.02 & 0.68\pm0.01 & 2.50\pm2.00 & 0.99^{+0.42}_{-0.46} &  36,25 \\ 
HAT-P-13 & 5653\pm90 & 1.32\pm0.06 & 1.76\pm0.05 & 5.00^{+2.50}_{-0.70} & 1.70\pm0.40 &  37,38 \\ 
HAT-P-14 & 6600\pm90 & 1.42\pm0.05 & 1.59\pm0.06 & 1.30\pm0.40 & 8.18\pm0.40 &  39,27 \\ 
HAT-P-16 & 6140\pm72 & 1.22\pm0.06 & 1.16\pm0.03 & 1.30\pm0.40 & 3.90\pm0.80 &  40,32 \\ 
HAT-P-17 & 5246\pm80 & 0.86\pm0.04 & 0.84\pm0.02 & 7.80\pm0.30 & 0.56^{+0.12}_{-0.14} &  41,42 \\ 
HAT-P-18 & 4870\pm50 & 0.77\pm0.03 & 0.72\pm0.03 & 12.40^{+4.40}_{-6.40} & 1.58\pm0.18 &  43,44 \\ 
HAT-P-20 & 4595\pm45 & 0.75\pm0.04 & 0.68\pm0.01 & 6.70^{+5.70}_{-3.80} & 1.85\pm0.27 &  45,46 \\ 
HAT-P-22 & 5314\pm50 & 0.94\pm0.04 & 1.06\pm0.05 & 12.40\pm2.40 & 1.65\pm0.26 &  45,25 \\ 
HAT-P-23 & 5885\pm72 & 1.10\pm0.05 & 1.09\pm0.03 & 4.00\pm1.00 & 7.80\pm1.60 &  45,32 \\ 
HAT-P-24 & 6373\pm80 & 1.19^{+0.04}_{-0.04} & 1.29^{+0.07}_{-0.06} & 2.80\pm0.60 & 11.20\pm0.90 &  47,29 \\ 
HAT-P-30 & 6338\pm42 & 1.24\pm0.04 & 1.22\pm0.05 & 1.00^{+0.80}_{-0.50} & 3.07\pm0.24 &  48 \\ 
HAT-P-32 & 6269\pm64 & 1.18\pm0.05 & 1.23\pm0.02 & 2.70\pm0.80 & 20.60\pm1.50 &  9,29 \\ 
HAT-P-34 & 6442\pm88 & 1.39\pm0.05 & 1.53^{+0.14}_{-0.10} & 1.70^{+0.40}_{-0.50} & 24.30\pm1.20 &  9,29 \\ 
HAT-P-36 & 5620\pm40 & 1.03\pm0.04 & 1.04\pm0.02 & 6.60^{+2.90}_{-1.80} & 3.12\pm0.75 &  9,49 \\ 
HAT-P-41 & 6390\pm100 & 1.42\pm0.05 & 1.68^{+0.06}_{-0.04} & 2.20\pm0.40 & 19.60\pm0.50 &  9,50 \\ 
HAT-P-56 & 6566\pm50 & 1.30\pm0.04 & 1.43\pm0.03 & 2.01\pm0.35 & 36.40\pm0.70 &  9,51 \\ 
HAT-P-69 & 7724^{+250}_{-360} & 1.70\pm0.03 & 1.85^{+0.04}_{-0.02} & 1.27^{+0.44}_{-0.28} & 77.43^{+0.64}_{-0.53} &  52 \\ 
HAT-P-70 & 8450^{+540}_{-690} & 1.89^{+0.01}_{-0.01} & 1.86^{+0.12}_{-0.09} & 0.60^{+0.38}_{-0.20} & 99.85^{+0.64}_{-0.61} &  52 \\ 
HATS-2 & 5227\pm95 & 0.88\pm0.04 & 0.90\pm0.02 & 9.70\pm2.90 & 1.50\pm0.50 &  9,53 \\ 
HATS-3 & 6351\pm76 & 1.21\pm0.04 & 1.40\pm0.03 & 3.20^{+0.60}_{-0.40} & 5.75\pm2.98 &  9,54 \\ 
HATS-70 & 7930^{+630}_{-820} & 1.78\pm0.12 & 1.88^{+0.06}_{-0.07} & 0.81^{+0.50}_{-0.33} & 40.61^{+0.32}_{-0.35} &  55 \\ 
HD 332231 & 6089^{+97}_{-96} & 1.13\pm0.08 & 1.28^{+0.04}_{-0.04} & 4.30^{+2.50}_{-1.90} & 5.63\pm0.11 &  56,57 \\ 
HD 3167 & 5261\pm60 & 0.84^{+0.05}_{-0.04} & 0.88^{+0.01}_{-0.01} & 7.80\pm4.30 & 2.10\pm0.40 &  58,59 \\ 
HD 17156 & 6079\pm56 & 1.30\pm0.07 & 1.49\pm0.04 & 3.37^{+0.20}_{-0.47} & 4.18\pm0.31 &  60,61 \\ 
HD 63433 & 5640\pm74 & \multicolumn2c{$-$} & \multicolumn2c{$-$} & 0.41\pm0.23 & 7.30\pm0.30 &  62 \\ 
HD 80606 & 5584\pm13 & 1.02\pm0.06 & 1.04\pm0.04 & 4.70\pm0.60 & 1.70\pm0.30 &  63,64 \\ 
HD 106315 & 6364\pm87 & 1.15\pm0.04 & 1.27\pm0.02 & 4.00\pm1.00 & 13.00\pm0.28 &  65,66 \\ 
HD 149026 & 6147\pm50 & 1.34\pm0.02 & 1.54^{+0.05}_{-0.04} & 2.90\pm0.30 & 7.70\pm0.80 &  21,29 \\ 
HD 189733 & 5050\pm50 & 0.84\pm0.04 & 0.75\pm0.03 & 6.20\pm3.40 & 3.25\pm0.02 &  21,67 \\ 
HD 209458 & 6117\pm50 & 1.15\pm0.04 & 1.16\pm0.01 & 4.00\pm1.20 & 4.80\pm0.20 &  21,68 \\ 
HIP 67522 & 5675\pm75 & 1.22\pm0.05 & 1.38\pm0.06 & 0.02\pm0.00 & 49.21^{+0.97}_{-0.95} &  69,70 \\ 
K2-25 & 3207\pm58 & 0.26\pm0.01 & 0.29\pm0.01 & 0.73\pm0.08 & 8.90\pm0.60 &  71,72 \\ 
K2-29 & 5358\pm38 & 0.94\pm0.02 & 0.86\pm0.01 & 2.60^{+1.20}_{-2.35} & 3.70\pm0.50 &  73 \\ 
K2-34 & 6131\pm47 & 1.20\pm0.03 & 1.38\pm0.05 & 2.88^{+0.26}_{-0.24} & 5.00^{+1.30}_{-1.40} &  74 \\ 
K2-140 & 5585\pm120 & 0.96^{+0.06}_{-0.04} & 1.06^{+0.07}_{-0.06} & 9.80^{+3.40}_{-4.60} & 2.51\pm0.38 &  75,76 \\ 
K2-232 & 6154\pm60 & 1.19^{+0.03}_{-0.03} & 1.16\pm0.02 & 1.43^{+0.82}_{-0.75} & 5.15\pm0.62 &  77,78 \\ 
K2-290 & 6302\pm120 & 1.19\pm0.07 & 1.51^{+0.08}_{-0.08} & 4.00^{+1.60}_{-0.80} & 6.90^{+0.50}_{-0.60} &  79 \\ 
KELT-1 & 6516\pm49 & 1.33\pm0.06 & 1.47^{+0.04}_{-0.04} & 1.75\pm0.25 & 56.00\pm2.00 &  80 \\ 
KELT-6 & 6272\pm61 & 1.13\pm0.06 & 1.53^{+0.14}_{-0.14} & 4.90^{+0.66}_{-0.46} & 4.12\pm0.26 &  81 \\ 
KELT-7 & 6789^{+50}_{-49} & 1.53^{+0.07}_{-0.05} & 1.73^{+0.04}_{-0.04} & 1.20\pm0.20 & 69.30\pm0.20 &  82,51 \\ 
KELT-9 & 9600\pm400 & 2.32\pm0.16 & 2.42\pm0.06 & 0.30\pm0.00 & 116.90\pm1.80 &  83,84 \\ 
KELT-17 & 7454\pm49 & 1.64^{+0.07}_{-0.06} & 1.65^{+0.06}_{-0.06} & 0.65\pm0.15 & 44.20^{+1.50}_{-1.30} &  85 \\ 
KELT-19 & 7500\pm110 & 1.62^{+0.25}_{-0.20} & 1.83\pm0.10 & 1.10\pm0.10 & 84.20\pm2.00 &  86 \\ 
KELT-21 & 7598^{+81}_{-84} & 1.46^{+0.03}_{-0.03} & 1.64\pm0.03 & 1.60\pm0.10 & 146.03\pm0.48 &  87 \\ 
KELT-25 & 8280^{+440}_{-180} & 2.18^{+0.12}_{-0.11} & 2.26^{+0.05}_{-0.05} & 0.46^{+0.12}_{-0.14} & 114.20\pm1.20 &  88 \\ 
KELT-26 & 8640^{+500}_{-240} & 1.92^{+0.14}_{-0.16} & 1.80^{+0.05}_{-0.05} & 0.43^{+0.31}_{-0.25} & 12.28^{+0.78}_{-0.82} &  88 \\ 
Kepler-8 & 6213\pm150 & 1.23\pm0.07 & 1.50\pm0.04 & 3.80\pm1.50 & 8.90\pm1.00 &  9,29 \\ 
Kepler-9 & 5774\pm60 & 1.02^{+0.03}_{-0.04} & 0.96\pm0.02 & 3.00\pm1.00 & 2.74\pm0.40 &  89,90 \\ 
Kepler-13 & 7650\pm250 & 1.72\pm0.10 & 1.71\pm0.04 & 0.71^{+0.18}_{-0.15} & 62.70\pm0.20 &  91,92 \\ 
Kepler-17 & 5781\pm85 & 1.07^{+0.05}_{-0.16} & 0.98^{+0.02}_{-0.05} & 2.90^{+1.50}_{-1.60} & 4.70\pm1.00 &  93 \\ 
Kepler-25 & 6270\pm79 & 1.17^{+0.03}_{-0.03} & 1.32^{+0.02}_{-0.01} & 2.75\pm0.30 & 8.20\pm0.20 &  94,95 \\ 
Kepler-30 & 5498\pm54 & 0.99\pm0.08 & 0.95\pm0.12 & 2.00\pm0.80 & 1.94\pm0.50 &  96,97 \\ 
Kepler-50 & 6225\pm66 & 1.24\pm0.05 & 1.58\pm0.02 & 3.80\pm0.80 & 8.00^{+1.20}_{-1.00} &  98 \\ 
Kepler-56 & 4840\pm97 & 1.32\pm0.13 & 4.23\pm0.15 & 3.50\pm1.30 & 1.70\pm1.00 &  99 \\ 
Kepler-63 & 5576\pm50 & 0.98^{+0.04}_{-0.04} & 0.90^{+0.03}_{-0.02} & 0.21\pm0.05 & 5.60\pm0.80 &  100 \\ 
Kepler-65 & 6211\pm66 & 1.25^{+0.02}_{-0.02} & 1.44^{+0.03}_{-0.03} & 2.90\pm0.70 & 10.40\pm0.60 &  98 \\ 
Kepler-89 & 6182\pm82 & 1.28\pm0.05 & 1.52\pm0.14 & 3.90^{+0.30}_{-0.20} & 7.30\pm0.60 &  101 \\ 
Kepler-408 & 6088\pm65 & 1.05\pm0.04 & 1.25\pm0.05 & 4.70\pm1.20 & 2.80\pm1.00 &  102 \\ 
Kepler-410 & 6375\pm44 & 1.21^{+0.03}_{-0.03} & 1.35\pm0.01 & 2.76\pm0.54 & 12.90\pm0.60 &  103 \\ 
Kepler-420 & 5520\pm80 & 0.99\pm0.05 & 1.13\pm0.14 & 9.30\pm0.30 & 7.30\pm0.60 &  104 \\ 
Kepler-432 & 5020\pm60 & 1.32^{+0.10}_{-0.07} & 4.06^{+0.12}_{-0.08} & 9.30^{+0.80}_{-1.00} & 2.70\pm0.50 &  9 \\ 
Kepler-448 & 6820\pm120 & 1.45\pm0.09 & 1.63\pm0.15 & 1.40\pm0.50 & 66.43^{+1.00}_{-0.95} &  105,50 \\ 
MASCARA-1 & 7554\pm150 & 1.72\pm0.07 & 2.10\pm0.20 & 1.00\pm0.20 & 109.00\pm3.00 &  106 \\ 
MASCARA-2 & 8730^{+250}_{-260} & 1.76^{+0.14}_{-0.20} & 1.56^{+0.06}_{-0.06} & 0.20^{+0.10}_{-0.05} & 114.00\pm3.00 &  107 \\ 
MASCARA-3 & 6508\pm49 & 1.46^{+0.06}_{-0.06} & 1.51\pm0.02 & 0.78^{+0.61}_{-0.42} & 19.76\pm0.16 &  108 \\ 
MASCARA-4 & 7800\pm200 & 1.75\pm0.05 & 1.92\pm0.11 & 0.70\pm-0.20 & 46.50\pm1.00 &  109 \\ 
NGTS-2 & 6450\pm50 & 1.30\pm0.03 & 1.62\pm0.09 & 2.70\pm0.20 & 15.91\pm0.49 &  110 \\ 
Qatar-1 & 4910\pm100 & 0.84^{+0.04}_{-0.04} & 0.80\pm0.02 & 8.90\pm3.70 & 1.70\pm0.30 &  21,111 \\ 
Qatar-2 & 4645\pm50 & 0.73\pm0.02 & 0.70\pm0.01 & 9.40\pm3.20 & 2.80\pm0.50 &  112,113 \\ 
TOI-1268 & 5257\pm40 & 0.90\pm0.13 & 0.86\pm0.02 & 280.00\pm90.00 & 4.30^{+0.55}_{-0.45} &  114 \\ 
TOI-1333 & 6274\pm97 & 1.46^{+0.08}_{-0.08} & 1.92^{+0.06}_{-0.06} & 2.33^{+0.71}_{-0.56} & 14.20\pm0.50 &  115 \\ 
TOI-1518 & 7300\pm100 & 1.79\pm0.26 & 1.95\pm0.05 & \multicolumn2c{$-$} & 74.40\pm2.30 &  116 \\ 
TOI-2109 & 6540\pm16 & 1.45\pm0.07 & 1.70^{+0.06}_{-0.06} & 1.77^{+0.88}_{-0.68} & 81.20\pm1.60 &  117 \\ 
TOI-251 & 5875^{+100}_{-190} & 1.04^{+0.01}_{-0.01} & 0.88^{+0.04}_{-0.05} & 0.18\pm0.14 & 11.50\pm1.00 &  118 \\ 
TOI-451 & 5550\pm56 & 0.95\pm0.02 & 0.88\pm0.03 & 0.13\pm0.01 & 7.90\pm0.50 &  119 \\ 
TOI-811 & 6107\pm77 & 1.32^{+0.05}_{-0.07} & 1.27^{+0.06}_{-0.09} & 0.12^{+0.04}_{-0.04} & 7.11\pm0.50 &  120 \\ 
TOI-852 & 5768^{+84}_{-81} & 1.32^{+0.05}_{-0.04} & 1.71\pm0.04 & 4.04^{+0.68}_{-0.76} & 14.50\pm0.50 &  120 \\ 
TOI-942 & 4928^{+125}_{-85} & 0.79^{+0.04}_{-0.03} & 1.02^{+0.02}_{-0.02} & 0.09\pm0.07 & 14.30\pm0.50 &  118 \\ 
TRAPPIST-1 & 2557\pm47 & 0.09\pm0.00 & 0.12\pm0.00 & 7.60\pm2.20 & 2.04\pm0.18 &  121,122 \\ 
TrES-1 & 5226\pm50 & 0.89\pm0.05 & 0.82\pm0.02 & 3.70^{+3.40}_{-2.80} & 1.30\pm0.30 &  9,123 \\ 
TrES-2 & 5850\pm50 & 0.99\pm0.06 & 0.96\pm0.02 & 5.00^{+2.70}_{-2.10} & 1.00\pm0.60 &  9,124 \\ 
TrES-4 & 6295\pm65 & 1.45\pm0.05 & 1.81\pm0.08 & 2.20\pm0.40 & 8.10\pm1.10 &  9,125 \\ 
V1298 Tau & 4970\pm120 & 1.13\pm0.05 & 1.41^{+0.00}_{-0.02} & 0.02\pm0.00 & 24.77\pm0.19 &  126,127,128 \\ 
WASP-3 & 6340\pm90 & 1.11^{+0.08}_{-0.06} & 1.30^{+0.05}_{-0.04} & 2.10\pm1.20 & 13.90\pm0.03 &  9,129 \\ 
WASP-4 & 5540\pm55 & 0.93\pm0.06 & 0.91\pm0.02 & 7.00\pm2.90 & 2.14^{+0.38}_{-0.35} &  7,130 \\ 
WASP-5 & 5770\pm65 & 1.03\pm0.05 & 1.09\pm0.04 & 5.60\pm2.20 & 3.20\pm0.30 &  9,131 \\ 
WASP-6 & 5375\pm65 & 0.84\pm0.07 & 0.86\pm0.03 & 11.00^{+3.00}_{-7.00} & 1.60^{+0.27}_{-0.17} &  9,132 \\ 
WASP-7 & 6520\pm70 & 1.32\pm0.07 & 1.48\pm0.09 & 2.40\pm1.00 & 14.00\pm2.00 &  133,134 \\ 
WASP-8 & 5690\pm36 & 1.09\pm0.03 & 0.98\pm0.02 & 4.00\pm1.00 & 1.90\pm0.05 &  135,136 \\ 
WASP-11 & 4900\pm65 & 0.81\pm0.04 & 0.77\pm0.01 & 7.60^{+5.80}_{-3.00} & 1.04\pm0.15 &  9,49 \\ 
WASP-12 & 6313\pm52 & 1.43^{+0.11}_{-0.09} & 1.66^{+0.05}_{-0.04} & 2.00^{+0.70}_{-2.00} & 1.60^{+0.80}_{-0.40} &  9,29 \\ 
WASP-13 & 6025\pm21 & 1.22\pm0.12 & 1.66\pm0.08 & 5.00^{+2.60}_{-1.70} & 5.70\pm0.40 &  9,137 \\ 
WASP-14 & 6462\pm75 & 1.30\pm0.06 & 1.32^{+0.10}_{-0.07} & 0.75\pm0.25 & 2.80\pm0.57 &  9,138 \\ 
WASP-15 & 6573\pm70 & 1.30\pm0.05 & 1.52\pm0.04 & 2.40^{+0.80}_{-1.00} & 4.27^{+0.26}_{-0.36} &  139,131 \\ 
WASP-16 & 5630\pm70 & 0.98\pm0.05 & 1.09\pm0.04 & 8.60^{+3.90}_{-3.80} & 1.20^{+0.40}_{-0.50} &  139 \\ 
WASP-17 & 6550\pm100 & 1.29\pm0.08 & 1.58\pm0.04 & 2.65\pm0.25 & 14.67^{+0.81}_{-0.57} &  9,131 \\ 
WASP-18 & 6400\pm70 & 1.29\pm0.06 & 1.25\pm0.03 & 1.00\pm0.50 & 11.20\pm0.60 &  9,29 \\ 
WASP-19 & 5460\pm90 & 0.94\pm0.04 & 1.02\pm0.01 & 9.95\pm2.49 & 4.40\pm0.90 &  7,29 \\ 
WASP-20 & 5987\pm20 & 1.11\pm0.03 & 1.24\pm0.04 & 7.00^{+2.00}_{-1.00} & 4.75\pm0.51 &  9,140 \\ 
WASP-21 & 5924\pm55 & 0.89\pm0.08 & 1.14\pm0.05 & 5.50\pm2.00 & 0.66\pm0.14 &  21,141 \\ 
WASP-22 & 6153\pm46 & 1.25^{+0.07}_{-0.03} & 1.25\pm0.03 & 4.30^{+1.60}_{-1.10} & 4.40\pm0.30 &  9,142 \\ 
WASP-24 & 6107\pm77 & 1.17\pm0.07 & 1.32\pm0.04 & 3.80^{+1.30}_{-1.20} & 7.00\pm0.90 &  9,143 \\ 
WASP-25 & 5736\pm35 & 1.05\pm0.04 & 0.92\pm0.02 & 0.02^{+3.96}_{-0.01} & 2.90\pm0.30 &  144,145 \\ 
WASP-26 & 6015\pm55 & 1.09\pm0.05 & 1.28\pm0.04 & 6.00\pm2.00 & 2.20\pm0.70 &  9,29 \\ 
WASP-28 & 6084\pm45 & 0.99\pm0.07 & 1.08\pm0.03 & 5.00^{+3.00}_{-2.00} & 3.25\pm0.34 &  140 \\ 
WASP-30 & 6190\pm50 & 1.25^{+0.03}_{-0.04} & 1.39^{+0.03}_{-0.03} & 3.40^{+0.30}_{-0.50} & 12.10^{+0.40}_{-0.50} &  146 \\ 
WASP-31 & 6175\pm70 & 1.16\pm0.03 & 1.25\pm0.03 & 1.00^{+3.00}_{-0.50} & 6.60\pm0.60 &  9,145 \\ 
WASP-32 & 6100\pm100 & 1.10\pm0.03 & 1.11\pm0.05 & 2.22^{+0.62}_{-0.73} & 3.90^{+0.40}_{-0.50} &  9,147 \\ 
WASP-33 & 7430\pm100 & 1.56^{+0.04}_{-0.08} & 1.51^{+0.02}_{-0.03} & 0.10^{+0.40}_{-0.09} & 86.63^{+0.32}_{-0.37} &  148,149 \\ 
WASP-38 & 6150\pm80 & 1.20\pm0.04 & 1.33^{+0.03}_{-0.03} & 3.29^{+0.42}_{-0.53} & 7.50^{+0.10}_{-0.20} &  9,147 \\ 
WASP-39 & 5485\pm50 & 0.91\pm0.05 & 0.94\pm0.02 & 9.00^{+3.00}_{-4.00} & 1.40\pm0.25 &  9,25 \\ 
WASP-41 & 5546\pm33 & 0.99\pm0.03 & 0.89\pm0.01 & 9.80^{+2.30}_{-3.90} & 1.60\pm1.10 &  9,150 \\ 
WASP-43 & 4520\pm120 & 0.72\pm0.03 & 0.67^{+0.01}_{-0.01} & 7.00\pm7.00 & 2.26\pm0.54 &  9,46 \\ 
WASP-47 & 5576\pm67 & 1.04\pm0.03 & 1.14\pm0.01 & 6.70^{+1.50}_{--1.10} & 1.80^{+0.24}_{-0.16} &  151,152 \\ 
WASP-52 & 5000\pm100 & 0.80\pm0.05 & 0.79\pm0.02 & 10.70^{+1.90}_{-4.50} & 2.62\pm0.07 &  9,141 \\ 
WASP-53 & 4950\pm60 & 0.84\pm0.05 & 0.80\pm0.02 & \multicolumn2c{$-$} & 0.86\pm0.21 &  153 \\ 
WASP-60 & 6105\pm50 & 1.23\pm0.03 & 1.40\pm0.07 & 1.70^{+0.90}_{-0.70} & 2.97\pm0.47 &  25 \\ 
WASP-61 & 6250\pm150 & 1.27\pm0.06 & 1.39\pm0.03 & 2.70^{+0.10}_{-0.60} & 11.10\pm0.70 &  154 \\ 
WASP-62 & 6230\pm80 & 1.25\pm0.05 & 1.28\pm0.05 & 0.80\pm0.60 & 9.30\pm0.20 &  154 \\ 
WASP-66 & 6600\pm150 & 1.30\pm0.07 & 1.75\pm0.09 & 3.70^{+0.70}_{-1.20} & 12.10\pm2.20 &  9,155 \\ 
WASP-69 & 4700\pm50 & 0.83\pm0.03 & 0.81\pm0.03 & 7.00\pm7.00 & 2.20\pm0.40 &  9,156 \\ 
WASP-71 & 6180\pm52 & 1.56\pm0.06 & 2.26\pm0.17 & 3.60^{+1.60}_{-1.00} & 7.80^{+0.30}_{-0.40} &  154 \\ 
WASP-72 & 6250\pm100 & 1.39\pm0.06 & 1.98\pm0.24 & 3.20\pm0.60 & 5.00^{+1.40}_{-1.20} &  157,158 \\ 
WASP-74 & 5984\pm57 & 1.19\pm0.04 & 1.54\pm0.03 & 3.49\pm0.65 & 5.85\pm0.50 &  159 \\ 
WASP-76 & 6329\pm65 & 1.46\pm0.02 & 1.76\pm0.07 & 1.82\pm0.27 & 1.48\pm0.28 &  160 \\ 
WASP-78 & 6100\pm150 & 1.39^{+0.09}_{-0.08} & 2.35^{+0.10}_{-0.09} & 2.80^{+1.60}_{-0.30} & 7.10\pm0.50 &  154 \\ 
WASP-79 & 6600\pm100 & 1.39\pm0.08 & 1.51^{+0.04}_{-0.03} & 1.40\pm0.30 & 17.41^{+0.20}_{-0.12} &  154,50 \\ 
WASP-80 & 4145\pm100 & 0.60\pm0.04 & 0.59\pm0.01 & 7.00\pm7.00 & 1.27^{+0.14}_{-0.17} &  9,161 \\ 
WASP-84 & 5280\pm80 & 0.85\pm0.06 & 0.77\pm0.02 & 2.10\pm1.60 & 2.56\pm0.08 &  162 \\ 
WASP-85 & 5685\pm65 & 1.09\pm0.08 & 0.94\pm0.02 & 0.50^{+0.30}_{-0.10} & 3.41\pm0.89 &  163,164 \\ 
WASP-87 & 6450\pm120 & 1.20\pm0.09 & 1.63\pm0.06 & 3.80\pm0.80 & 9.80\pm0.60 &  165,155 \\ 
WASP-94A & 6170\pm80 & 1.45\pm0.09 & 1.62^{+0.05}_{-0.04} & 2.70\pm0.60 & 4.20\pm0.50 &  9,166 \\ 
WASP-100 & 6940\pm120 & 1.47^{+0.06}_{-0.05} & 1.67^{+0.18}_{-0.11} & 1.57^{+0.30}_{-0.20} & 12.80^{+2.30}_{-2.20} &  9,158 \\ 
WASP-103 & 6110\pm160 & 1.21^{+0.10}_{-0.12} & 1.41^{+0.04}_{-0.05} & 4.00\pm1.00 & 6.50\pm2.00 &  9,155 \\ 
WASP-107 & 4425\pm70 & 0.68^{+0.02}_{-0.02} & 0.67\pm0.02 & 6.90^{+3.70}_{-3.40} & 2.50\pm0.80 &  167,168 \\ 
WASP-109 & 6520\pm140 & 1.20\pm0.09 & 1.35\pm0.04 & 2.60\pm0.90 & 18.90^{+2.40}_{-2.30} &  158 \\ 
WASP-111 & 6400\pm150 & 1.50\pm0.11 & 1.85\pm0.10 & 2.60\pm0.60 & 11.12\pm0.77 &  165 \\ 
WASP-117 & 6040\pm90 & 1.13\pm0.03 & 1.17^{+0.07}_{-0.06} & 4.60\pm2.00 & 1.46\pm0.14 &  9,169 \\ 
WASP-121 & 6586\pm59 & 1.38\pm0.02 & 1.44\pm0.03 & 1.50\pm1.00 & 13.56^{+0.69}_{-0.68} &  170 \\ 
WASP-127 & 5750\pm100 & 0.95\pm0.02 & 1.33\pm0.03 & 9.66\pm1.00 & 0.53^{+0.07}_{-0.05} &  171 \\ 
WASP-134 & 5700\pm100 & 1.13\pm0.04 & 1.18\pm0.05 & 5.10\pm1.60 & 2.08\pm0.26 &  172 \\ 
WASP-148 & 5437\pm21 & 0.97^{+0.05}_{-0.06} & 0.90^{+0.01}_{-0.01} & \multicolumn2c{$-$} & 2.30^{+0.38}_{-0.34} &  173 \\ 
WASP-166 & 6050\pm50 & 1.19\pm0.06 & 1.22\pm0.06 & 2.10\pm0.90 & 5.10\pm0.30 &  174 \\ 
WASP-167 & 7043^{+89}_{-68} & 1.59\pm0.08 & 1.79\pm0.05 & 1.54\pm0.40 & 49.94\pm0.04 &  175 \\ 
WASP-174 & 6400\pm100 & 1.24\pm0.04 & 1.35\pm0.02 & 1.65\pm0.85 & 16.50\pm0.50 &  176 \\ 
WASP-180 & 6600\pm200 & 1.30\pm0.10 & 1.19\pm0.06 & 1.22\pm0.99 & 19.60\pm0.60 &  177 \\ 
WASP-189 & 8000\pm80 & 2.03\pm0.07 & 2.36\pm0.03 & 0.73\pm0.13 & 100.00\pm5.00 &  178 \\ 
WASP-190 & 6400\pm100 & 1.35\pm0.05 & 1.60\pm0.10 & 2.80\pm0.40 & 13.30\pm0.60 &  179 \\ 
XO-2 & 5332\pm57 & 0.96\pm0.05 & 1.00^{+0.03}_{-0.03} & 7.80^{+1.20}_{-1.30} & 1.07\pm0.09 &  9,180 \\ 
XO-3 & 6429\pm75 & 1.21\pm0.06 & 1.41\pm0.06 & 2.82^{+0.58}_{-0.82} & 18.40\pm0.80 &  9,181 \\ 
XO-4 & 6397\pm70 & 1.29^{+0.03}_{-0.03} & 1.55^{+0.04}_{-0.03} & 2.10\pm0.60 & 8.90\pm0.50 &  182,183 \\ 
XO-6 & 6720\pm100 & 1.47\pm0.06 & 1.93\pm0.18 & 1.88^{+0.90}_{-0.20} & 48.00\pm3.00 &  184 \\ 
pi Men & 5998\pm62 & 1.07\pm0.04 & 1.17\pm0.02 & 2.98^{+1.40}_{-0.30} & 3.16\pm0.27 &  185,186 \\ 

\enddata
\end{deluxetable*}
\tablecomments{ All data is taken from \href{https://www.astro.keele.ac.uk/jkt/tepcat/}{TEPCat} \citep{Southworth2011} or the following references:  
 1 \citep{Plavchan+2020}, 2 \citep{Hirano+2020}, 3 \citep{Kervella+2004}, 4 \citep{Lacour+2021}, 5 \citep{Zuckerman+2001}, 6 \citep{Royer+2007}, 7 \citep{MaxtedSerenelliSouthworth2015}, 8 \citep{Czesla+2012}, 9 \citep{Bonomo+2017}, 10 \citep{Triaud+2009}, 11 \citep{Gandolfi+2010}, 12 \citep{Gandolfi+2012}, 13 \citep{Hebrard+2011}, 14 \citep{Newton+2019}, 15 \citep{Zhou+2020}, 16 \citep{Yu+2018}, 17 \citep{Bailer-Jones+2018}, 18 \citep{Stefansson+2022}, 19 \citep{Bonfils+2012}, 20 \citep{Bourrier+2018}, 21 \citep{Bonfanti+2015}, 22 \citep{Johnson+2008}, 23 \citep{Pal+2010}, 24 \citep{Loeillet+2008}, 25 \citep{Mancini+2018}, 26 \citep{Kovacs+2007}, 27 \citep{Winn+2011}, 28 \citep{Noyes+2008}, 29 \citep{AlbrechtWinnJohnson+2012}, 30 \citep{Lund+2014}, 31 \citep{Latham+2009}, 32 \citep{Moutou+2011}, 33 \citep{Shporer+2009}, 34 \citep{Bakos+2010}, 35 \citep{Winn+2010_hatp11}, 36 \citep{Hartman+2009}, 37 \citep{Bakos+2009}, 38 \citep{Winn+2010_hat-p-13}, 39 \citep{Torres+2010}, 40 \citep{Buchhave+2010}, 41 \citep{Howard+2012}, 42 \citep{Fulton+2013}, 43 \citep{Hartman+2011}, 44 \citep{Esposito+2014}, 45 \citep{Bakos+2011}, 46 \citep{Esposito+2017}, 47 \citep{Kipping2010}, 48 \citep{Johnson+2011}, 49 \citep{Mancini+2015b}, 50 \citep{Johnson+2017}, 51 \citep{Zhou+2016}, 52 \citep{Zhou+2019}, 53 \citep{Mohler-Fischer+2013}, 54 \citep{Addison+2014}, 55 \citep{Zhou+2019b}, 56 \citep{Dalba+2020}, 57 \citep{KnudstrupAlbrecht2022}, 58 \citep{Christiansen+2017}, 59 \citep{Dalal+2019}, 60 \citep{Nutzman+2011}, 61 \citep{Narita+2009_HD17156}, 62 \citep{Mann+2020}, 63 \citep{Liu+2018}, 64 \citep{Hebrard+2010}, 65 \citep{Crossfield+2017}, 66 \citep{Zhou+2018}, 67 \citep{Cegla+2016}, 68 \citep{Santos+2020}, 69 \citep{Rizzuto+2020}, 70 \citep{Heitzmann+2021}, 71 \citep{Mann+2016}, 72 \citep{Stefansson+2020}, 73 \citep{Santerne+2016}, 74 \citep{Hirano+2016}, 75 \citep{Korth+2019}, 76 \citep{Rice+2021}, 77 \citep{Brahm+2018}, 78 \citep{Wang+2021}, 79 \citep{Hjorth+2019}, 80 \citep{Siverd+2012}, 81 \citep{Damasso+2015}, 82 \citep{Bieryla+2015}, 83 \citep{Gaudi+2017}, 84 \citep{Wyttenbach+2020}, 85 \citep{Zhou+2016b}, 86 \citep{Siverd+2018}, 87 \citep{Johnson+2018}, 88 \citep{Rodriguez+2020}, 89 \citep{Holman+2010}, 90 \citep{Wang+2018}, 91 \citep{Szabo+2011}, 92 \citep{HowarthMorello2017}, 93 \citep{Desert+2011}, 94 \citep{Benomar+2014}, 95 \citep{AlbrechtWinnMarcy+2013}, 96 \citep{Sanchis-Ojeda+2012}, 97 \citep{Fabrycky+2012b}, 98 \citep{Chaplin+2013}, 99 \citep{Huber+2013}, 100 \citep{Sanchis-Ojeda+2013}, 101 \citep{Hirano+2012}, 102 \citep{KamikaBenomarSuto+2019}, 103 \citep{VanEylen+2014}, 104 \citep{Santerne+2014}, 105 \citep{Bourrier+2015}, 106 \citep{Talens+2017b}, 107 \citep{Lund+2017}, 108 \citep{Rodriguez+2019}, 109 \citep{Dorval+2020}, 110 \citep{Anderson+2018c}, 111 \citep{Covino+2013}, 112 \citep{MocnikSouthworthHellier2017}, 113 \citep{Bryan+2012}, 114 \citep{Dong+2022}, 115 \citep{Rodriguez+2021}, 116 \citep{Cabot+2021}, 117 \citep{Wong+2021}, 118 \citep{Zhou+2021}, 119 \citep{Newton+2021}, 120 \citep{Carmichael+2021}, 121 \citep{BurgasserMamajek2017}, 122 \citep{Hirano+2020b}, 123 \citep{Narita+2007}, 124 \citep{Winn+2008}, 125 \citep{Narita+2010}, 126 \citep{Davide+2019b}, 127 \citep{Johnson+2022}, 128 \citep{David+2019}, 129 \citep{Miller+2010}, 130 \citep{Sanchis-Ojeda+2011}, 131 \citep{Triaud+2010}, 132 \citep{Gillion+2009}, 133 \citep{Southworth+2011}, 134 \citep{AlbrechtWinnButler+2012}, 135 \citep{Queloz+2010}, 136 \citep{Bourrier+2017}, 137 \citep{Brothwell+2014}, 138 \citep{Johnson+2009}, 139 \citep{Southworth+2013}, 140 \citep{Anderson2015}, 141 \citep{Chen+2020}, 142 \citep{Anderson+2011}, 143 \citep{Simpson+2011}, 144 \citep{Enoch+2011}, 145 \citep{Brown+2012}, 146 \citep{Triaud+2013}, 147 \citep{Brown+2012b}, 148 \citep{Moya+2011}, 149 \citep{Johnson+2015}, 150 \citep{Southworth+2016}, 151 \citep{Vanderburg+2017}, 152 \citep{Sanchis-Ojeda+2015}, 153 \citep{Triaud+2017}, 154 \citep{Brown+2017}, 155 \citep{Addison+2016}, 156 \citep{Casasayas-Barris+2017}, 157 \citep{Gillon+2013}, 158 \citep{Addison+2018}, 159 \citep{Luque+2020}, 160 \citep{Ehrenreich2020}, 161 \citep{Triaud+2015}, 162 \citep{Anderson+2015}, 163 \citep{Brown+2014}, 164 \citep{Mocnik+2016}, 165 \citep{Anderson+2014}, 166 \citep{Neveu-VanMalle+2014}, 167 \citep{Piaulet+2021}, 168 \citep{Anderson+2017}, 169 \citep{Carone+2021}, 170 \citep{Bourrier+2020}, 171 \citep{Allart+2020}, 172 \citep{Anderson+2018}, 173 \citep{Wang+2021_wasp148}, 174 \citep{Hellier+2019}, 175 \citep{Temple+2017}, 176 \citep{Temple+2018}, 177 \citep{Temple+2019}, 178 \citep{Anderson+2018_wasp189}, 179 \citep{Temple+2019b}, 180 \citep{Damasso+2015b}, 181 \citep{Hirano+2011c}, 182 \citep{McCullough+2008}, 183 \citep{Narita+2010b}, 184 \citep{Crouzet+2017}, 185 \citep{Huang+2018}, 186 \citep{Kunovac-Hodzic+2021}
}
\startlongtable
\begin{deluxetable*}{p{0.12\textwidth} DDD D DDD p{0.03\textwidth}}
\tablecaption{\label{tab:planets} Key properties of planets for which $\lambda$, $i$, or both angles
were determined. In the reference column, the boldface number refers to the work from
which we drew the $\lambda$ measurement. The other numbers refer to
works reporting additional measurements of $\lambda$ or other system parameters not taken from \href{https://www.astro.keele.ac.uk/jkt/tepcat/}{TEPCat}  \citep{Southworth2011}. 
Stellar parameters are given in Table~\ref{tab:stars}. For a few systems, obliquities have been measured with respect to more
than one planetary orbit, but in this table we display only a single measurement. The multi-planetary systems are separately listed in Table \S~\ref{tab:multitransits}.}
\tablewidth{0pt}
\tablehead{
\colhead{Planet} 
& \multicolumn2c{$a/R$} & \multicolumn2c{m} & \multicolumn2c{r}
& \multicolumn2c{$e$} 
& \multicolumn2c{$\lambda$} & \multicolumn2c{$i$} & \multicolumn2c{$\psi$} 
& \colhead{References} \\
\colhead{ }   
& \multicolumn2c{ } & \multicolumn2c{(M$_{\rm Jupiter}$)} & \multicolumn2c{(R$_{\rm Jupiter}$)} 
& \multicolumn2c{ }  
& \multicolumn2c{($^{\circ})$}& \multicolumn2c{($^{\circ})$} & \multicolumn2c{($^{\circ})$}
& \colhead{} \\
}
\decimalcolnumbers
\startdata
AU Micb & 18.92^{+2.15}_{-2.42} & 0.05\pm0.01 & 0.38\pm0.02 & 0.100^{+0.170}_{-0.090} & 4.7^{+6.4}_{-6.8} & 90.0^{+0.0}_{-19.5} & 12.1^{+11.3}_{-7.5} &  {\bf 1},2,3,4,5 \\ 
Beta PICb & 1141.92\pm3.47 & 11.90^{+2.93}_{-3.04} & 1.46\pm0.01 & 0.103\pm0.030 & 3.0\pm5.0 & \multicolumn2c{$-$} & \multicolumn2c{$-$} &  6,7,{\bf 8} \\ 
CoRoT-2b & 6.77\pm0.18 & 3.57\pm0.15 & 1.46\pm0.03 & 0.000^{+0.024}_{-0.000} & 1.0^{+7.7}_{-6.0} & 90.0^{+0.0}_{-10.8} & 8.9^{+6.7}_{-5.1} &  9,{\bf 10},2,11,12,13 \\ 
CoRoT-3b & 7.90^{+0.49}_{-0.48} & 21.96\pm0.70 & 1.04\pm0.07 & 0.000^{+0.016}_{-0.000} & 37.6^{+10.0}_{-22.3} & \multicolumn2c{$-$} & \multicolumn2c{$-$} &  9,{\bf 14} \\ 
CoRoT-11b & 6.89\pm0.40 & 2.34\pm0.39 & 1.43\pm0.06 & 0.000^{+0.360}_{-0.000} & 0.1\pm2.6 & \multicolumn2c{$-$} & \multicolumn2c{$-$} &  9,{\bf 15} \\ 
CoRoT-18b & 7.01^{+0.28}_{-0.31} & 3.30\pm0.19 & 1.15^{+0.04}_{-0.05} & 0.000^{+0.025}_{-0.000} & 10.0\pm20.0 & 90.0^{+0.0}_{-23.2} & 25.4^{+13.6}_{-13.1} &  9,{\bf 16},2 \\ 
DS Tucb & 16.08^{+0.32}_{-0.28} & 0.09^{+0.11}_{-0.04} & 0.50\pm0.02 & \multicolumn2c{$0$} & 2.9^{+0.9}_{-0.9} & 90.0^{+0.0}_{-7.3} & 4.0^{+4.6}_{-1.6} &  {\bf 17},18,2,19 \\ 
EPIC 246851721b & 9.59^{+0.23}_{-0.23} & 3.00^{+1.10}_{-1.20} & 1.00^{+0.05}_{-0.05} & \multicolumn2c{$0$} & 1.5\pm0.9 & 90.0^{+0.0}_{-12.4} & 2.9^{+10.8}_{-1.6} &  {\bf 20},2 \\ 
GJ 3470 b & 13.84^{+0.51}_{-0.46} & 0.04\pm0.01 & 0.35\pm0.03 & 0.126^{+0.042}_{-0.041} & 101.0^{+29.0}_{-14.0} & 51.0^{+25.0}_{-21.0} & 97.0^{+16.0}_{-11.0} &  {\bf 21},22 \\ 
GJ 436b & 14.56^{+0.84}_{-0.81} & 0.08^{+0.01}_{-0.01} & 0.37\pm0.01 & 0.162^{+0.004}_{-0.003} & 72.0^{+33.0}_{-24.0} & 37.3^{+16.2}_{-12.5} & 80.0^{+21.0}_{-18.0} &  {\bf 23} \\ 
HAT-P-1b & 10.19^{+0.27}_{-0.28} & 0.53\pm0.02 & 1.32\pm0.02 & 0.000^{+0.011}_{-0.000} & 3.7\pm2.1 & \multicolumn2c{$-$} & \multicolumn2c{$-$} &  9,{\bf 24} \\ 
HAT-P-2b & 8.63\pm0.78 & 8.74\pm0.27 & 1.19\pm0.12 & 0.508^{+0.001}_{-0.001} & 0.2^{+12.2}_{-12.5} & \multicolumn2c{$-$} & \multicolumn2c{$-$} &  9,{\bf 25},26,27 \\ 
HAT-P-3b & 9.81\pm0.31 & 0.59\pm0.02 & 0.91\pm0.03 & 0.000^{+0.010}_{-0.000} & 21.2\pm8.7 & \multicolumn2c{$-$} & \multicolumn2c{$-$} &  9,{\bf 28} \\ 
HAT-P-4b & 6.00^{+0.47}_{-0.19} & 0.68^{+0.05}_{-0.03} & 1.34^{+0.08}_{-0.04} & 0.000^{+0.007}_{-0.000} & 4.9\pm11.9 & \multicolumn2c{$-$} & \multicolumn2c{$-$} &  9,{\bf 29} \\ 
HAT-P-6b & 7.43\pm0.37 & 1.06\pm0.06 & 1.40\pm0.08 & 0.000^{+0.044}_{-0.000} & 165.0\pm6.0 & \multicolumn2c{$-$} & \multicolumn2c{$-$} &  9,{\bf 27},30 \\ 
HAT-P-7b & 4.05\pm0.04 & 1.87\pm0.03 & 1.53\pm0.01 & 0.000^{+0.004}_{-0.000} & 142.0^{+12.0}_{-16.0} & 0.0^{+36.0}_{-0.0} & 97.0\pm14.0 &  9,{\bf 31},32,33,27,34 \\ 
HAT-P-8b & 6.40\pm0.20 & 1.27\pm0.05 & 1.32\pm0.04 & 0.000^{+0.006}_{-0.000} & 17.0^{+11.5}_{-9.2} & \multicolumn2c{$-$} & \multicolumn2c{$-$} &  9,{\bf 35},36 \\ 
HAT-P-9b & 8.50\pm0.56 & 0.78\pm0.08 & 1.38\pm0.10 & 0.000^{+0.160}_{-0.000} & 16.0\pm8.0 & \multicolumn2c{$-$} & \multicolumn2c{$-$} &  9,{\bf 35} \\ 
HAT-P-11b & 16.56\pm0.29 & 0.09\pm0.01 & 0.39\pm0.01 & 0.264\pm0.001 & 103.0^{+26.0}_{-10.0} & 67.0^{+2.0}_{-4.0} & 97.0^{+8.0}_{-4.0} &  {\bf 37},38,39,40,41 \\ 
HAT-P-12b & 11.93\pm0.30 & 0.20\pm0.01 & 0.92\pm0.02 & 0.000^{+0.035}_{-0.000} & 54.0^{+13.0}_{-41.0} & \multicolumn2c{$-$} & \multicolumn2c{$-$} &  9,{\bf 28} \\ 
HAT-P-13b & 5.37\pm0.16 & 0.91\pm0.03 & 1.49\pm0.04 & 0.009^{+0.004}_{-0.002} & 1.9\pm8.6 & \multicolumn2c{$-$} & \multicolumn2c{$-$} &  42,{\bf 43} \\ 
HAT-P-14b & 8.26\pm0.31 & 2.27\pm0.08 & 1.22\pm0.06 & 0.107^{+0.008}_{-0.008} & 170.9\pm5.1 & \multicolumn2c{$-$} & \multicolumn2c{$-$} &  9,{\bf 29} \\ 
HAT-P-16b & 7.67\pm0.20 & 4.19\pm0.13 & 1.19\pm0.04 & 0.046^{+0.003}_{-0.002} & 10.0\pm16.0 & \multicolumn2c{$-$} & \multicolumn2c{$-$} &  9,{\bf 35},27 \\ 
HAT-P-17b & 22.63\pm0.67 & 0.53\pm0.02 & 1.01\pm0.03 & 0.342\pm0.004 & 19.0^{+14.0}_{-16.0} & \multicolumn2c{$-$} & \multicolumn2c{$-$} &  9,{\bf 44} \\ 
HAT-P-18b & 16.77\pm0.64 & 0.20\pm0.01 & 0.95\pm0.04 & 0.000^{+0.087}_{-0.000} & 132.0\pm15.0 & \multicolumn2c{$-$} & \multicolumn2c{$-$} &  9,{\bf 45} \\ 
HAT-P-20b & 11.36\pm0.14 & 7.22\pm0.36 & 1.02\pm0.05 & 0.016^{+0.002}_{-0.002} & 8.0\pm6.9 & 52.3^{+16.0}_{-11.0} & 36.9^{+12.4}_{-17.6} &  9,{\bf 46},2 \\ 
HAT-P-22b & 8.45\pm0.40 & 2.19\pm0.08 & 1.06\pm0.07 & 0.000^{+0.002}_{-0.000} & 2.1\pm3.0 & 64.4^{+20.8}_{-10.1} & 6.7^{+29.4}_{-3.8} &  9,{\bf 28},2 \\ 
HAT-P-23b & 4.55\pm0.13 & 2.07\pm0.12 & 1.22\pm0.04 & 0.000^{+0.052}_{-0.000} & 15.0\pm22.0 & \multicolumn2c{$-$} & \multicolumn2c{$-$} &  9,{\bf 35} \\ 
HAT-P-24b & 7.71^{+0.43}_{-0.38} & 0.68^{+0.03}_{-0.03} & 1.24^{+0.07}_{-0.06} & 0.000^{+0.038}_{-0.000} & 20.0\pm16.0 & \multicolumn2c{$-$} & \multicolumn2c{$-$} &  9,{\bf 27} \\ 
HAT-P-30b & 7.42\pm0.32 & 0.71\pm0.03 & 1.34\pm0.07 & 0.000^{+0.016}_{-0.000} & 73.5\pm9.0 & \multicolumn2c{$-$} & \multicolumn2c{$-$} &  9,{\bf 47} \\ 
HAT-P-32b & 6.05\pm0.12 & 0.80\pm0.14 & 1.81\pm0.03 & 0.000^{+0.044}_{-0.000} & 85.0\pm1.5 & \multicolumn2c{$-$} & \multicolumn2c{$-$} &  9,{\bf 27} \\ 
HAT-P-34b & 9.48^{+0.84}_{-0.64} & 3.33\pm0.21 & 1.20^{+0.13}_{-0.09} & 0.432^{+0.029}_{-0.027} & 0.0\pm14.0 & \multicolumn2c{$-$} & \multicolumn2c{$-$} &  9,{\bf 27} \\ 
HAT-P-36b & 4.93\pm0.10 & 1.85\pm0.10 & 1.30\pm0.03 & 0.000^{+0.059}_{-0.000} & 14.0\pm18.0 & 90.0^{+0.0}_{-28.3} & 28.6^{+16.0}_{-14.6} &  9,{\bf 48},2 \\ 
HAT-P-41b & 5.44^{+0.20}_{-0.13} & 0.80\pm0.10 & 1.69^{+0.08}_{-0.05} & 0.000^{+0.220}_{-0.000} & 22.1^{+6.0}_{-0.8} & \multicolumn2c{$-$} & \multicolumn2c{$-$} &  9,{\bf 49} \\ 
HAT-P-56b & 6.37\pm0.15 & 2.18\pm0.25 & 1.47\pm0.04 & 0.000^{+0.290}_{-0.000} & 8.0\pm2.0 & \multicolumn2c{$-$} & \multicolumn2c{$-$} &  9,{\bf 50} \\ 
HAT-P-69b & 7.69^{+0.18}_{-0.10} & 3.73^{+0.61}_{-0.59} & 1.63^{+0.03}_{-0.03} & \multicolumn2c{$0$} & 30.3^{+6.1}_{-7.3} & \multicolumn2c{$-$} & \multicolumn2c{$-$} &  {\bf 51} \\ 
HAT-P-70b & 5.48^{+0.35}_{-0.30} & 0.00^{+6.78}_{-0.00} & 1.87^{+0.15}_{-0.10} & \multicolumn2c{$0$} & 113.1^{+5.1}_{-3.4} & \multicolumn2c{$-$} & \multicolumn2c{$-$} &  {\bf 51} \\ 
HATS-2b & 5.51\pm0.14 & 1.34\pm0.15 & 1.17\pm0.03 & 0.000^{+0.290}_{-0.000} & 8.0\pm8.0 & 64.7^{+25.3}_{-12.3} & 20.6^{+23.6}_{-10.4} &  9,{\bf 52},2 \\ 
HATS-3b & 7.43^{+0.17}_{-0.18} & 1.07\pm0.14 & 1.38\pm0.04 & 0.000^{+0.300}_{-0.000} & 3.0\pm25.0 & \multicolumn2c{$-$} & \multicolumn2c{$-$} &  9,{\bf 53} \\ 
HATS-70b & 4.15^{+0.16}_{-0.18} & 12.90^{+1.80}_{-1.60} & 1.38^{+0.08}_{-0.07} & 0.000^{+0.180}_{-0.000} & 8.9^{+5.6}_{-4.5} & \multicolumn2c{$-$} & \multicolumn2c{$-$} &  {\bf 54},55 \\ 
HD 332231b & 24.50^{+0.30}_{-0.20} & 0.24\pm0.02 & 0.86\pm0.02 & 0.000^{+0.050}_{-0.000} & 2.0\pm6.0 & \multicolumn2c{$-$} & \multicolumn2c{$-$} &  56,{\bf 57},58 \\ 
HD 3167c & 43.86^{+0.82}_{-0.86} & 0.03\pm0.00 & 0.27^{+0.04}_{-0.03} & 0.000^{+0.267}_{-0.000} & 98.0\pm23.0 & \multicolumn2c{$-$} & \multicolumn2c{$-$} &  {\bf 59} \\ 
HD 17156b & 23.67\pm0.79 & 3.26\pm0.11 & 1.06\pm0.04 & 0.670^{+0.001}_{-0.001} & 10.0\pm5.1 & \multicolumn2c{$-$} & \multicolumn2c{$-$} &  9,{\bf 60},61,62,63 \\ 
HD 63433c & 38.00^{+4.60}_{-1.70} & 0.02\pm0.02 & 2.71\pm0.14 & 0.000\pm0.000 & 11.0^{+32.0}_{-35.0} & 90.0^{+0.0}_{-18.7} & 25.6^{+22.5}_{-15.3} &  {\bf 64},65,2 \\ 
HD 80606b & 94.64\pm3.66 & 4.11\pm0.15 & 1.00\pm0.03 & 0.932^{+0.001}_{-0.001} & 42.0\pm8.0 & \multicolumn2c{$-$} & \multicolumn2c{$-$} &  9,{\bf 66},67,68,69 \\ 
HD 106315c & 24.79^{+0.39}_{-0.43} & 0.04\pm0.01 & 0.39\pm0.01 & 0.220\pm0.150 & 10.9^{+3.8}_{-3.6} & \multicolumn2c{$-$} & \multicolumn2c{$-$} &  {\bf 70} \\ 
HD 149026b & 6.02^{+0.20}_{-0.18} & 0.37^{+0.01}_{-0.01} & 0.81^{+0.03}_{-0.03} & 0.000^{+0.013}_{-0.000} & 12.0\pm7.0 & \multicolumn2c{$-$} & \multicolumn2c{$-$} &  9,{\bf 27},71 \\ 
HD 189733b & 8.98\pm0.33 & 1.15\pm0.04 & 1.15\pm0.04 & 0.000^{+0.004}_{-0.000} & 0.4\pm0.2 & 90.0^{+0.0}_{-14.9} & 2.3^{+13.5}_{-1.6} &  9,{\bf 72},2,73,14,74 \\ 
HD 209458b & 8.78\pm0.15 & 0.71\pm0.02 & 1.38\pm0.02 & 0.000^{+0.008}_{-0.000} & 0.6\pm0.4 & 61.0^{+13.3}_{-8.8} & 28.2^{+9.7}_{-13.5} &  9,{\bf 75},2,76,27,77 \\ 
HIP 67522b & 11.71^{+0.09}_{-0.16} & 0.00^{+5.00}_{-0.00} & 0.93\pm0.04 & 0.000^{+0.300}_{-0.000} & 5.8^{+2.8}_{-5.7} & 90.0^{+0.0}_{-5.0} & 20.2^{+8.7}_{-10.3} &  {\bf 78},79 \\ 
K2-25b & 21.05\pm1.10 & 0.08^{+0.02}_{-0.02} & 0.31\pm0.01 & 0.428^{+0.050}_{-0.049} & 3.0\pm16.0 & 90.0^{+0.0}_{-11.8} & 13.8^{+10.1}_{-7.8} &  {\bf 80},2,81 \\ 
K2-29b & 10.54\pm0.14 & 0.73\pm0.04 & 1.19\pm0.02 & 0.066^{+0.022}_{-0.066} & 1.5\pm8.7 & 68.4^{+18.2}_{-8.9} & 19.3^{+13.7}_{-11.1} &  {\bf 82},2 \\ 
K2-34b & 6.73\pm0.23 & 1.65\pm0.10 & 1.22\pm0.05 & 0.000^{+0.022}_{-1.000} & 1.0^{+9.0}_{-10.0} & \multicolumn2c{$-$} & \multicolumn2c{$-$} &  {\bf 83} \\ 
K2-140b & 12.88^{+0.00}_{-0.00} & 1.13^{+0.12}_{-0.11} & 1.20\pm0.08 & 0.069^{+0.042}_{-0.028} & 0.5\pm9.7 & \multicolumn2c{$-$} & \multicolumn2c{$-$} &  {\bf 84},85 \\ 
K2-232b & 19.25^{+0.27}_{-0.31} & 0.38^{+0.05}_{-0.05} & 1.10\pm0.02 & 0.247^{+0.020}_{-0.021} & 11.1\pm6.6 & \multicolumn2c{$-$} & \multicolumn2c{$-$} &  86,{\bf 87} \\ 
K2-290c & 43.50\pm1.20 & 1.01\pm0.05 & 0.77\pm0.05 & 0.000^{+0.241}_{-0.000} & 153.0\pm8.0 & 39.0\pm7.0 & 124.0\pm6.0 &  88,{\bf 89} \\ 
KELT-1b & 3.61^{+0.12}_{-0.10} & 27.38\pm0.93 & 1.12^{+0.04}_{-0.03} & 0.010^{+0.010}_{-0.007} & 2.0\pm16.0 & \multicolumn2c{$-$} & \multicolumn2c{$-$} &  {\bf 90} \\ 
KELT-6b & 11.17^{+1.05}_{-1.01} & 0.43^{+0.04}_{-0.05} & 1.19^{+0.13}_{-0.08} & 0.000^{+0.036}_{-0.000} & 36.0\pm11.0 & \multicolumn2c{$-$} & \multicolumn2c{$-$} &  9,{\bf 91} \\ 
KELT-7b & 5.48^{+0.16}_{-0.16} & 1.28\pm0.18 & 1.53^{+0.05}_{-0.05} & \multicolumn2c{$0$} & 2.7\pm0.6 & \multicolumn2c{$-$} & \multicolumn2c{$-$} &  {\bf 50},92 \\ 
KELT-9b & 3.00\pm0.10 & 2.88\pm0.35 & 1.94\pm0.05 & 0.000^{+0.035}_{-0.000} & 85.0\pm0.2 & 52.0^{+8.0}_{-7.0} & 87.0^{+10.0}_{-11.0} &  {\bf 93},94,95,96 \\ 
KELT-17b & 6.38^{+0.25}_{-0.23} & 1.31^{+0.28}_{-0.29} & 1.52^{+0.07}_{-0.06} & 0.000^{+0.990}_{-0.000} & 115.9\pm4.1 & 90.0^{+0.0}_{-21.3} & 114.4^{+4.3}_{-4.6} &  {\bf 97},2 \\ 
KELT-19b & 7.49^{+0.54}_{-0.51} & 0.00^{+4.07}_{-0.00} & 1.91\pm0.11 & \multicolumn2c{$0$} & 179.7^{+3.8}_{-3.7} & \multicolumn2c{$-$} & \multicolumn2c{$-$} &  {\bf 98} \\ 
KELT-21b & 6.86^{+0.15}_{-0.15} & 0.00^{+3.90}_{-0.00} & 1.59^{+0.04}_{-0.04} & \multicolumn2c{$0$} & 5.6^{+1.9}_{-1.7} & \multicolumn2c{$-$} & \multicolumn2c{$-$} &  {\bf 99} \\ 
KELT-25b & 6.46^{+0.20}_{-0.15} & 0.00^{+64.00}_{-0.00} & 1.64^{+0.04}_{-0.04} & 0.000^{+1.000}_{-0.000} & 23.4^{+3.2}_{-2.3} & \multicolumn2c{$-$} & \multicolumn2c{$-$} &  {\bf 100} \\ 
KELT-26b & 6.49\pm0.18 & 1.41^{+0.43}_{-0.51} & 1.94^{+0.06}_{-0.06} & 0.000^{+1.000}_{-0.000} & 91.3^{+6.5}_{-6.3} & \multicolumn2c{$-$} & \multicolumn2c{$-$} &  {\bf 100} \\ 
Kepler-8b & 6.98\pm0.24 & 0.59\pm0.12 & 1.38\pm0.04 & \multicolumn2c{$0$} & 5.0\pm7.0 & 58.0^{+15.5}_{-10.1} & 31.1^{+10.9}_{-15.7} &  {\bf 27},2,101 \\ 
Kepler-9b & 32.05\pm0.74 & 0.14^{+0.01}_{-0.01} & 0.74\pm0.01 & 0.064\pm0.000 & 13.0\pm16.0 & 71.9^{+18.1}_{-8.8} & 28.1^{+13.0}_{-13.6} &  {\bf 102},2 \\ 
Kepler-13b & 6.15^{+0.16}_{-0.21} & 6.00\pm3.00 & 1.41\pm0.04 & \multicolumn2c{$0$} & 59.2\pm0.1 & 81.1\pm0.0 & 60.2\pm0.1 &  {\bf 103},104,105,106 \\ 
Kepler-17b & 5.70^{+0.14}_{-0.41} & 2.34^{+0.09}_{-0.24} & 1.31^{+0.02}_{-0.07} & \multicolumn2c{$0$} & 0.0\pm15.0 & 90.0^{+0.0}_{-24.5} & 19.7^{+14.4}_{-10.9} &  {\bf 107},2 \\ 
Kepler-25c & 18.62\pm0.24 & 0.05^{+0.00}_{-0.01} & 0.47^{+0.01}_{-0.01} & 0.061^{+0.005}_{-0.004} & 0.5\pm5.7 & 90.0^{+0.0}_{-21.3} & 5.7^{+4.2}_{-3.2} &  {\bf 108},109,2,110 \\ 
Kepler-30c & 67.86\pm8.57 & 1.69\pm0.02 & 1.07\pm0.03 & 0.011\pm0.001 & 4.0\pm10.0 & \multicolumn2c{$-$} & \multicolumn2c{$-$} &  {\bf 111} \\ 
Kepler-50b & 10.48^{+1.64}_{-2.73} & 0.00^{+0.02}_{-0.00} & 0.15^{+0.00}_{-0.01} & 0.000^{+0.100}_{-0.000} & \multicolumn2c{$-$} & 82.0^{+8.0}_{-7.0} & \multicolumn2c{$-$} &  112 \\ 
Kepler-56b & 5.23\pm0.26 & 0.07^{+0.01}_{-0.01} & 0.58\pm0.03 & \multicolumn2c{$0$} & \multicolumn2c{$-$} & 47.0\pm6.0 & \multicolumn2c{$-$} &  113 \\ 
Kepler-63b & 19.09^{+0.75}_{-0.67} & 0.00^{+0.38}_{-0.00} & 0.54\pm0.02 & 0.000^{+0.450}_{-0.000} & 110.0^{+14.0}_{-22.0} & 138.0\pm7.0 & 104.0^{+9.0}_{-14.0} &  {\bf 114} \\ 
Kepler-65b & 5.24^{+0.32}_{-0.18} & 0.01^{+0.01}_{-0.01} & 0.13^{+0.00}_{-0.00} & 0.028^{+0.031}_{-0.020} & \multicolumn2c{$-$} & 81.0^{+9.0}_{-16.0} & \multicolumn2c{$-$} &  112 \\ 
Kepler-89d & 23.82\pm2.22 & 0.33\pm0.04 & 1.00\pm0.10 & 0.022^{+0.038}_{-0.033} & 6.0^{+11.0}_{-13.0} & \multicolumn2c{$-$} & \multicolumn2c{$-$} &  {\bf 115},108 \\ 
Kepler-408b & 6.23\pm0.00 & 0.00^{+0.01}_{-0.00} & 0.08\pm0.00 & \multicolumn2c{$0$} & \multicolumn2c{$-$} & 42.0^{+5.0}_{-4.5} & -99.0\pm-99.0 &  116,117 \\ 
Kepler-410b & 19.50\pm0.76 & 0.00\pm-1.00 & 0.25\pm0.00 & 0.170^{+0.070}_{-0.060} & \multicolumn2c{$-$} & 82.5^{+7.5}_{-2.5} & \multicolumn2c{$-$} &  118 \\ 
Kepler-420b & 72.70\pm9.08 & 1.45\pm0.35 & 0.94\pm0.12 & 0.772\pm0.045 & 74.0^{+32.0}_{-46.0} & \multicolumn2c{$-$} & \multicolumn2c{$-$} &  {\bf 119} \\ 
Kepler-432b & 15.94^{+0.97}_{-0.66} & 5.41^{+0.32}_{-0.18} & 1.15^{+0.04}_{-0.04} & 0.477\pm0.007 & \multicolumn2c{$-$} & 90.0^{+0.0}_{-8.0} & \multicolumn2c{$-$} &  9,120 \\ 
Kepler-448b & 19.92\pm1.88 & 5.00\pm5.00 & 1.44\pm0.13 & 0.340^{+0.080}_{-0.070} & 7.1^{+2.8}_{-4.2} & 90.0^{+0.0}_{-16.8} & 10.0^{+10.4}_{-4.5} &  121,{\bf 49},2,122 \\ 
MASCARA-1b & 4.40\pm0.66 & 3.70\pm0.90 & 1.50\pm0.30 & \multicolumn2c{$0$} & 69.5\pm3.0 & 55.5^{+2.3}_{-2.9} & 72.1^{+2.5}_{-2.4} &  {\bf 123},124 \\ 
MASCARA-2b & 7.47^{+0.34}_{-0.42} & 0.00^{+3.50}_{-0.00} & 1.74^{+0.07}_{-0.07} & \multicolumn2c{$0$} & 3.4\pm2.1 & \multicolumn2c{$-$} & \multicolumn2c{$-$} &  {\bf 125},126,96 \\ 
MASCARA-3b & 9.95^{+0.19}_{-0.20} & 5.18^{+0.22}_{-0.21} & 1.27^{+0.02}_{-0.02} & 0.077^{+0.024}_{-0.026} & 2.6^{+5.1}_{-3.6} & \multicolumn2c{$-$} & \multicolumn2c{$-$} &  {\bf 127},128 \\ 
MASCARA-4b & 5.26\pm0.54 & 3.10\pm0.90 & 1.53^{+0.07}_{-0.04} & \multicolumn2c{$0$} & 112.5^{+1.7}_{-1.5} & 47.4^{+22.9}_{-13.6} & 104.0^{+7.0}_{-13.0} &  {\bf 129},130 \\ 
NGTS-2b & 7.75\pm0.45 & 0.67\pm0.09 & 1.54\pm0.06 & \multicolumn2c{$0$} & 11.3\pm4.8 & \multicolumn2c{$-$} & \multicolumn2c{$-$} &  {\bf 131} \\ 
Qatar-1b & 6.25^{+0.16}_{-0.16} & 1.29^{+0.05}_{-0.05} & 1.14^{+0.03}_{-0.03} & 0.000^{+0.012}_{-0.000} & 8.4\pm7.1 & 90.0^{+0.0}_{-26.1} & 16.9^{+15.6}_{-9.0} &  9,{\bf 132},2 \\ 
Qatar-2b & 6.53\pm0.10 & 2.47\pm0.06 & 1.11\pm0.01 & 0.000^{+0.011}_{-0.000} & 0.0\pm8.0 & 90.0^{+0.0}_{-19.5} & 11.4^{+12.3}_{-7.1} &  9,{\bf 133},2,130,46 \\ 
TOI-1268b & 17.04^{+0.26}_{-0.39} & 0.29\pm0.04 & 0.75\pm0.02 & 0.130^{+0.270}_{-0.130} & 24.9^{+13.0}_{-13.1} & \multicolumn2c{$-$} & 30.4\pm11.1 &  {\bf 134},135 \\ 
TOI-1333b & 6.98^{+0.24}_{-0.23} & 2.37\pm0.24 & 1.40^{+0.06}_{-0.05} & 0.073^{+0.092}_{-0.052} & \multicolumn2c{$-$} & 51.3^{+3.5}_{-3.3} & \multicolumn2c{$-$} &  136 \\ 
TOI-1518b & 4.29^{+0.06}_{-0.06} & 0.00^{+2.30}_{-0.00} & 1.20^{+1.88}_{-0.05} & 0.000\pm0.010 & 119.7^{+1.0}_{-0.9} & \multicolumn2c{$-$} & \multicolumn2c{$-$} &  {\bf 137} \\ 
TOI-2109b & 2.27\pm0.02 & 5.02\pm0.75 & 1.35\pm0.05 & 0.000^{+0.035}_{-0.000} & 1.7\pm1.7 & \multicolumn2c{$-$} & \multicolumn2c{$-$} &  {\bf 138} \\ 
TOI-251b & 14.02^{+0.82}_{-0.61} & 0.00^{+1.00}_{-0.00} & 0.24\pm0.02 & 0.000\pm0.000 & \multicolumn2c{$-$} & 78.0^{+7.0}_{-14.0} & \multicolumn2c{$-$} &  139 \\ 
TOI-451b & 6.93^{+0.11}_{-0.16} & 0.02^{+0.02}_{-0.02} & 0.17\pm0.01 & 0.000\pm0.000 & \multicolumn2c{$-$} & 69.0^{+11.0}_{-8.0} & \multicolumn2c{$-$} &  140 \\ 
TOI-811b & 31.60^{+1.30}_{-1.20} & 59.90^{+13.00}_{-8.60} & 1.26\pm0.06 & 0.509\pm0.075 & \multicolumn2c{$-$} & 19.6^{+3.6}_{-1.8} & \multicolumn2c{$-$} &  141 \\ 
TOI-852b & 7.92^{+0.24}_{-0.23} & 53.70^{+1.40}_{-1.30} & 0.83^{+0.04}_{-0.04} & 0.004^{+0.004}_{-0.003} & \multicolumn2c{$-$} & 73.1^{+11.9}_{-9.9} & \multicolumn2c{$-$} &  141 \\ 
TOI-942b & 10.12^{+0.13}_{-0.18} & 0.00^{+2.60}_{-0.00} & 0.43^{+0.02}_{-0.00} & 0.000\pm0.000 & 1.0^{+41.0}_{-33.0} & 76.0^{+9.0}_{-11.0} & 2.0^{+27.0}_{-33.0} &  139,{\bf 142} \\ 
TRAPPIST-1b & 20.04^{+0.72}_{-0.69} & 0.00^{+0.00}_{-0.00} & 0.10^{+0.00}_{-0.00} & 0.000^{+0.081}_{-0.000} & 15.0^{+26.0}_{-30.0} & 90.0^{+0.0}_{-17.4} & 23.3^{+17.0}_{-13.6} &  {\bf 143},2 \\ 
TrES-1b & 10.37\pm0.33 & 0.76\pm0.05 & 1.10\pm0.04 & 0.000^{+0.012}_{-0.000} & 30.0\pm21.0 & \multicolumn2c{$-$} & \multicolumn2c{$-$} &  9,{\bf 144} \\ 
TrES-2b & 7.96\pm0.21 & 1.21\pm0.05 & 1.19\pm0.02 & 0.000^{+0.003}_{-0.000} & 9.0\pm12.0 & \multicolumn2c{$-$} & \multicolumn2c{$-$} &  9,{\bf 145} \\ 
TrES-4b & 6.13\pm0.28 & 0.49\pm0.04 & 1.84^{+0.08}_{-0.09} & 0.000^{+0.015}_{-0.000} & 6.3\pm4.7 & \multicolumn2c{$-$} & \multicolumn2c{$-$} &  9,{\bf 146} \\ 
V1298 Taub & 26.06\pm0.46 & 0.00^{+0.38}_{-0.00} & 0.92^{+0.05}_{-0.05} & 0.000\pm0.300 & 4.0^{+7.0}_{-10.0} & 51.0^{+25.0}_{-21.0} & 8.0^{+4.0}_{-7.0} &  {\bf 147},148,22,149 \\ 
WASP-3b & 5.05^{+0.25}_{-0.20} & 1.77^{+0.11}_{-0.09} & 1.35\pm0.06 & 0.000^{+0.006}_{-0.000} & 5.0^{+6.0}_{-5.0} & \multicolumn2c{$-$} & \multicolumn2c{$-$} &  9,{\bf 150},151,152,153 \\ 
WASP-4b & 5.48\pm0.15 & 1.25\pm0.05 & 1.36\pm0.03 & 0.000^{+0.003}_{-0.000} & 1.0^{+12.0}_{-14.0} & 90.0^{+0.0}_{-27.3} & 20.0^{+15.2}_{-11.1} &  9,{\bf 40},2,154 \\ 
WASP-5b & 5.42\pm0.22 & 1.59\pm0.05 & 1.18\pm0.06 & 0.000^{+0.012}_{-0.000} & 12.1^{+8.0}_{-10.0} & 71.5^{+17.0}_{-7.5} & 22.2^{+11.5}_{-10.3} &  9,{\bf 154},2 \\ 
WASP-6b & 10.30\pm0.40 & 0.48\pm0.03 & 1.23\pm0.04 & 0.000^{+0.070}_{-0.000} & 7.2\pm3.7 & 62.3^{+19.5}_{-10.4} & 13.1^{+22.9}_{-3.8} &  9,{\bf 155},2,156 \\ 
WASP-7b & 9.08\pm0.56 & 0.98\pm0.13 & 1.37\pm0.09 & 0.000^{+0.049}_{-0.000} & 86.0\pm6.0 & 51.4^{+26.1}_{-14.0} & 87.1^{+5.1}_{-5.3} &  9,{\bf 157},2 \\ 
WASP-8b & 18.00\pm0.43 & 2.22\pm0.05 & 1.17\pm0.03 & 0.306\pm0.005 & 143.0^{+1.5}_{-1.6} & 36.3^{+4.0}_{-3.6} & 118.2^{+3.2}_{-3.0} &  9,{\bf 158},2,159 \\ 
WASP-11b & 12.19\pm0.31 & 0.49\pm0.02 & 0.99\pm0.02 & 0.000^{+0.030}_{-0.000} & 7.0\pm5.0 & \multicolumn2c{$-$} & \multicolumn2c{$-$} &  9,{\bf 48} \\ 
WASP-12b & 3.04^{+0.11}_{-0.10} & 1.47^{+0.08}_{-0.07} & 1.90^{+0.06}_{-0.04} & 0.000^{+0.020}_{-0.000} & 59.0^{+15.0}_{-20.0} & 8.3^{+3.8}_{-3.1} & 85.5^{+6.8}_{-7.8} &  9,{\bf 27},2 \\ 
WASP-13b & 7.23\pm0.42 & 0.51\pm0.06 & 1.53\pm0.08 & 0.000^{+0.016}_{-0.000} & 8.0^{+13.0}_{-12.0} & \multicolumn2c{$-$} & \multicolumn2c{$-$} &  9,{\bf 160} \\ 
WASP-14b & 6.04^{+0.46}_{-0.37} & 7.59^{+0.24}_{-0.23} & 1.24^{+0.12}_{-0.10} & 0.078^{+0.001}_{-0.001} & 33.1\pm7.4 & \multicolumn2c{$-$} & \multicolumn2c{$-$} &  9,{\bf 161},162 \\ 
WASP-15b & 7.30\pm0.23 & 0.59\pm0.02 & 1.41\pm0.05 & 0.000^{+0.055}_{-0.000} & 139.6^{+5.2}_{-4.3} & \multicolumn2c{$-$} & \multicolumn2c{$-$} &  9,{\bf 154} \\ 
WASP-16b & 8.21\pm0.35 & 0.83\pm0.04 & 1.22\pm0.04 & 0.000^{+0.018}_{-0.000} & 4.2^{+13.9}_{-11.0} & \multicolumn2c{$-$} & \multicolumn2c{$-$} &  9,{\bf 163},27 \\ 
WASP-17b & 6.98\pm0.23 & 0.48\pm0.03 & 1.93\pm0.05 & 0.000^{+0.020}_{-0.000} & 148.5^{+5.4}_{-4.2} & \multicolumn2c{$-$} & \multicolumn2c{$-$} &  9,{\bf 154},164,165,166 \\ 
WASP-18b & 3.52\pm0.09 & 10.52\pm0.32 & 1.20\pm0.03 & 0.008\pm0.001 & 13.0\pm7.0 & \multicolumn2c{$-$} & \multicolumn2c{$-$} &  9,{\bf 27} \\ 
WASP-19b & 3.45\pm0.07 & 1.14\pm0.04 & 1.41\pm0.02 & 0.000^{+0.006}_{-0.000} & 1.0\pm1.2 & 90.0^{+0.0}_{-29.2} & 3.7^{+28.0}_{-2.3} &  9,{\bf 167},2,27,168,169 \\ 
WASP-20b & 10.13\pm0.38 & 0.40\pm0.05 & 1.38\pm0.06 & 0.000^{+0.039}_{-0.000} & 12.7\pm4.2 & \multicolumn2c{$-$} & \multicolumn2c{$-$} &  9,{\bf 170} \\ 
WASP-21b & 9.45\pm0.51 & 0.28\pm0.02 & 1.16\pm0.05 & 0.000^{+0.048}_{-0.000} & 8.0^{+26.0}_{-27.0} & \multicolumn2c{$-$} & \multicolumn2c{$-$} &  9,{\bf 171} \\ 
WASP-22b & 8.38\pm0.26 & 0.62^{+0.03}_{-0.02} & 1.20^{+0.05}_{-0.03} & 0.000^{+0.021}_{-0.000} & 22.0\pm16.0 & \multicolumn2c{$-$} & \multicolumn2c{$-$} &  9,{\bf 172} \\ 
WASP-24b & 5.94\pm0.22 & 1.11\pm0.05 & 1.30\pm0.05 & 0.000^{+0.005}_{-0.000} & 4.7\pm4.0 & \multicolumn2c{$-$} & \multicolumn2c{$-$} &  9,{\bf 36} \\ 
WASP-25b & 11.22\pm0.26 & 0.60\pm0.05 & 1.25\pm0.03 & 0.000^{+0.083}_{-0.000} & 14.6\pm6.7 & \multicolumn2c{$-$} & \multicolumn2c{$-$} &  9,{\bf 163} \\ 
WASP-26b & 6.64\pm0.21 & 1.02\pm0.03 & 1.22\pm0.05 & 0.000^{+0.004}_{-0.000} & 34.0^{+26.0}_{-36.0} & \multicolumn2c{$-$} & \multicolumn2c{$-$} &  9,{\bf 27} \\ 
WASP-28b & 8.78\pm0.28 & 0.89\pm0.06 & 1.22\pm0.03 & 0.000^{+0.058}_{-0.000} & 8.0\pm18.0 & \multicolumn2c{$-$} & \multicolumn2c{$-$} &  9,{\bf 170},173 \\ 
WASP-30b & 8.57^{+0.22}_{-0.17} & 62.50\pm1.20 & 0.95^{+0.03}_{-0.02} & 0.000^{+0.004}_{-0.000} & 7.0^{+19.0}_{-27.0} & \multicolumn2c{$-$} & \multicolumn2c{$-$} &  {\bf 174} \\ 
WASP-31b & 8.00\pm0.22 & 0.48\pm0.03 & 1.55\pm0.05 & 0.000^{+0.047}_{-0.000} & 2.8\pm3.1 & \multicolumn2c{$-$} & \multicolumn2c{$-$} &  9,{\bf 163},27 \\ 
WASP-32b & 7.63\pm0.35 & 3.60\pm0.07 & 1.18\pm0.07 & 0.000^{+0.004}_{-0.000} & 10.5^{+6.4}_{-6.5} & 54.5^{+17.9}_{-11.0} & 35.8^{+10.2}_{-17.2} &  9,{\bf 175},2,160 \\ 
WASP-33b & 3.69^{+0.05}_{-0.10} & 2.16\pm0.20 & 1.68^{+0.02}_{-0.03} & \multicolumn2c{$0$} & 112.9^{+0.2}_{-0.2} & 37.5^{+6.8}_{-5.7} & 104.1^{+2.8}_{-2.8} &  {\bf 176},2,95,177,178 \\ 
WASP-38b & 12.15^{+0.30}_{-0.26} & 2.69\pm0.06 & 1.09^{+0.03}_{-0.03} & 0.028^{+0.003}_{-0.003} & 7.5^{+4.7}_{-6.1} & \multicolumn2c{$-$} & \multicolumn2c{$-$} &  9,{\bf 175},36 \\ 
WASP-39b & 11.06\pm0.32 & 0.28\pm0.03 & 1.28\pm0.04 & 0.000^{+0.048}_{-0.000} & 0.0\pm11.0 & \multicolumn2c{$-$} & \multicolumn2c{$-$} &  9,{\bf 28} \\ 
WASP-41b & 9.95\pm0.18 & 0.98\pm0.03 & 1.18\pm0.02 & 0.000^{+0.120}_{-0.000} & 6.0\pm11.0 & 62.0^{+28.0}_{-16.5} & 22.6^{+28.9}_{-11.9} &  9,{\bf 179},2,180 \\ 
WASP-43b & 4.92^{+0.09}_{-0.10} & 2.03^{+0.05}_{-0.05} & 1.04\pm0.02 & 0.000^{+0.006}_{-0.000} & 3.5\pm6.8 & 90.0^{+0.0}_{-27.8} & 12.7^{+19.7}_{-7.7} &  9,{\bf 46},2 \\ 
WASP-47b & 9.67\pm0.15 & 1.14\pm0.02 & 1.12\pm0.01 & 0.028^{+0.004}_{-0.002} & 0.0\pm24.0 & \multicolumn2c{$-$} & \multicolumn2c{$-$} &  {\bf 181} \\ 
WASP-52b & 7.23\pm0.21 & 0.43\pm0.02 & 1.25\pm0.03 & 0.000^{+0.092}_{-0.000} & 1.1\pm1.1 & 90.0^{+0.0}_{-11.1} & 2.8^{+9.8}_{-1.8} &  9,{\bf 171},2,182,183 \\ 
WASP-53b & 11.05^{+0.39}_{-0.40} & 2.13^{+0.09}_{-0.09} & 1.07\pm0.04 & 0.000^{+0.030}_{-0.000} & 1.0\pm12.0 & \multicolumn2c{$-$} & \multicolumn2c{$-$} &  {\bf 184} \\ 
WASP-60b & 8.51\pm0.41 & 0.56\pm0.04 & 1.23\pm0.07 & 0.000^{+0.064}_{-0.000} & 129.0\pm17.0 & \multicolumn2c{$-$} & \multicolumn2c{$-$} &  9,{\bf 28} \\ 
WASP-61b & 8.06\pm0.21 & 2.06\pm0.16 & 1.22\pm0.03 & 0.000^{+0.074}_{-0.000} & 4.0^{+17.1}_{-18.4} & \multicolumn2c{$-$} & \multicolumn2c{$-$} &  9,{\bf 185} \\ 
WASP-62b & 9.53\pm0.39 & 0.57\pm0.04 & 1.39\pm0.06 & 0.006\pm0.001 & 19.4^{+5.1}_{-4.9} & 73.0^{+12.2}_{-6.4} & 25.0^{+6.6}_{-6.2} &  186,{\bf 185},2 \\ 
WASP-66b & 6.71\pm0.36 & 2.32\pm0.13 & 1.39\pm0.09 & 0.000^{+0.046}_{-0.000} & 4.0\pm22.0 & \multicolumn2c{$-$} & \multicolumn2c{$-$} &  9,{\bf 187} \\ 
WASP-69b & 11.97\pm0.44 & 0.26\pm0.02 & 1.06\pm0.05 & 0.000^{+0.110}_{-0.000} & 0.4^{+2.0}_{-1.9} & 90.0^{+0.0}_{-21.5} & 3.8^{+18.5}_{-2.7} &  9,{\bf 188},2 \\ 
WASP-71b & 4.40\pm0.34 & 2.24\pm0.08 & 1.46\pm0.13 & 0.000^{+0.019}_{-0.000} & 1.9^{+7.5}_{-7.1} & \multicolumn2c{$-$} & \multicolumn2c{$-$} &  9,{\bf 185},189 \\ 
WASP-72b & 4.03\pm0.49 & 1.46^{+0.06}_{-0.06} & 1.27\pm0.20 & 0.000^{+0.017}_{-0.000} & 7.0^{+12.0}_{-11.0} & \multicolumn2c{$-$} & \multicolumn2c{$-$} &  9,{\bf 190} \\ 
WASP-74b & 4.82\pm0.09 & 0.83\pm0.02 & 1.40\pm0.02 & 0.000^{+0.030}_{-0.000} & 0.8\pm1.0 & \multicolumn2c{$-$} & \multicolumn2c{$-$} &  9,{\bf 191} \\ 
WASP-76b & 4.02\pm0.16 & 0.89^{+0.02}_{-0.01} & 1.85^{+0.08}_{-0.08} & 0.000^{+0.050}_{-0.000} & 61.3^{+7.6}_{-5.1} & 9.1^{+2.4}_{-2.1} & 85.7^{+2.5}_{-2.4} &  {\bf 192},2,185 \\ 
WASP-78b & 3.36^{+0.16}_{-0.15} & 0.86\pm0.08 & 2.06\pm0.10 & 0.000^{+0.092}_{-0.000} & 6.4\pm5.9 & \multicolumn2c{$-$} & \multicolumn2c{$-$} &  9,{\bf 185} \\ 
WASP-79b & 7.39^{+0.23}_{-0.19} & 0.86\pm0.08 & 1.53\pm0.04 & 0.000^{+0.066}_{-0.000} & 99.1^{+3.9}_{-4.1} & \multicolumn2c{$-$} & \multicolumn2c{$-$} &  9,{\bf 49},185,193 \\ 
WASP-80b & 12.62\pm0.35 & 0.56\pm0.03 & 0.99\pm0.02 & 0.000^{+0.020}_{-0.000} & 14.0\pm14.0 & \multicolumn2c{$-$} & \multicolumn2c{$-$} &  9,{\bf 194} \\ 
WASP-84b & 21.70\pm0.72 & 0.69\pm0.03 & 0.98\pm0.03 & 0.000^{+0.077}_{-0.000} & 0.3\pm1.7 & 71.4^{+13.7}_{-7.1} & 18.6^{+4.4}_{-15.7} &  9,{\bf 195},2 \\ 
WASP-85b & 8.97\pm0.32 & 1.26\pm0.07 & 1.24\pm0.03 & \multicolumn2c{$0$} & 0.0\pm14.0 & 90.0^{+0.0}_{-31.3} & 22.1^{+17.4}_{-12.1} &  {\bf 196},2 \\ 
WASP-87b & 3.89\pm0.18 & 2.18\pm0.15 & 1.39\pm0.06 & 0.000^{+0.099}_{-0.000} & 8.0\pm11.0 & \multicolumn2c{$-$} & \multicolumn2c{$-$} &  {\bf 187} \\ 
WASP-94Ab & 7.30^{+0.26}_{-0.22} & 0.45^{+0.04}_{-0.03} & 1.72^{+0.06}_{-0.05} & 0.000^{+0.130}_{-0.000} & 151.0^{+16.0}_{-23.0} & 33.1^{+8.8}_{-7.1} & 116.6^{+9.9}_{-9.1} &  {\bf 197},2 \\ 
WASP-100b & 5.54^{+0.88}_{-0.53} & 2.03\pm0.12 & 1.40^{+0.20}_{-0.30} & 0.000^{+0.044}_{-0.000} & 79.0^{+19.0}_{-10.0} & \multicolumn2c{$-$} & \multicolumn2c{$-$} &  9,{\bf 190} \\ 
WASP-103b & 3.01^{+0.12}_{-0.14} & 1.47^{+0.11}_{-0.13} & 1.65^{+0.05}_{-0.06} & 0.000^{+0.150}_{-0.000} & 3.0\pm33.0 & \multicolumn2c{$-$} & \multicolumn2c{$-$} &  9,{\bf 187} \\ 
WASP-107b & 17.75\pm0.67 & 0.10\pm0.01 & 0.92\pm0.02 & 0.060\pm0.040 & 112.6^{+24.9}_{-20.6} & 28.2^{+40.4}_{--20.0} & 92.6^{+30.7}_{-1.8} &  198,{\bf 199},200 \\ 
WASP-109b & 7.40\pm0.30 & 0.91\pm0.13 & 1.44\pm0.05 & 0.000^{+0.320}_{-0.000} & 99.0^{+10.0}_{-9.0} & \multicolumn2c{$-$} & \multicolumn2c{$-$} &  {\bf 190} \\ 
WASP-111b & 4.55\pm0.27 & 1.83\pm0.15 & 1.44\pm0.09 & 0.000^{+0.100}_{-0.000} & 5.0\pm16.0 & \multicolumn2c{$-$} & \multicolumn2c{$-$} &  {\bf 201} \\ 
WASP-117b & 17.39^{+1.01}_{-0.89} & 0.28\pm0.01 & 1.02^{+0.08}_{-0.07} & 0.300\pm0.026 & 46.9^{+4.8}_{-5.5} & \multicolumn2c{$-$} & \multicolumn2c{$-$} &  9,{\bf 202},203 \\ 
WASP-121b & 3.80^{+0.11}_{-0.11} & 1.18^{+0.06}_{-0.06} & 1.86\pm0.04 & 0.000^{+0.070}_{-0.000} & 87.2^{+0.4}_{-0.4} & 39.8^{+7.4}_{-6.3} & 88.1\pm0.2 &  {\bf 204},205,206 \\ 
WASP-127b & 8.39\pm0.19 & 0.17^{+0.02}_{-0.02} & 1.31^{+0.03}_{-0.03} & \multicolumn2c{$0$} & 128.4^{+5.5}_{-5.6} & \multicolumn2c{$-$} & \multicolumn2c{$-$} &  {\bf 207},208 \\ 
WASP-134b & 17.50\pm0.85 & 1.41\pm0.07 & 0.99\pm0.06 & 0.145\pm0.009 & 43.7\pm9.9 & \multicolumn2c{$-$} & \multicolumn2c{$-$} &  {\bf 209} \\ 
WASP-148b & 19.80\pm1.50 & 0.35^{+0.05}_{-0.05} & 0.80^{+0.02}_{-0.02} & 0.351^{+0.060}_{-0.064} & 8.2^{+9.7}_{-8.7} & \multicolumn2c{$-$} & \multicolumn2c{$-$} &  210,{\bf 211} \\ 
WASP-166b & 11.32\pm0.56 & 0.10\pm0.00 & 0.63\pm0.03 & 0.000^{+0.070}_{-0.000} & 3.0\pm5.0 & 90.0^{+0.0}_{-23.8} & 10.1^{+16.7}_{-5.9} &  {\bf 212},2 \\ 
WASP-167b & 4.38\pm0.14 & 0.00^{+8.00}_{-0.00} & 1.56\pm0.05 & \multicolumn2c{$0$} & 165.0\pm5.0 & 35.0^{+5.2}_{-4.6} & 123.8^{+11.6}_{-10.6} &  {\bf 213},2 \\ 
WASP-174b & 8.79\pm0.15 & 0.33\pm0.09 & 1.44\pm0.05 & \multicolumn2c{$0$} & 31.0\pm1.0 & \multicolumn2c{$-$} & \multicolumn2c{$-$} &  {\bf 214} \\ 
WASP-180b & 8.67\pm0.47 & 0.80\pm0.10 & 1.29\pm0.07 & \multicolumn2c{$0$} & 162.0\pm5.0 & \multicolumn2c{$-$} & \multicolumn2c{$-$} &  {\bf 215} \\ 
WASP-189b & 4.60\pm0.11 & 1.99^{+0.16}_{-0.14} & 1.62\pm0.02 & \multicolumn2c{$0$} & 89.3\pm1.4 & 75.5^{+3.1}_{-2.2} & 87.4^{+5.0}_{-4.0} &  {\bf 216},217,2,218 \\ 
WASP-190b & 8.91\pm0.57 & 1.00\pm0.10 & 1.15\pm0.09 & \multicolumn2c{$0$} & 21.0\pm6.0 & \multicolumn2c{$-$} & \multicolumn2c{$-$} &  {\bf 219} \\ 
XO-2b & 7.79^{+0.36}_{-0.59} & 0.60\pm0.02 & 1.02\pm0.03 & 0.000^{+0.006}_{-0.000} & 7.0\pm11.0 & 64.6^{+19.2}_{-9.6} & 26.5^{+11.8}_{-13.7} &  9,{\bf 220},2,221 \\ 
XO-3b & 6.91\pm0.30 & 11.83\pm0.38 & 1.25\pm0.05 & 0.276^{+0.001}_{-0.001} & 37.3\pm3.0 & \multicolumn2c{$-$} & \multicolumn2c{$-$} &  9,{\bf 222},223,224 \\ 
XO-4b & 7.68\pm0.11 & 1.61\pm0.10 & 1.32^{+0.04}_{-0.03} & 0.000^{+0.004}_{-0.000} & 46.7^{+6.1}_{-8.1} & \multicolumn2c{$-$} & \multicolumn2c{$-$} &  {\bf 225},9 \\ 
XO-6b & 8.08\pm1.03 & 2.01\pm0.71 & 2.08\pm0.18 & \multicolumn2c{$0$} & 20.7\pm2.3 & 62.3^{+17.4}_{-9.3} & 23.3^{+14.3}_{-2.9} &  {\bf 226},2 \\ 
pi Men b & 12.50\pm0.26 & 0.04\pm0.00 & 0.19\pm0.00 & \multicolumn2c{$0$} & 24.0\pm4.1 & 79.7^{+10.3}_{-11.4} & 26.9^{+5.8}_{-4.7} &  227,{\bf 228} \\ 

\enddata
\end{deluxetable*}
\tablecomments{All data is taken from \href{https://www.astro.keele.ac.uk/jkt/tepcat/}{TEPCat}  \citep{Southworth2011} or the following references: 
 1 \citep{Hirano+2020}, 2 \citep{Albrecht+2021}, 3 \citep{Martioli+2020}, 4 \citep{Palle+2020}, 5 \citep{Addison+2021}, 6 \citep{Lacour+2021}, 7 \citep{Chilcote+2017}, 8 \citep{Kraus+2020}, 9 \citep{Bonomo+2017}, 10 \citep{Czesla+2012}, 11 \citep{Bouchy+2008}, 12 \citep{Nutzman+2011}, 13 \citep{Gillon+2010}, 14 \citep{Triaud+2009}, 15 \citep{Gandolfi+2012}, 16 \citep{Hebrard+2011}, 17 \citep{Zhou+2020}, 18 \citep{ChenKipping2017}, 19 \citep{Benatti+2021}, 20 \citep{Yu+2018}, 21 \citep{Stefansson+2022}, 22 \citep{Biddle+2014}, 23 \citep{Bourrier+2018}, 24 \citep{Johnson+2008}, 25 \citep{Loeillet+2008}, 26 \citep{Winn+2007}, 27 \citep{AlbrechtWinnJohnson+2012}, 28 \citep{Mancini+2018}, 29 \citep{Winn+2011}, 30 \citep{Hebrard+2011b}, 31 \citep{Masuda2015}, 32 \citep{Lund+2014}, 33 \citep{Winn+2009}, 34 \citep{Narita+2009_HAT-P-7}, 35 \citep{Moutou+2011}, 36 \citep{Simpson+2011}, 37 \citep{Winn+2010_hatp11}, 38 \citep{Sanchis-OjedaWinn2011}, 39 \citep{Hirano+2011_HAT-P-11}, 40 \citep{Sanchis-Ojeda+2011}, 41 \citep{Deming+2011}, 42 \citep{Hardy+2017}, 43 \citep{Winn+2010_hat-p-13}, 44 \citep{Fulton+2013}, 45 \citep{Esposito+2014}, 46 \citep{Esposito+2017}, 47 \citep{Johnson+2011}, 48 \citep{Mancini+2015b}, 49 \citep{Johnson+2017}, 50 \citep{Zhou+2016}, 51 \citep{Zhou+2019}, 52 \citep{Mohler-Fischer+2013}, 53 \citep{Addison+2014}, 54 \citep{Zhou+2019b}, 55 \citep{Bello-Arufe+2022}, 56 \citep{Dalba+2020}, 57 \citep{KnudstrupAlbrecht2022}, 58 \citep{Sedaghati+2021}, 59 \citep{Dalal+2019}, 60 \citep{Narita+2009_HD17156}, 61 \citep{Narita+2008_HD17156}, 62 \citep{Cochran+2008}, 63 \citep{Barbieri+2009}, 64 \citep{Dai+2020}, 65 \citep{Mann+2020}, 66 \citep{Hebrard+2010}, 67 \citep{Moutou+2009}, 68 \citep{Pont+2009}, 69 \citep{Winn+2009_HD80606}, 70 \citep{Zhou+2018}, 71 \citep{Wolf+2007}, 72 \citep{Cegla+2016}, 73 \citep{CollierCameron+2010}, 74 \citep{Winn+2006}, 75 \citep{Santos+2020}, 76 \citep{Queloz+2000}, 77 \citep{Winn+2005}, 78 \citep{Heitzmann+2021}, 79 \citep{Rizzuto+2020}, 80 \citep{Stefansson+2020}, 81 \citep{Gaidos+2020}, 82 \citep{Santerne+2016}, 83 \citep{Hirano+2016}, 84 \citep{Rice+2021}, 85 \citep{Korth+2019}, 86 \citep{Brahm+2018}, 87 \citep{Wang+2021}, 88 \citep{Hjorth+2019}, 89 \citep{Hjorth+2021}, 90 \citep{Siverd+2012}, 91 \citep{Damasso+2015}, 92 \citep{Bieryla+2015}, 93 \citep{Wyttenbach+2020}, 94 \citep{Ahlers+2020b}, 95 \citep{Stephan+2022}, 96 \citep{Borsa+2019}, 97 \citep{Zhou+2016b}, 98 \citep{Siverd+2018}, 99 \citep{Johnson+2018}, 100 \citep{Rodriguez+2020}, 101 \citep{Jenkins+2010}, 102 \citep{Wang+2018}, 103 \citep{HowarthMorello2017}, 104 \citep{BarnesLinscottShporer2011}, 105 \citep{Johnson+2014}, 106 \citep{Herman+2018}, 107 \citep{Desert+2011}, 108 \citep{AlbrechtWinnMarcy+2013}, 109 \citep{Campante+2016}, 110 \citep{Benomar+2014}, 111 \citep{Sanchis-Ojeda+2012}, 112 \citep{Chaplin+2013}, 113 \citep{Huber+2013}, 114 \citep{Sanchis-Ojeda+2013}, 115 \citep{Hirano+2012}, 116 \citep{Marcy+2014}, 117 \citep{KamikaBenomarSuto+2019}, 118 \citep{VanEylen+2014}, 119 \citep{Santerne+2014}, 120 \citep{Quinn+2015}, 121 \citep{Masuda2017}, 122 \citep{Bourrier+2015}, 123 \citep{Talens+2017b}, 124 \citep{Hooton+2021}, 125 \citep{Lund+2017}, 126 \citep{Talens+2018}, 127 \citep{Rodriguez+2019}, 128 \citep{Hjorth+2019_MASCARA-3}, 129 \citep{Dorval+2020}, 130 \citep{Ahlers+2020}, 131 \citep{Anderson+2018c}, 132 \citep{Covino+2013}, 133 \citep{MocnikSouthworthHellier2017}, 134 \citep{Dong+2022}, 135 \citep{Subjak+2022}, 136 \citep{Rodriguez+2021}, 137 \citep{Cabot+2021}, 138 \citep{Wong+2021}, 139 \citep{Zhou+2021}, 140 \citep{Newton+2021}, 141 \citep{Carmichael+2021}, 142 \citep{Wirth+2021}, 143 \citep{Hirano+2020b}, 144 \citep{Narita+2007}, 145 \citep{Winn+2008}, 146 \citep{Narita+2010}, 147 \citep{Johnson+2022}, 148 \citep{David+2019}, 149 \citep{Gaidos+2022}, 150 \citep{Miller+2010}, 151 \citep{Simpson+2010}, 152 \citep{Tripathi+2010}, 153 \citep{Oshagh+2013}, 154 \citep{Triaud+2010}, 155 \citep{Tregloan-Reed+2015}, 156 \citep{Gillion+2009}, 157 \citep{AlbrechtWinnButler+2012}, 158 \citep{Bourrier+2017}, 159 \citep{Queloz+2010}, 160 \citep{Brothwell+2014}, 161 \citep{Johnson+2009}, 162 \citep{Joshi+2009}, 163 \citep{Brown+2012}, 164 \citep{Anderson+2011_WASP-17}, 165 \citep{Bayliss+2010}, 166 \citep{Anderson+2010}, 167 \citep{Tregloan-reedSouthworthTappert2012}, 168 \citep{Hellier+2011}, 169 \citep{Sedaghati+2021_WASP-19}, 170 \citep{Anderson2015}, 171 \citep{Chen+2020}, 172 \citep{Anderson+2011}, 173 \citep{MocnikHellierAnderson2017}, 174 \citep{Triaud+2013}, 175 \citep{Brown+2012b}, 176 \citep{Johnson+2015}, 177 \citep{CollierCameron+2010_WASP-33}, 178 \citep{Watanabe+2022}, 179 \citep{Southworth+2016}, 180 \citep{Neveu-VanMalle+2016}, 181 \citep{Sanchis-Ojeda+2015}, 182 \citep{Hebrard+2013}, 183 \citep{Mancini+2017}, 184 \citep{Triaud+2017}, 185 \citep{Brown+2017}, 186 \citep{Garhart+2020}, 187 \citep{Addison+2016}, 188 \citep{Casasayas-Barris+2017}, 189 \citep{Smith+2012}, 190 \citep{Addison+2018}, 191 \citep{Luque+2020}, 192 \citep{Ehrenreich2020}, 193 \citep{Addison+2013}, 194 \citep{Triaud+2015}, 195 \citep{Anderson+2015}, 196 \citep{Mocnik+2016}, 197 \citep{Neveu-VanMalle+2014}, 198 \citep{Piaulet+2021}, 199 \citep{Rubenzahl+2021}, 200 \citep{DaiWinn2017}, 201 \citep{Anderson+2014}, 202 \citep{Carone+2021}, 203 \citep{Lendl+2014}, 204 \citep{Bourrier+2020}, 205 \citep{Delrez+2016}, 206 \citep{Borsa+2021_WASP-121}, 207 \citep{Allart+2020}, 208 \citep{Cristo+2022}, 209 \citep{Anderson+2018}, 210 \citep{Hebrard+2021}, 211 \citep{Wang+2021_wasp148}, 212 \citep{Hellier+2019}, 213 \citep{Temple+2017}, 214 \citep{Temple+2018}, 215 \citep{Temple+2019}, 216 \citep{Anderson+2018_wasp189}, 217 \citep{Lendl+2020}, 218 \citep{Deline+2022}, 219 \citep{Temple+2019b}, 220 \citep{Damasso+2015b}, 221 \citep{Narita+2011}, 222 \citep{Hirano+2011c}, 223 \citep{Hebrard+2008}, 224 \citep{Winn+2009_X03}, 225 \citep{Narita+2010b}, 226 \citep{Crouzet+2017}, 227 \citep{Hatzes+2022}, 228 \citep{Kunovac-Hodzic+2021}
}

%\bibliographystyle{aasjournal-hyperref}
%\bibliography{refs,refs_tools,refs_data}

\end{document}